\documentclass[apj,twocolumn]{emulatearxiv}
\usepackage{amsmath}

\begin{document}

\title{Explosive Nucleosynthesis in Near-Chandrasekhar-mass White Dwarf Models for Type Iax Supernovae: Dependence on Model Parameters}

\author{Shing-Chi Leung\thanks{Email address: scleung@caltech.edu}}
 
\affiliation{Kavli Institute for the Physics and 
Mathematics of the Universe (WPI),The University 
of Tokyo Institutes for Advanced Study, The 
University of Tokyo, Kashiwa, Chiba 277-8583, Japan} 
 
\affiliation{TAPIR, Walter Burke Institute for Theoretical Physics, 
Mailcode 350-17, Caltech, Pasadena, CA 91125, USA} 
 
\author{Ken'ichi. Nomoto\thanks{Email address: scleung@phy.cuhk.edu.hk}}

\affiliation{Kavli Institute for the Physics and 
Mathematics of the Universe (WPI),The University 
of Tokyo Institutes for Advanced Study, The 
University of Tokyo, Kashiwa, Chiba 277-8583, Japan}

\date{\today}

\begin{abstract}

The recently observed diversity of Type Ia supernovae (SNe Ia) has
motivated us to conduct the theoretical modeling of SNe Ia for a wide
parameter range.  In particular, the origin of Type Iax supernovae
(SNe Iax) has been obscure.  Following our earlier work on the
parameter dependence of SN Ia models, we focus on SNe Iax in the
present study.  For a model of SNe Iax, we adopt the currently leading
model of pure turbulent deflagration (PTD) of near-Chandrasekhar
mass C+O white dwarfs (WDs).  We carry out 2-dimensional
hydrodynamical simulations of the propagation of deflagration wave,
which leaves a small WD remnant behind and eject nucleosynthesis
materials.  We show how the explosion properties, such as
nucleosynthesis and explosion energy, depend on the model parameters
such as central densities and compositions of the WDs (including the
hybrid WDs), and turbulent flame prescription and initial flame
geometry.  We extract the associated observables in our models, and
compare with the recently discovered low-mass WDs with unusual surface
abundance patterns and the abundance patterns of some SN remnants.  We
provide the nucleosynthesis yield tables for applications to stellar
archaeology and galactic chemical evolution.  Our results are compared
with the representative models in the literature.


\end{abstract}




\section{Introduction}
\label{sec:intro}

\subsection{Type-Ia Supernovae Physics}

Type Ia supernova (SN Ia) is important in the cosmological and
chemical context. Their standardized light curves with Phillip's
relation \citep{Phillips1993} lead to the use of SNe Ia as standard
candles and the discovery of cosmic acceleration and its implications
of dark energy \citep{Riess1996, Perlmutter1997}.  SNe Ia are also the
major channel for iron-peak element production
\citep[e.g.,][]{Nomoto2017b}. The role of SN Ia can be seen in the
galactic chemical evolution, including the [$\alpha$/Fe] versus [Fe/H]
relation \citep{Nomoto2013} and the [Mn/Fe] versus [Fe/H] relation
\citep{Seitenzahl2013, Leung2019subChand, Kobayashi2019}.


The diversified observational properties of SNe Ia \citep[e.g.,][]{Li2001,
 Foley2016, Taubenberger2017, Jiang2018} suggest the multiple
explosion mechanisms \citep[see, e.g.,][for recent reviews]{Hoeflich2017,
 Roepke2017,GarciaBerro2017,Pakmor2017,Nomoto2017b}.

In our earlier work with 2-dimensional hydrodynamical simulations, we
have studied the turbulent deflagration model with or without
deflagration-detonation transition (DDT) for the near-Chandrasekhar
mass C+O white dwarfs (WDs) \citep{Leung2018Chand} and the double
detonation (DD) models for both sub-Chandrasekhar mass WDs and
near-Chandrasekhar mass WDs \citep{Leung2019subChand}.  We have shown
how the explosion properties, such as nucleosynthesis, explosion
energy, and asphericity, depend on the model parameters of the flame
and WDs.

In the present work, we focus on Type Iax supernovae (SNe Iax)
\citep[e.g.,][for a review]{Jha2017}.  A few well observed SNe Iax are
SN 2002cx \citep[e.g.][]{Li2003}, SN 2005hk
\citep[e.g.,][]{Phillips2007}, SN 2008ha \citep[e.g.,][]{Foley2009}
and SN 2014dt \citep{Kawabata2018}.
SNe Iax are peculiar SNe Ia, having lower luminosities as well as
lower ejecta velocities and masses than normal SNe Ia, Moreover, SNe
Iax show quite a large variation of light curve and ejecta properties.
SNe Iax are peculiar but may not be rare, as indicated by its
statistics.  It contributes to $\sim 10\%$ of the SN Ia population
\citep{Li2001}.

Among the various proposed models, we adopt a pure deflagration model,
which has been shown to be a promising model to explain the main
properties of SNe Iax \citep[e.g.,][]{Branch2004,Phillips2007,
  Sahu2008,Foley2009,Jordan2012,Kromer2013a,Fink2014}, The outcome of
the pure deflagration depends mainly on the subsonic flame speed
\citep[e.g.,][]{Nomoto1976,Nomoto1984,Gamezo2004,Roepke2007a,
  Jordan2012,Fink2014}, and the weak explosions with low kinetic
energy and small production of $^{56}$Ni can be consistent with the
observed properties of SNe Iax.

In our earlier works on the hydrodynamics and nucleosynthesis of SNe Ia
\citep{Leung2018Chand,Leung2019subChand}, we have shown that the
abundance pattern of nucleosynthesis yields are important to constrain
the supernova model from the light curves and spectra.  In the present
study, we perform 2-dimensional (2-D) simulations of PTD and calculate
nucleosynthesis for near-Chandrasekhar mass WDs with various
parameters such as composition (C+O and C+O+Ne) and central densities
of WDs and flame physics such as initial flame geometry.

We provide tables of our nucleosynthesis yields, and compare them with
the unusual abundance patterns of some WDs and SN remnants to infer
their origins.  
Our results could be important to the chemical evolution of some
dwarf spheroidal galaxies (dSphs).
In \cite{Kobayashi2015}, the SN Iax model can be an explanation to the
early drop in [O/Fe] and the early rise in [Mn/Fe] relations.

In section \ref{sec:methods}, we describe the progenitor stars and
WDs, and briefly review the methods and input physics to calculate the
PTD.  In section 3, we present typical hydrodynamical and
thermodynamical properties of the PTD models.  We investigate how the
chemical yields of SNe Iax change with the model parameters.
We compare our yields with some metal-enriched low-mass WDs and SNRs.
and discuss its application to chemical evolution of galaxies.

\section{Initial Models and Input Physics}
\label{sec:methods}

\subsection{Initial Models}

We first construct an isothermal WD in hydrostatic equilibrium with a
central density $\rho_c$, temperature $T_{{\rm ini}}$ and metallicity
$Z$.  In the equation of state, Coulomb effects are taken into account.
The characteristic model described in the next section assumes a
C+O WD of mass $1.37 ~M_{\odot}$ with a central density $3 \times 10^9$
g cm$^{-3}$ and solar metallicity. In the simulation
solar metallicity is represented by the characteristic isotope $^{22}$Ne
with the mass fraction of $X(^{22}$Ne) = 0.025 \citep{Nomoto1984,Leung2018Chand}.
The composition is different from the 
detailed solar composition \citep{Asplund2009}.
However, we remark that the nucleosynthesis results are insensitive
to the exact representation because the nucleosynthesis depends 
primarily on the electron mole fraction $Y_{\rm e}$
of the burnt matter, which is 
dominated by electron capture after the matter is 
swept by the deflagration wave. The initial electron mole fraction
and composition
play much smaller roles unlike the canonical deflagration
with deflagration-detonation transition due to the lack of
nucleosynthesis by detonation at low densities.

Such a C+O WD is assumed to be formed in the single degenerate channel
of binary evolution \citep[e.g.,][]{Nomoto1982a,Nomoto2017b,Nomoto2018}.
Thus the central density of the WD mainly depends on the accretion rate.
It also depends on the spin of the white dwarf
\citep[e.g.,][]{Benvenuto2015} for which the central density at the
deflagration is as high as 5 - 6 $\times 10^9$ g cm$^{-3}$ and even
higher depending on the time scale of the angular momentum loss.
Thus we parametrized the central density.

We also consider possible hybrid WDs as demonstrated by
\cite{Denissenkov2013}.  Here carbon is ignited in the outer layer of
the C+O core of near $\sim 10 ~M_\odot$ star during its AGB phase
phase \citep[e.g.,][]{Nomoto1984b}.  If
the carbon flame propagates through the center, the C+O core is
converted into an O+Ne+Mg core \citep[e.g.,][]{Nomoto1987,Nomoto2017a}.  However,
if the overshooting of convection across the carbon-flame down to the
C+O layer is large, it would reduce the nuclear fuel to prevent the
carbon-flame from propagation to the center 
\cite[see, however, e.g.,][]{Lecoanet2016,Brooks2017}.  Then the degenerate core is
composed of C+O in the central region and O+Ne+Mg in the outer layer,
where the composition changes at a designed transition mass $M_{{\rm
    core}}$.

When such a core becomes a WD, it is called a hybrid WD.  When the
mass of the hybrid WD becomes close to the near-Chandrasekhar limit,
carbon is ignited in the central region which produces a convection
zone. During the simmering phase, the convection would mix the central
C+O core and the outer O+Ne+Mg layer before the deflagration is
initiated.  It produces a WD made of C+O+Ne mixture.

In the present study, we assume that such hybrid WD is formed and the
C+O-rich core is completely mixed with the outer O+Ne-rich envelope
for simplicity.  After mixing during the simmering phase, we assume
that the WD has a uniform composition with the mass fractions of
$^{12}$C, $^{16}$O, and $^{20}$Ne are 0.2, 0.5, and 0.3 for simplicity
\citep{Kromer2015} and in view of the uncertainties in the
overshooting parameter. The effect of metallicity is 
included by combining the above composition with $^{22}$Ne as
the key isotope for metallicity.

If the carbon burning reaches the center, the WD becoems an
oxygen-neon-magnesium (O+Ne+Mg) WD.  These WDs undergo electron
capture and are very likely collapse to form neutron stars
\citep{Zha2019,Leung2019DMAIC}.  
For a possible case of partial mass
ejection, the nucleosynthesis yields are given in Appendix C.

\subsection{Methods and Input Physics}

For the adopted initial models of the WDs, we perform two-dimensional
hydrodynamics simulations of the propagation of the PTD by specifying the
initial deflagration structure by hand.

Deflagration
\citep{Nomoto1976,Nomoto1982b,Timmes1992a,Niemeyer1997,Woosley1997} is
the nuclear flame, where thermonuclear runaway takes place with a
timescale shorter than the dynamical timescale.  The flame propagates
at a sub-sonic speed by energy transport across the flame due to
electron conduction and convection as described below
\citep[e.g.,][]{Nomoto2017a,Nomoto2017b,Leung2017MmSAI,Leung2019PASA,Leung2019ECSN}.

Deflagration is subject to hydrodynamical instabilities including
Rayleigh-Taylor (RT) instabilities, Kelvin-Helmholtz (KH)
instabilities and Landau-Darrieus (LD) instabilities present during
its propagation \citep[e.g.,][]{Timmes1992b, Livne1993b,
  Niemeyer1995b,Roepke2004a,Roepke2004b,Bell2004a,Bell2004b}.  The
flame width ($\sim 10^{-6} - 10^{-3}$ cm) is in general much smaller
than that resolved by simulation
\citep[e.g.,][]{Timmes1992a,Niemeyer1995a}.  Special flame front
tracking technique, such as the level-set method, \citep{Osher1988},
is necessary for a consistent description of the interaction between
flame and the dynamics \citep{Reinecke1999a}. Sub-grid scale
turbulence model is often necessary \citep{Clement1993,Schmidt2006b}.
The flame structure is coupled with turbulence, where the turbulent
motion emerges in the length scale down to the Kolmogorov scale. Eddy
motions can alter the deflagration front by fluid advection above the
Gibson's scale. Below that scale, the irregularities are polished
\citep[e.g.,][]{Niemeyer1995a,Jackson2014}.  Early multi-dimensional
realizations have been done in
\cite[e.g.,][]{Reinecke1999b,Reinecke2002a,Reinecke2002b,
  Roepke2004a,Roepke2004b,Roepke2005a,Roepke2005b,Roepke2006a,Roepke2006b}.

The turbulent deflagration is tracked by the level-set method
\citep{Osher1988,Reinecke1999a} with the sub-grid turbulence scheme
given in \cite{Clement1993,Niemeyer1995a}.  The turbulent flame
formula used the analytical results in \cite{Pocheau1994}.  The
connections between the local turbulence strength and the effective
turbulent flame propagation speed are taken from \cite{Schmidt2006b},
but with variations shown in \cite{Hicks2015}. DDT is artificially
suppressed throughout the simulation. 

Our simulation code is a general hydrodynamics code embedded with
supernova physics \citep{Leung2015a}. This code has been applied to
previous SN Ia parameter surveys for the near-Chandrasekhar mass WDs
\citep{Leung2017,Leung2018Chand} and also sub-Chandrasekhar mass WDs
\citep{Leung2019subChand}. The code has been recently extended to
study other types of supernova, including the electron capture
supernova \citep{Leung2019ECSN} and jet-driven supernova
\citep{Tsuruta2018}.  The simulation code uses a simplified 7-isotope
network \citep{Timmes2000a} to trace the isotopic abundances and a
three-step burning network \citep{Townsley2007} to mimic the nuclear
reactions in deflagration.

In the three-step burning scheme, the fast reaction of carbon-burning
is controlled by the level-set method. The advanced burning
and O-burning are set to be dependent on the corresponding
burning timescales \citep{Calder2007}. For matter with a sufficiently
high density ($> 10^8$ g cm$^{-3}$) and temperature ($> 5 \times 10^9$ K)
such that the matter is burnt in nuclear statistical equilibrium (NSE), 
electron captures take place. The corresponding rates for the 
matter in the NSE are computed by summing up the individual rates
of many isotopes included in the nuclear reaction network
used by the post-processing. Updated weak interaction rates are obtained
from the literature including \cite{Nabi2004,Seitenzahl2009}.
We refer the further details in Appendix A. 

Pure deflagration cannot completely unbind the star
if the flame is quenched early by the WD expansion \citep{Nomoto1976,Livne1993a}.
Its sub-sonic nature provides sufficient time for 
electron capture to take place before the WD expands \citep[e.g.,][]{Iwamoto1999,Seitenzahl2009}, 
which is a key to produce neutron-rich isotopes in the ejecta.
Such conditions are hardly achieved by other types of SNe.

After the hydrodynamical simulations, we use the thermodynamics
history of the tracer particles to reconstruct the detailed
nucleosynthesis \citep{Travaglio2004,Seitenzahl2009,Townsley2016}.

\section{Characteristic Model for SN Iax}
\label{sec:benchmark}

We first examine the typical behaviour of the SN Iax based on the 
C+O WD as described in Section \ref{sec:methods}.  Then, we examine the
stellar properties, thermodynamics, energetics and chemical
properties.  A central flame $c3$ (corresponding to three-finger, see
\cite{Reinecke1999b} for illustrations) is placed at the beginning.

\subsection{Thermodynamics}

\begin{figure}
\centering
\includegraphics*[width=8cm,height=5.7cm]{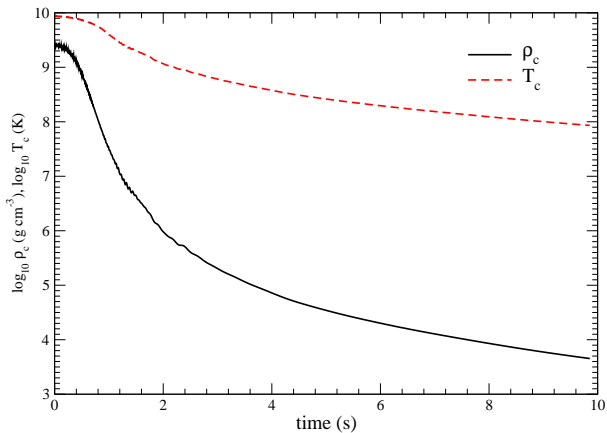}
\caption{Central density (black dashed line) 
and central temperature (red dashed line) against
time for the characteristic model.}
\label{fig:rhoc_benchmark_plot}
\end{figure}

In Figure \ref{fig:rhoc_benchmark_plot} we plot the 
central density (black dashed line) and the central
temperature (red dashed line) of the characteristic model
as a function of time. After the core is burnt, 
the core remains static for the first 0.2 s. Afterwards, it 
expands and the central density drops by
2 orders of magnitude in the first 2 s. Then 
the core expansion slows down. It reaches an 
asymptotic value $\sim 10^{4.5}$ g cm$^{-3}$ 
beyond 8 s, showing that the core stops
expanding because it has transferred all its
momentum to the outer material, which has a 
relatively lower density and hence smaller inertia
to be ejected. The temperature shows a similar
evolution but with a difference that the core 
temperature continues to drop even at the end
of simulation. The temperature is
too low for any important nuclear reactions to take place
after it has left NSE (defined at 
$T > 5 \times 10^9$ K), which occurs at $\sim 1$ s
after explosion. 

\begin{figure}
\centering
\includegraphics*[width=8cm,height=5.7cm]{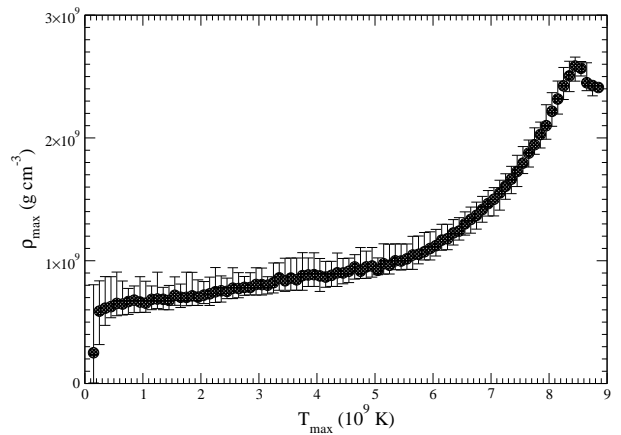}
\caption{The maximum density against
maximum temperature for the characteristic model
using the thermodynamics history of the 
tracer particles. The error bars and the 
data points correspond to the range of 
density and the mean density of the tracer
particles obtained in the thermodynamics
history.}
\label{fig:traj_stat_benchmark_plot}
\end{figure}

To further clarify the effects of nuclear reactions in the 
star, we plot in Figure \ref{fig:traj_stat_benchmark_plot} the
tracer particle summary. In this figure, for each tracer
particle we search for its maximum density and 
temperature achieved in the simulations, and then
we bin them into different temperature ranges.  
The variety of particle maximum density within the same temperature
range is represented 
be the "error bars" in the figure. Notice that
for SN Iax, the maximum $\rho_c$ and $T_c$
are obtained at the same time because there is 
no shock wave triggered in the event.

There are two groups of the particles
developed in the simulation. The first group
corresponds to the particle which is directly burnt,
this consists of particles with $\rho_{{\rm max}}$ 
$> \sim 6 \times 10^9$ K. There is a monotonic relation between $\rho_{{\rm max}}$
and $T_{{\rm max}}$ and the fluctuation is very small.
The star remains close to static and the asymmetry
of the deflagration wave is not large enough to 
create significant time difference for burning matter
near the same radius. 
The second group corresponds to the particles being
burnt by the deflagration wave but after the star
has expanded and the flame becomes aspherical. 
Matter with the same $\rho_{{\rm max}}$ can have
different $T_{\rm max}$. This means that they have the 
same initial density but they experience different
levels of expansion before the deflagration arrives.  

\subsection{Energetics}

\begin{figure}
\centering
\includegraphics*[width=8cm,height=5.7cm]{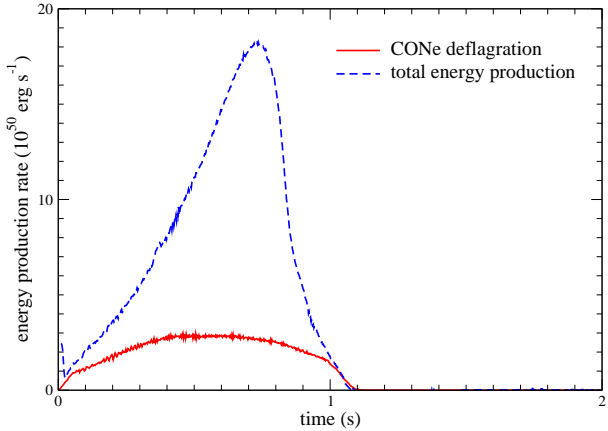}
\caption{The energy production rate (black solid line) and its 
component (red dashed line) against time for the 
characteristic model. Here, the energy production rate by CONe deflagration is extracted
for comparison.}
\label{fig:lumin_benchmark_plot}
\end{figure}

\begin{figure}
\centering
\includegraphics*[width=8cm,height=5.7cm]{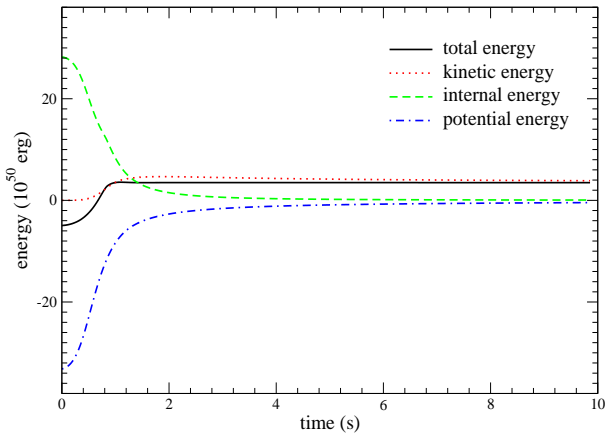}
\caption{The total energy (black solid line), kinetic energy red dotted line),
internal energy (green dashed line) and gravitational
energy (blue long dashed line) against time for the 
characteristic model.}
\label{fig:energy_benchmark_plot}
\end{figure}

In Figure \ref{fig:lumin_benchmark_plot} we
plot the energy production rate against time
for the characteristic model. The energy production
rate is defined by $\Delta Q/\Delta t$, where
$\Delta Q$ is the amount of energy gained 
by the system through nuclear reaction at that
current step with a time step size $\Delta t$. 
The amount of energy, as described in Section
\ref{sec:methods}, is done by the level-set methods
coupled with our simplified network.

In the first 0.4 s, the system releases energy
mainly by nuclear deflagration. The energy production
rate is low, $\sim 10^{50}$ erg s$^{-1}$. After that
from 0.4 s to 1 s, the energy production switches
to advance burning and NSE burning when the density at the 
deflagration front becomes low ($\sim 10^7$ g cm$^{-3}$). 
The energy production rate 
is high and reach $\sim 10^{51}$ erg s$^{-1}$
at maximum. At that time nuclear deflagration 
no longer produces any observable amount of energy. 
We remark that different from standard SNe Ia,
there can be enhancement from mixing of fresh 
$^{16}$O and $^{20}$Ne into the ash when the 
ash expands. 

In Figure \ref{fig:energy_benchmark_plot} we also
plot the energy evolution similar to Figure 
\ref{fig:lumin_benchmark_plot}. 
The total energy, kinetic, internal and potential
energy are plotted.  

Similar to the energy production rate, the total
energy quickly rises at the beginning and reaches
its equilibrium value at $\sim 1$ s after the 
flame has propagated. When the expansion quenches
the flame and cools down the ash, 
the total energy no longer changes. The kinetic energy 
shows a similar behaviour but with a small bump about 1.3 s. 
This is because during the expansion of the flame,
it creates non-local acceleration of 
matter, especially the hot matter. 

The total internal energy on the contrary
is dominated by the initial internal energy.
It constantly decreases, showing that the 
star expands and loses energy through its 
expansion work done. Unlike SN Ia, it has no
bump in its evolution, which means that the flame
does not produce any significant shock compression
to the matter, including the low-density matter 
on the surface. This is consistent with the 
idea that the deflagration is sub-sonic.
The system has always sufficient time
to adjust its structure to accommodate the 
energy input by nuclear burning. 
It reaches its asymptotic energy $\sim 0$
beyond 3 s. 

For the gravitational energy, since there is 
no contraction during the whole evolution, 
it first quickly rises in the first 2 s, and then 
it approaches its asymptotic value near zero, slowly
as the star expands. But it reaches its asymptotic
value much slower than the internal energy,
showing that the system is expanding much slower
than ordinary SNe Ia.

\subsection{Flame Propagation}

\begin{figure*}
\centering
\includegraphics*[width=8cm,height=6cm]{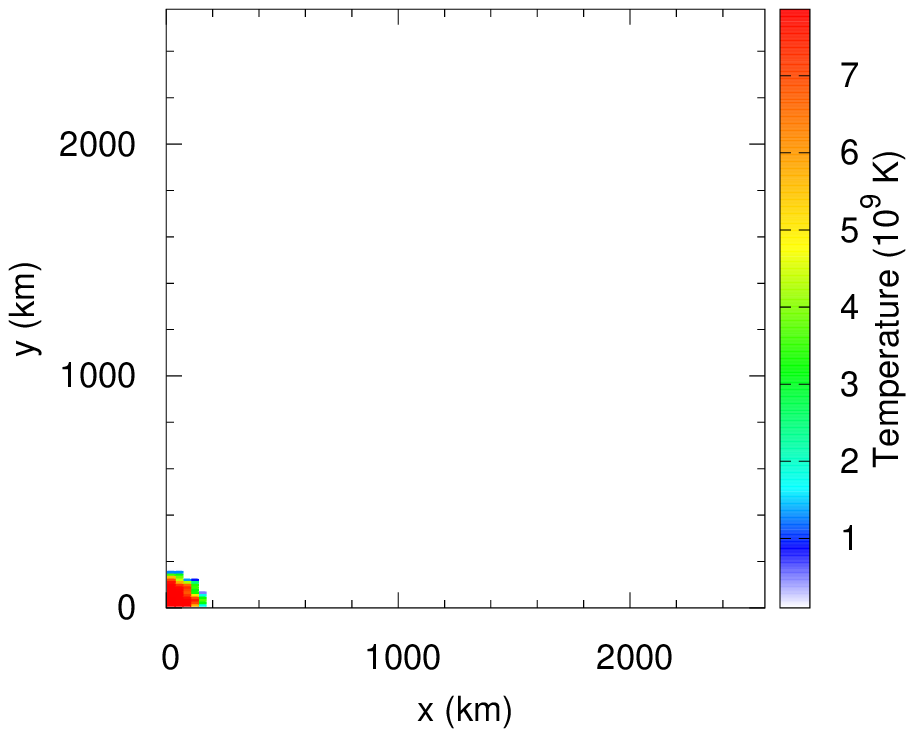}
\includegraphics*[width=8cm,height=6cm]{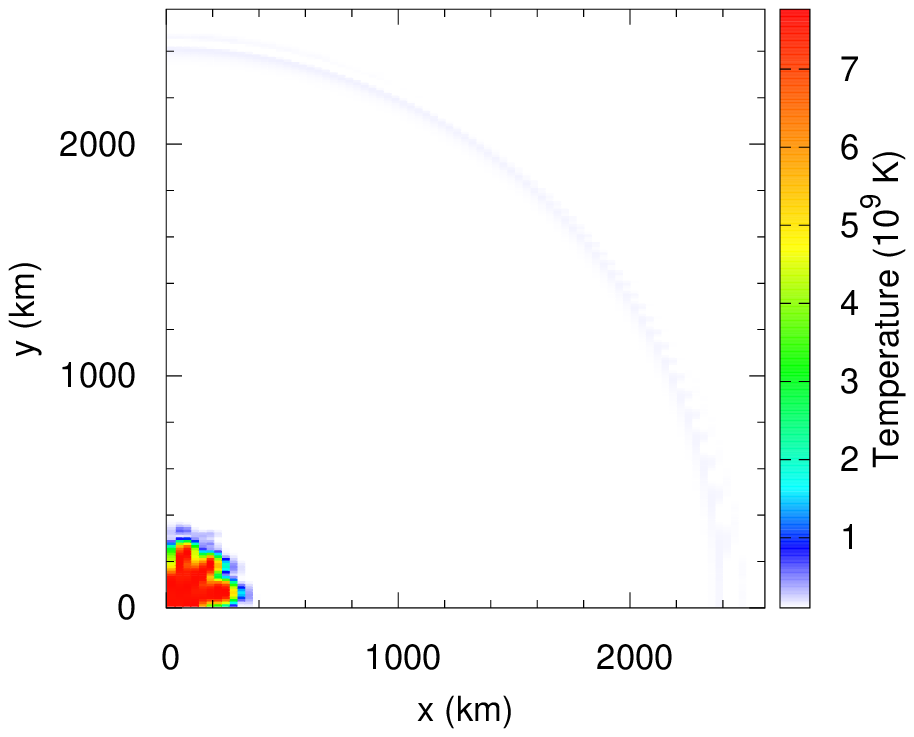}
\includegraphics*[width=8cm,height=6cm]{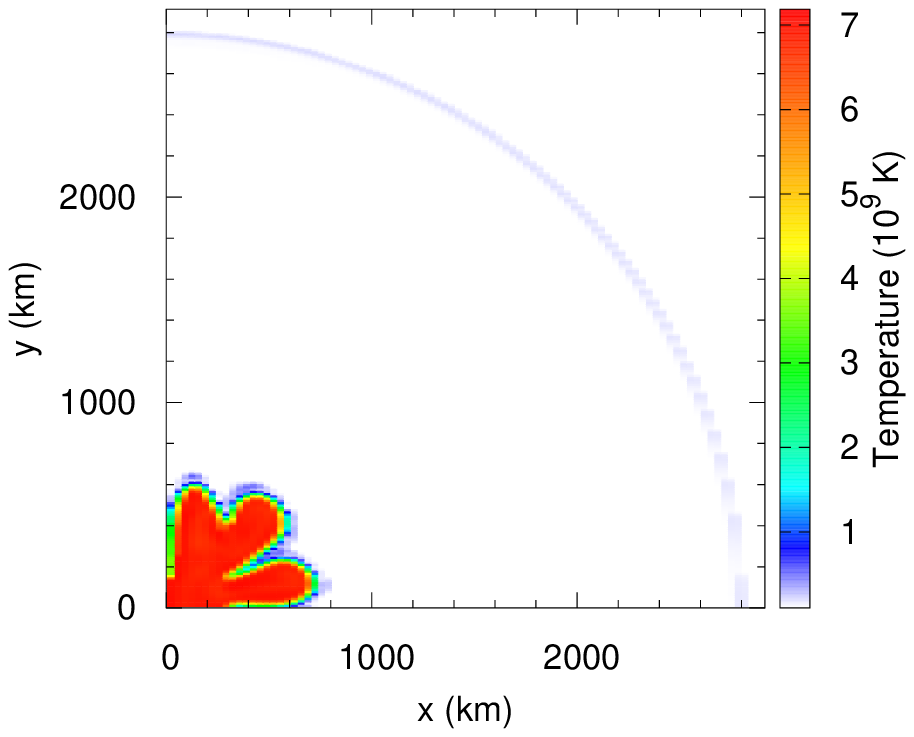}
\includegraphics*[width=8cm,height=6cm]{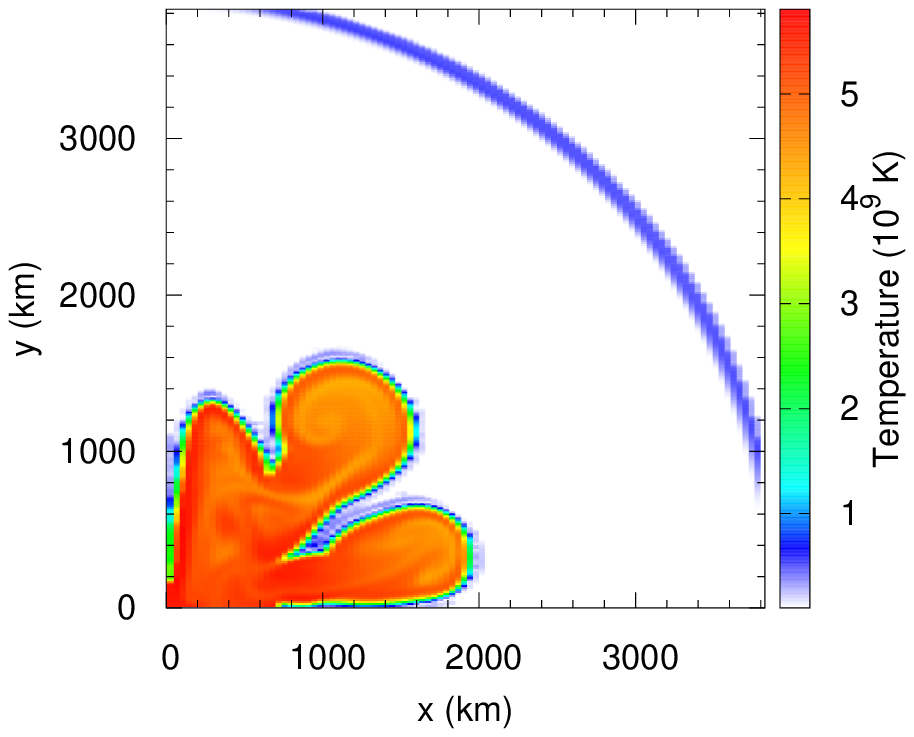}
\includegraphics*[width=8cm,height=6cm]{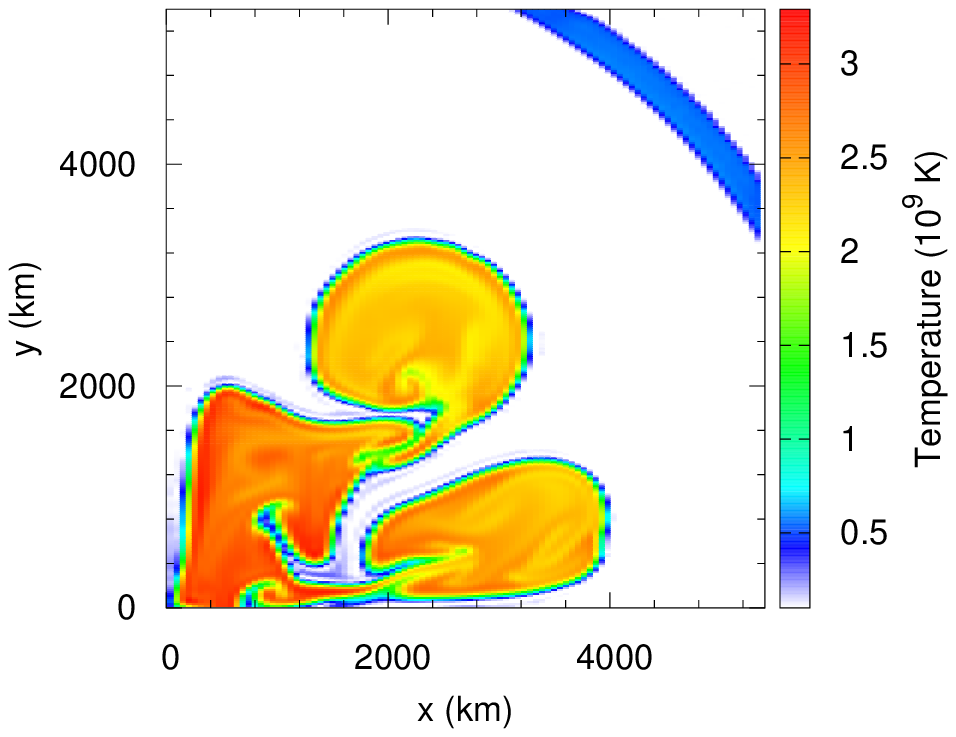}
\includegraphics*[width=8cm,height=6cm]{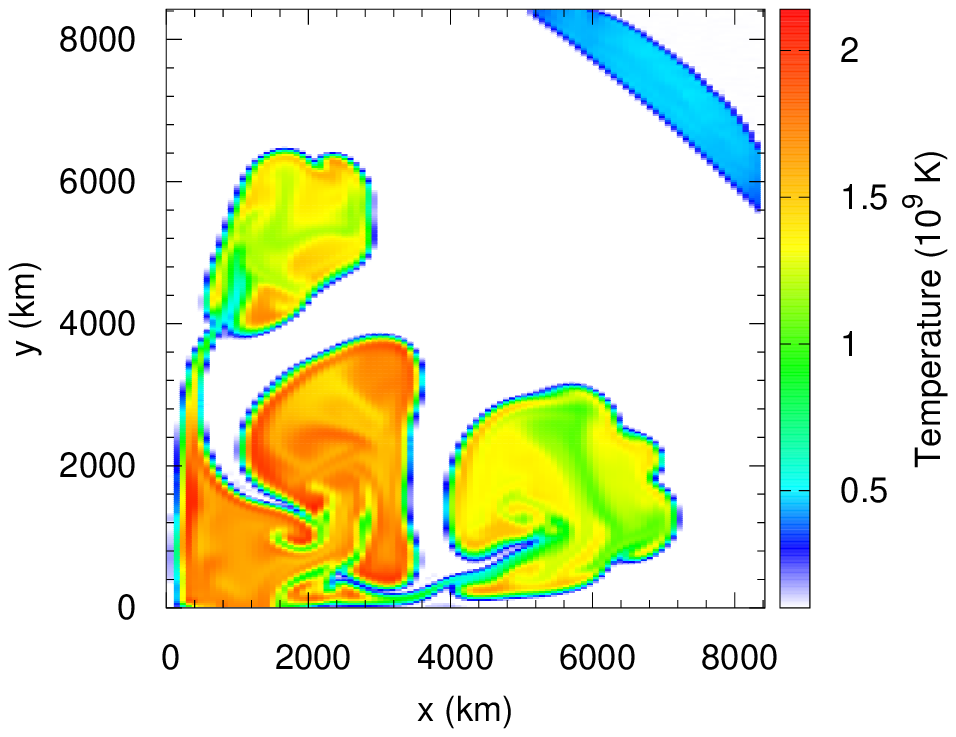}
\caption{The flame structure and the temperature colour 
plots of the characteristic model from the beginning to 
2.5 s after deflagration has started at an interval of 0.5 s.}
\label{fig:flame_benchmark_plot}
\end{figure*}

In Figure \ref{fig:flame_benchmark_plot} we plot the 
flame structure of our representative model from the 
beginning to 2.5 s at an interval of 0.5 s. 
We use the $c3$ flame which we have 
applied in previous SNe Ia surveys.
This flame mimics and promotes the early growth of 
RT instabilities. We note that
putting spherical flame at the center can 
give enhanced flame propagation
along the symmetry axis, while the WD
does not have a preferred direction due to its
static initial profile. 
Therefore, to avoid
the development of such unphysical structure,
we use the $c3$ flame so that 
the off-axis flame development dominates
the growth of flame. 

We also simulate PTD in the hybrid WD which is composed of C+O+Ne matter.
At 0.5 s the flame only burns the innermost $\sim 200$ km
of the star. One of the "finger" features near the 
$y$-axis is suppressed, 
while the other two "fingers" continue to grow
further. The curly colour pattern in its temperature 
distribution demonstrates the turbulent motion inside the ash
at 1.5 s. At that time
the flame has already cools down to about 
$\sim 5 \times 10^9$ K. After that, both NSE burning
and nuclear deflagration no longer 
supply extra energy. The star gradually expands.
The peak temperature reaches $\sim 3 \times 10^9$ K at 2.0 s and $\sim 2 \times 10^9$ K at 2.5 s.
From 2.0 s onward, when the flame expands upwards due to 
its buoyancy, the unburned 
matter floats downward to fill up the space.
The opposite direction of the flow current
creates the KH instability, where the curly structure can be seen along 
the "fingers". On top of the flame, the 
smaller scale "mushroom" shapes emerge as the
Rayleigh-Taylor instabilities. This occurs when the lower density ash creates a pressure inversion with 
the higher density fuel against the gravitational force. 
Also the inverse "mushroom" shape can be seen at the inner
part of the ash, showing the injection of 
fuel into the ash. The colour map of the 
temperature further shows instabilities in the 
smaller length scale inside the ash.

\subsection{Isotopic abundance}

\begin{figure*}
\centering
\includegraphics*[width=8cm,height=5.7cm]{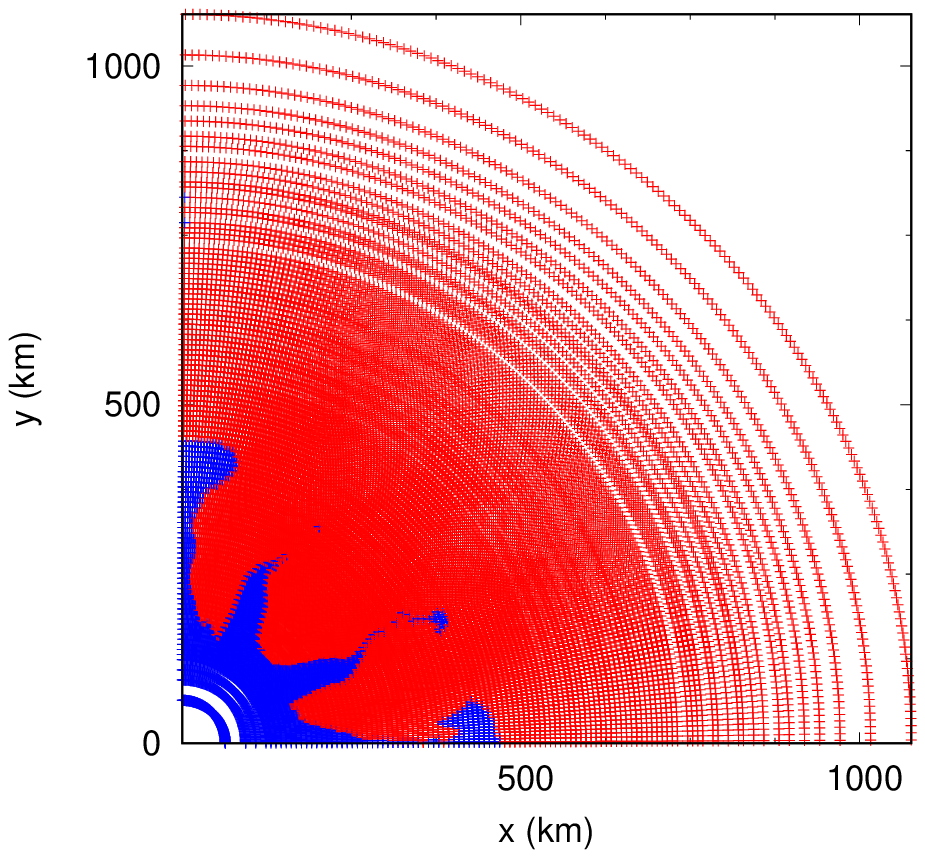}
\includegraphics*[width=8cm,height=5.7cm]{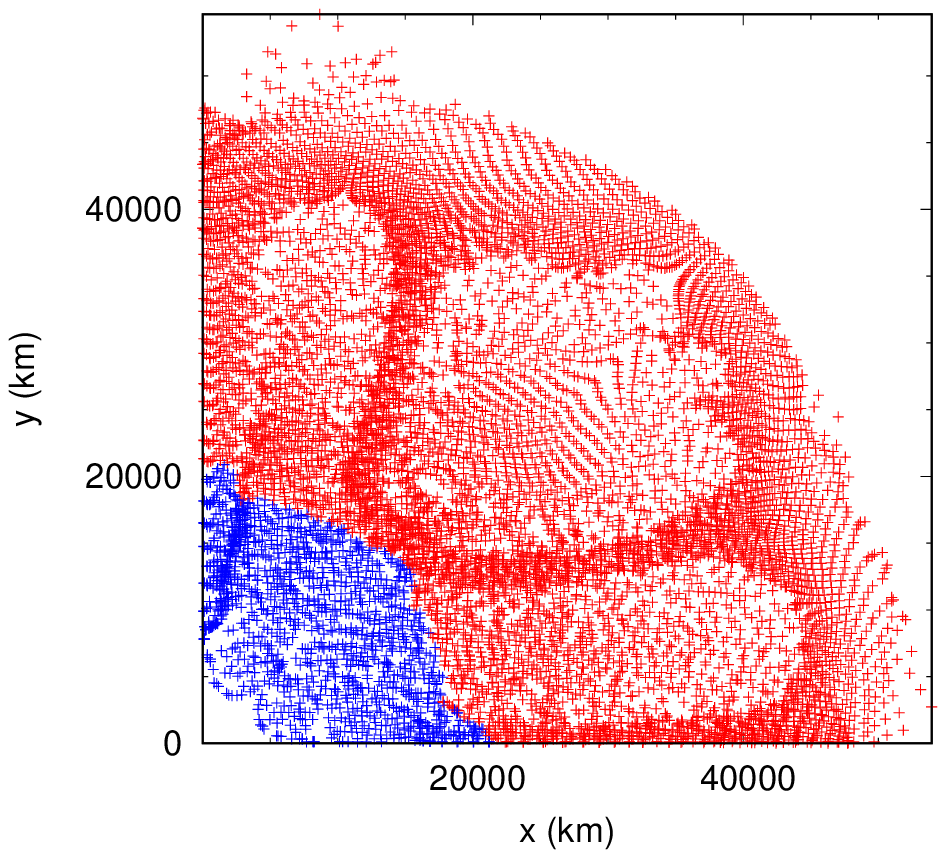}
\caption{The tracer particle distribution for the particles
which can escape (red) and which are gravitationally bounded
(blue). The escape criterion is decided by the final
energy of the tracer particles. The initial particle
distribution (left panel) and the final particle distribution
(right panel) are shown for comparison.}
\label{fig:particlepos0_benchmark_plot}
\end{figure*}

\begin{figure}
\centering
\includegraphics*[width=8cm,height=5.7cm]{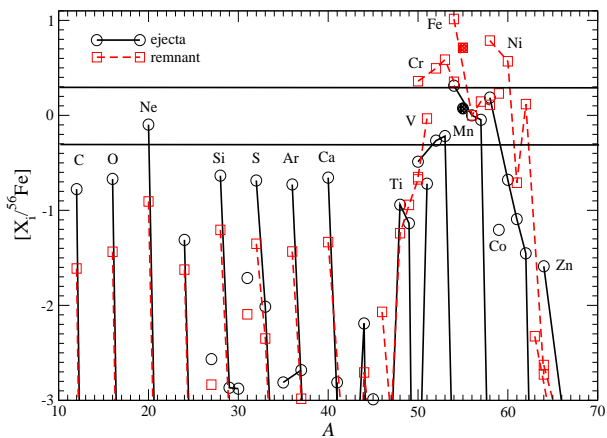}
\caption{$[X_i/^{56}$Fe] against
mass number for the characteristic model. The remnant composition
and ejecta composition are included.}
\label{fig:final_benchmark_plot}
\end{figure}

Unlike normal SNe Ia, the low explosion energy means
that the star is not completely disrupted by the 
nuclear flame. At the beginning, the burnt matter
has the largest momentum which can escape from
the star. However, during its upward motion,
it transferred part of its momentum to the 
matter of lower density in the outer part of
the star. Part of the burnt matter thus becomes
bounded during this momentum transfer.
Instead, the surface matter is expelled.
To clarify which part of the star can be ejected,
and which part remains bounded in the star, 
we use the kinematic properties of the tracer
particles. For each tracer particle, 
we obtain its specific kinetic energy $|\vec{v}|^2 / 2$,
and its current gravitational energy $\phi(\vec{r})$,
from the end of simulations.
We assume that tracer particles can escape
when $|\vec{v}|^2 / 2 + \phi(\vec{r}) > 0$. 
We notice that the tracer particles satisfying 
this relation remain unchanged beyond a few seconds
after explosion. 

In Figure \ref{fig:particlepos0_benchmark_plot} we plot the 
tracer particles which can escape (red) and which are bounded
by its self gravity (blue). We plot the tracer particle distribution
according to its initial position (left panel) and 
its final position (right panel). 
The initial profile demonstrates which part
of the star is being ejected after deflagration. 
Compared with Figure \ref{fig:flame_benchmark_plot},
most of the innermost part of the ash is trapped
in the star. Instead, most material within 200 km for its original 
position is trapped, while the particles between 200 -- 500 km along the diagonal are partially ejected. The trapped matter in this region is consistent
with the flame structure seen in Figure \ref{fig:flame_benchmark_plot}.
We plot the final 
position of the ejected particles in the right panel at 10 s after the simulation. The mixing effects
can no longer be seen. The ejected matter locates at the outermost part of the star while the 
trapped matter falls back to form the remnant. 
The trapped matter is already settled down 
in the innermost 20000 km.
A careful examination at the density of the particles
reveals some differences between the ejected ash and ejected fuel. The ejected ash has a lower
tracer particle density compared to the ejected fuel. This is related to the thermal expansion 
of the ash when it arrives the region filled with the cold fuel. 

In Figure \ref{fig:final_benchmark_plot} we plot 
$[X_i/^{56}$Fe] for the characteristic
model. Here, 
\begin{equation}
[X_i/^{56}{\rm Fe}] = \log_{10} [(X_i/X(^{56}{\rm Fe}))/(X_i/X(^{56}{\rm Fe}))_{\odot}]
\end{equation}
is the mass fraction ratio to $^{56}$Fe of the stable isotopes, 
relative to the solar ratio. 
Here, the stable isotopes means that all typical short lifetime
radioactive isotopes have decayed. After the post-processing nucleosynthesis
yield is obtained, we allows the yield product to decay for $\sim 10^6$ years,
such that most radioactive isotopes $^{56}$Ni ($\sim 8$ days), $^{57}$Ni ($\sim 60$ days), 
$^{59}$Ni (60000 years) have decayed. However, isotopes with a longer 
half life such as $^{27}$Al and $^{60}$Fe may not decayed completely.
Despite that, these isotopes are not mostly produced in SN Ia. 
Since the star is partially disrupted, 
we separate the ejecta and the remnant compositions
for comparison. As shown in previous figures, 
the ejecta are obtained from the tracer particles
which have a positive total energy. 

The ejecta mostly comes from matter
in the surface, where the abundances of O and Ne are abundant. 
It has very low masses of intermediate mass
elements (IMEs), such as Si, S, Ar and Ca. 
The lower part iron-peak elements (IPEs), i.e. Ti, V and 
Cr are also underproduced. On the contrary, 
upper part of IPEs, i.e. 
Mn, Fe and Ni are well produced. Zn is underproduced.

In the remnant, it has a similar pattern 
in O and Ne but is lower by one order of magnitude at $1 - 10 \%$ of 
solar values. 
The abundances of IMEs in the deflagration ash are
also low in the remnant, and they are still underproduced.
The IPEs have a more interesting pattern in the 
remnant. As discussed above, most tracer particles
in the inner core failed to escape from the star
during the momentum transfer process. 
The higher $Y_{\rm e}$ isotopes including $^{54}$Fe
and $^{58}$Ni can be 10 and 8 times higher than 
the solar ratios. 
As a result, the lighter part of IPEs, especially the neutron rich
ones including $^{51}$V, $^{53,54}$Cr
and $^{55}$Mn are higher than the solar ratios.
In particular, 
[$^{55}$Mn/$^{56}$Fe] and [$^{58}$Ni/$^{56}$Fe] can reach $\sim 0.5$ 
and 0.6 respectively. The final remnant
WD has a very different composition from 
standard C+O WD of similar mass ($\sim 0.3 ~M_{\odot}$), where the contamination 
by IMEs and IPEs is significant.

\section{Model Summary}

In this section, we examine the hydrodynamics and nucleosynthesis of 
our SN Iax models using PTD in C+O
WDs and hybrid C+O+Ne WDs.  

In Table \ref{table:models} we tabulate the models computed
in this work, and their corresponding exploding energetics and
global chemical properties. We name each model
according to the model parameters. For example the model 
300-137-1-c3-06 means a C+O WD model with a central 
density $3.00 \times 10^9$ g cm$^{-3}$, 
CO-rich matter of mass of 1.37 $M_{\odot}$, 1 $Z_{\odot}$,
$c3$-flame and C/O $= 0.6$ in mass fraction,
i.e. $X(^{12}$C) $= 0.366$, $X(^{16}$O) $ = 0.609$
and $X(^{22}$Ne) $= 0.025$.
For hybrid C+O+Ne WDs, the second entry is the mass of CO-rich matter
before we mix the composition by hand. 
The last entry does not apply to hybrid C+O+Ne WDs.

\begin{table*}
\begin{center}

\caption{The model parameters of the models studied in this work. 
"Type" corresponds to the classification of the WD being 
a carbon-oxygen (C+O) WD or a hybrid carbon-oxygen-neon (C+O+Ne) WD.
$M$, $M_{{\rm CO}}$, $M_{{\rm ONe}}$, $M_{{\rm Ni}}$, $M_{{\rm ej}}$, 
$M_{{\rm rem}}$ are the mass of the initial WD, the C+O part, the 
O+Ne part, the total $^{56}$Ni produced in the ejecta, the ejecta
mass and the remnant mass in units of $M_{\odot}$. $R$ and $R_{{\rm core}}$
are the radius of the initial WD and the core (if applicable), in units
of km. $E_{{\rm tot}}$ and $E_{{\rm nuc}}$ are the final total energy
and the nuclear energy produced by the deflagration in units of 
$10^{50}$ erg. $Z$ is the metallicity in units of $Z_{\odot}$.
$\rho_c$ and $\rho_{{\rm core}}$ are the initial central density
and core-envelope interface density (if applicable) in units 
of $10^9$ g cm$^{-3}$. Notation "flame" corresponds to the initial
geometry of the flame, including $c3$ (three-"finger" structure) and
the $b1$ (one bubble) structure. "Others" includes setting specific 
to that corresponding type of WD. 
}
\label{table:models}
\begin{tabular}{|c|c|c|c|c|c|c|c|c|c|c|c|c|c|c|}
\hline
Type & Model & $\rho_c$ & $Z$ & $M$ & $M_{{\rm CO}}$ & $M_{{\rm ONe}}$ & flame & $R$ & $E_{{\rm tot}}$ & $E_{{\rm nuc}}$ & $M_{{\rm ej}}$ & $M_{{\rm rem}}$ & $M_{{\rm Ni}}$ & others \\ \hline
CO WD & 050-130-1-c3-1 & 0.50 & 1 & 1.30 & 1.30 & N/A & $c3$ & 3070 & -1.61 & 2.36 & 0.00 & 1.30 & 0.00 & C/O = 1 \\
CO WD & 100-133-1-c3-1 & 1.00 & 1 & 1.33 & 1.33 & N/A & $c3$ & 2580 & 2.23 & 6.71 & 0.92 & 0.41 & 0.23 & C/O = 1 \\
CO WD & 200-135-1-c3-1 & 2.00 & 1 & 1.35 & 1.35 & N/A & $c3$ & 2170 & 3.69 & 8.52 & 1.18 & 0.17 & 0.24 & C/O = 1 \\
CO WD & 300-137-1-c3-1 & 3.00 & 1 & 1.37 & 1.37 & N/A & $c3$ & 1950 & 4.54 & 9.69 & 1.26 & 0.11 & 0.34 & C/O = 1 \\
CO WD & 500-138-1-c3-1 & 5.00 & 1 & 1.38 & 1.38 & N/A & $c3$ & 1710 & 5.13 & 10.5 & 1.29 & 0.09 & 0.32 & C/O = 1 \\  
CO WD & 550-138-1-c3-1 & 5.50 & 1 & 1.38 & 1.38 & N/A & $c3$ & 1670 & 5.81 & 11.0 & 1.30 & 0.08 & 0.31 & C/O = 1 \\ 
CO WD & 600-138-1-c3-1 & 6.00 & 1 & 1.38 & 1.38 & N/A & $c3$ & 1620 & 6.03 & 11.2 & 1.31 & 0.07 & 0.30 & C/O = 1 \\ 
CO WD & 750-139-1-c3-1 & 7.50 & 1 & 1.39 & 1.39 & N/A & $c3$ & 1540 & 6.25 & 11.5 & 1.33 & 0.06 & 0.32 & C/O = 1 \\ 
CO WD & 800-139-1-c3-1 & 8.00 & 1 & 1.39 & 1.39 & N/A & $c3$ & 1500 & 7.51 & 12.7 & 1.34 & 0.05 & 0.31 & C/O = 1 \\ 
CO WD & 900-140-1-c3-1 & 9.00 & 1 & 1.40 & 1.40 & N/A & $c3$ & 1460 & 7.81 & 13.1 & 1.36 & 0.04 & 0.34 & C/O = 1 \\ \hline

CO WD & 100-133-1-b1-1 & 1.00 & 1 & 1.33 & 1.33 & N/A & $b1$ & 2580 & 2.99 & 7.47 & 1.03 & 0.30 & 0.23 & C/O = 1 \\
CO WD & 300-137-1-b1-1 & 3.00 & 1 & 1.37 & 1.37 & N/A & $b1$ & 1950 & 4.82 & 9.97 & 1.19 & 0.19 & 0.26 & C/O = 1 \\
CO WD & 500-138-1-b1-1 & 5.00 & 1 & 1.38 & 1.38 & N/A & $b1$ & 1710 & 6.50 & 11.9 & 1.20 & 0.18 & 0.30 & C/O = 1 \\
CO WD & 550-138-1-b1-1 & 5.50 & 1 & 1.38 & 1.38 & N/A & $b1$ & 1670 & 6.50 & 11.9 & 1.20 & 0.18 & 0.29 & C/O = 1 \\ \hline

CONe WD & 100-043-1-c3 & 1.00 & 1 & 1.33 & 0.43 & 0.90 & $c3$ & 2580 & 1.89 & 6.30 & 0.96 & 0.37 & 0.18 & \\
CONe WD & 200-045-1-c3 & 2.00 & 1 & 1.35 & 0.45 & 0.90 & $c3$ & 2160 & 2.93 & 7.74 & 1.11 & 0.24 & 0.24 & \\
CONe WD & 300-047-1-c3 & 3.00 & 1 & 1.37 & 0.47 & 0.90 & $c3$ & 1950 & 4.54 & 9.69 & 1.15 & 0.22 & 0.28 & \\
CONe WD & 500-048-1-c3 & 5.00 & 1 & 1.38 & 0.48 & 0.90 & $c3$ & 1710 & 5.13 & 10.5 & 1.26 & 0.12 & 0.36 & \\ 
CONe WD & 550-048-1-c3 & 5.50 & 1 & 1.38 & 0.48 & 0.90 & $c3$ & 1670 & 4.42 & 9.59 & 1.26 & 0.12 & 0.33 & \\ 
CONe WD & 750-049-1-c3 & 7.50 & 1 & 1.39 & 0.49 & 0.90 & $c3$ & 1540 & 5.39 & 10.6 & 1.29 & 0.10 & 0.35 & \\ 
CONe WD & 900-050-1-c3 & 9.00 & 1 & 1.40 & 0.50 & 0.90 & $c3$ & 1460 & 5.92 & 11.1 & 1.33 & 0.07 & 0.36 & \\ \hline

CONe WD & 100-043-1-b1 & 1.00 & 1 & 1.33 & 0.43 & 0.90 & $b1$ & 2580 & 2.69 & 7.15 & 0.97 & 0.36 & 0.20 & \\
CONe WD & 200-045-1-b1 & 1.00 & 1 & 1.35 & 0.45 & 0.90 & $b1$ & 2160 & 3.78 & 8.68 & 1.10 & 0.25 & 0.23 & \\
CONe WD & 300-047-1-b1 & 3.00 & 1 & 1.37 & 0.47 & 0.90 & $b1$ & 1950 & 4.66 & 9.78 & 1.12 & 0.25 & 0.31 & \\
CONe WD & 500-048-1-b1 & 5.00 & 1 & 1.38 & 0.48 & 0.90 & $b1$ & 1710 & 5.13 & 10.5 & 1.17 & 0.21 & 0.32 & \\ 
CONe WD & 550-048-1-b1 & 5.50 & 1 & 1.38 & 0.48 & 0.90 & $b1$ & 1670 & 5.81 & 11.0 & 1.18 & 0.20 & 0.32 & \\ 
CONe WD & 750-049-1-b1 & 7.50 & 1 & 1.39 & 0.49 & 0.90 & $b1$ & 1540 & 5.72 & 11.3 & 1.21 & 0.18 & 0.32 & \\ 
CONe WD & 900-050-1-b1 & 9.00 & 1 & 1.40 & 0.50 & 0.90 & $b1$ & 1460 & 7.81 & 13.1 & 1.30 & 0.10 & 0.35 & \\ \hline



\end{tabular}
\end{center}
\end{table*}

\section{Pure Turbulent Deflagration in C+O White Dwarfs}

In this section we study in details the nucleosynthesis 
yields of C+O WDs which explode as SNe Iax. 
For the C+O WDs with the initial central density $\rho_{\rm c, ini}$
and metallicity $Z$ as the model parameters, we put in the 
initial centered or off-center flame. We follow the flame propagation
and the expansion of the star until the star develops into homologous
expansion. After that, we use the thermodynamical histories
of the tracer particles to calculate the post-process nucleosythesis.
In Table \ref{table:Isotope1}, \ref{table:Isotope1b} and 
\ref{table:Decay1} we tabulate the nucleosynthesis yield and
the radioactive isotopes in the ejecta.

\subsection{Ejecta of C+O White Dwarf}

\subsubsection{Dependence on Central Density of White Dwarf}

\begin{figure}
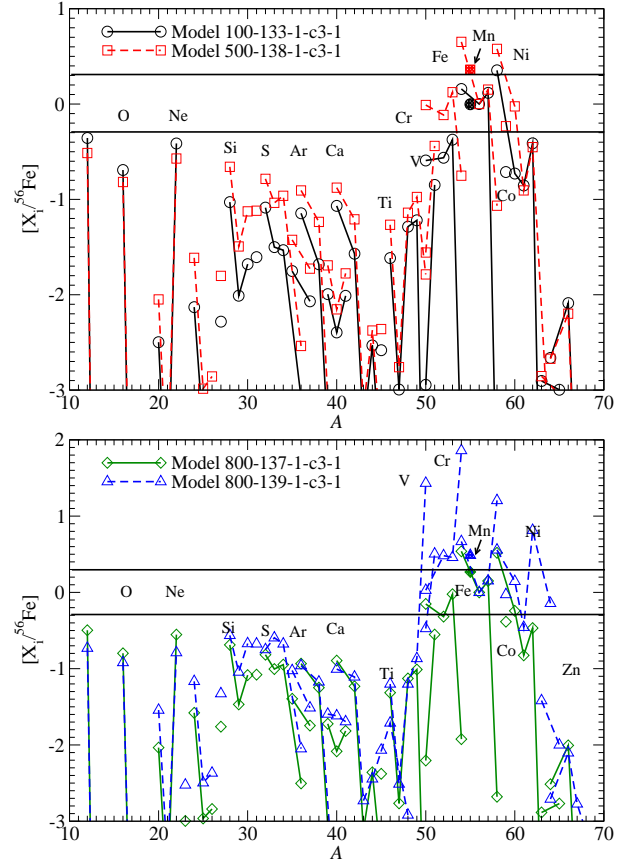

\centering
\includegraphics*[width=8cm,height=5.7cm]{fig8a.eps}
\includegraphics*[width=8cm,height=5.7cm]{fig8b.eps}
\caption{$[X_i/^{56}$Fe] against
 mass number for the ejecta of C+O WD models for 
Models
100-133-1-c3-1 ($\rho_c = 1 \times 10^9$ g cm$^{-3}$) and 
500-138-1-c3-1 ($\rho_c = 5 \times 10^9$ g cm$^{-3}$) in the top panel,
and Models 300-137-1-c3-1 ($\rho_c = 3 \times 10^9$ g cm$^{-3}$) 
and 800-138-1-c3-1 ($\rho_c = 8 \times 10^9$ g cm$^{-3}$) in the 
bottom panel. All models assume no O+Ne-rich matter,
$X(^{22}$Ne) = 0.025, $c3$ initial flame and C/O ratio $= 1$.}
\label{fig:final_esc_M_PTD_plot}
\end{figure}

In Figure \ref{fig:final_esc_M_PTD_plot} we plot the abundance ratios 
[X/$^{56}$Fe] in the ejecta of Models 100-133-1-c3-1 ($\rho_c = 10^9$ g cm$^{-3}$), 
300-137-1-c3-1 ($\rho_c = 3 \times 10^9$ g cm$^{-3}$), 
500-137-1-c3-1 ($\rho_c = 5 \times 10^9$ g cm$^{-3}$)
and 800-137-1-c3-1 ($\rho_c = 8 \times 10^9$ g cm$^{-3}$). 
These are the models based on C+O WDs with different $\rho_c$.
In the top panel we plot the abundance patterns for Models
100-133-1-c3-1 and 500-133-1-c3-1, while in the bottom panel 
the other two. 
More IMEs appear in the ejecta of higher $\rho_c$ except for
the very high $\rho_c = 8 \times 10^9$ g cm$^{-3}$. 
At low density (1 - 5 $\times 10^9$ g cm$^{-3}$) the IPEs are
comparable to the solar ratios with larger [X/Fe] for higher 
$\rho_c$. Isotope ratios including $^{50,52,53}$Cr, $^{54}$Fe, $^{55}$Mn
and $^{58}$Ni are larger for higher $\rho_c$. At high $\rho_c$, the 
increase in $^{54}$Fe, $^{55}$Mn and $^{58}$Ni levels off.
Instead, neutron-rich isotopes including $^{51}$V, $^{54}$Cr,
$^{60}$Fe and $^{62}$Ni become severely overproduced relative to the solar ratio.

\subsubsection{Dependence on Initial Flame Structure}

\begin{figure}
\centering
\includegraphics*[width=8cm,height=5.7cm]{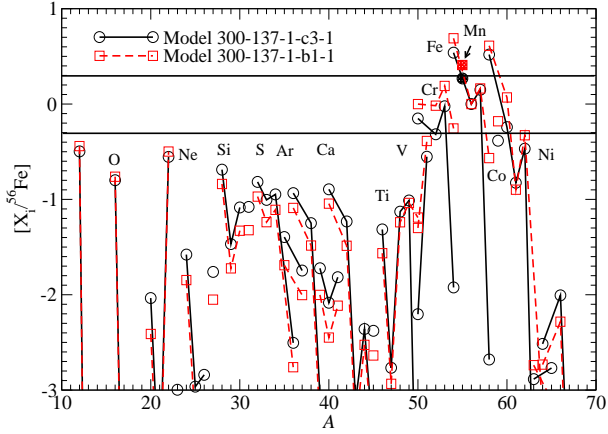}
\caption{Scaled mass fraction $[X_i/^{56}$Fe] against
mass number for the ejecta 
of C+O WD models for Models
300-137-1-c3-1 ($c3$ flame) and 
300-137-1-b1-1 ($b1$ flame).
All models assume $\rho_c = 3 \times 10^9$ g cm$^{-3}$, 
no hybrid O+Ne+Mg-rich matter,
$X(^{22}$Ne) $= 0.025$, and C/O ratio $= 1$.}
\label{fig:final_flame_PTD_plot}
\end{figure}

We examine the effects of initial flame structure 
to the nucleosynthesis yield for C+O WD models.
In Figure \ref{fig:final_flame_PTD_plot} we plot [X/$^{56}$Fe] in the ejecta
for Models 300-137-1-c3-1 ($c3$ flame) and 300-137-1-b1-1 ($b1$ flame). 
The two
WD models share the same configuration but with different
initial flame geometry.

The differences between the two models are small. IMEs show
systematic downward shifts when the initial flame moves from center
to off-center. This suggests that the mass of $^{56}$Fe as the denominator 
changes instead of the change in individual isotope yields.
Similar changes can be observed for the IPEs too.
On the contrary, the off-center burning facilitates 
the production of IPEs. It is because the IPEs produced 
in the off-center flame can be more readily ejected than those produced at the center.

However we remark that the effects in three-dimensional model
can be larger, as shown in \cite{Roepke2006a,Seitenzahl2013,Fink2014}. Three-dimensional
simulations can accommodate a more complex flame structure 
and hence more diversified distributions in the ejecta abundance.

\subsubsection{Dependence on C/O ratio}

We examine the effects of the initial C/O ratio on
the nucleosynthesis yields of C+O WD models.
The uncertainty in the C/O ratio mainly
stems from the uncertainties in the 
$^{12}$C$(\alpha,\gamma)^{16}$O rate and the
convective overshooting during He burning in the progenitor.

\begin{figure}
\centering
\includegraphics*[width=8cm,height=5.7cm]{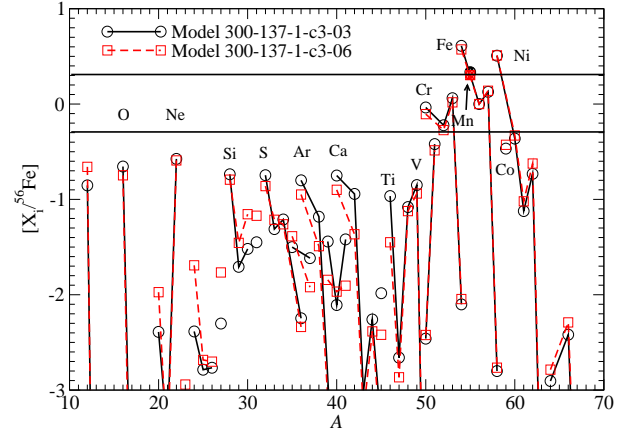}

\caption{
$[X_i/^{56}$Fe] against
mass number for the ejecta 
of C+O WD models for Models
300-137-1-c3-06 (C/O = 0.6) and 
300-137-1-c3-03 (C/O = 0.3).
All models assume $\rho_c = 3 \times 10^9$ g cm$^{-3}$, no 
O+Ne+Mg-rich matter,
$X(^{22}$Ne) $= 0.025$ and $c3$ initial flame.}
\label{fig:final_CO_PTD_plot}
\end{figure}

In Figure \ref{fig:final_CO_PTD_plot} we plot 
[X/$^{56}$Fe] in
the ejecta 
for Models 300-137-1-c3-06 ($C/O = 0.6$) and 
300-137-1-c3-03 ($C/O = 0.3$). The two
WD models share the same configuration but with a different
C/O ratio.


The effects of C/O ratio are much smaller than the previous 
two parameters. We find almost no change in IPEs 
and 
small enhancements in the IMEs for the low C/O ratio.
This is because slower flame for the low C/O ratio
produces lower energy,
thus the amount of matter experiencing incomplete burning 
increases. 

\subsubsection{Dependence on turbulent flame speed}

\begin{figure}
\centering
\includegraphics*[width=8cm,height=5.7cm]{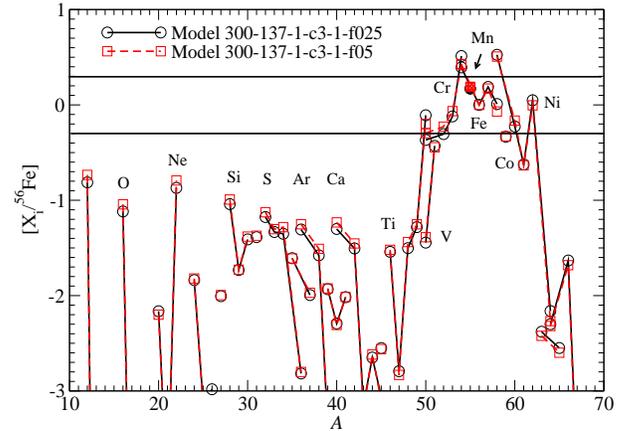}
\caption{$[X_i/^{56}$Fe] against
mass number for the ejecta for Models
300-137-1-c3-1-f05 ($v_{{\rm flame}} = 0.5 ~v_{{\rm flame,0}}$) and 
300-137-1-c3-1-f025 ($v_{{\rm flame}} = 0.25 ~v_{{\rm flame,0}}$).
All models assume $\rho_c = 3 \times 10^9$ g cm$^{-3}$, no hybrid O+Ne+Mg layer,
$X(^{22}$Ne) $= 0.025$, $c3$ initial flame and C/O ratio $= 1$.}
\label{fig:final_vflame_PTD_plot}
\end{figure}

We examine the effects of turbulent flame speed on
the nucleosynthesis yield of the C+O WD models.
In Figure \ref{fig:final_vflame_PTD_plot} we plot the final
abundance pattern of the ejecta 
for Models 300-137-1-c3-1-f05 and 300-137-1-c3-1-f025. The two
WD models share the same configuration but with different
asymptotic turbulent flame speed (at 50 \% and 25\% of the 
standard prescription). That means, we alter $C_A$ in the 
turbulent flame speed formula
\begin{equation}
v_{\rm flame} = v_{\rm lam}(\rho) \sqrt{1 + C_A (v'/v_{\rm lam})^2},
\end{equation}
where $v_{\rm lam}(\rho)$ and $v'$ are the laminar flame 
propagation speed and the velocity fluctuations due to 
turbulent motion. We can see that at $v' \rightarrow 0$, 
$v_{\rm flame} = v_{\rm lam}(\rho)$. This means that the flame
propagates like a laminar wave when the flow is not turbulent.
Otherwise, when $v' \gg v_{\rm lam}$, $v_{\rm flame} \rightarrow \sqrt{C_A} v'$.
This means that the flame burns with a speed following the 
turbulent motion. 

We remark that the connection between turbulent velocity 
fluctuations $v'$ and the corresponding flame velocity is not yet well
constrained because the corresponding WD condition and
environment cannot be reproduced in any laboratory. It is unclear
how the flame speed scales with $v'$ in the turbulent 
regime, where in a WD the Reynolds number can be as high as $10^{14}$.
Formula based on theoretical 
arguments can be found in, e.g., \cite{Hicks2015}.

In contrast to our previous works, the flame velocity has
almost no effect on the chemical yields of our models.
There are very minor differences in IMEs.

\subsection{Remnant of PTD in C+O White Dwarf}
\subsubsection{Dependence on Central Density}

\begin{figure}
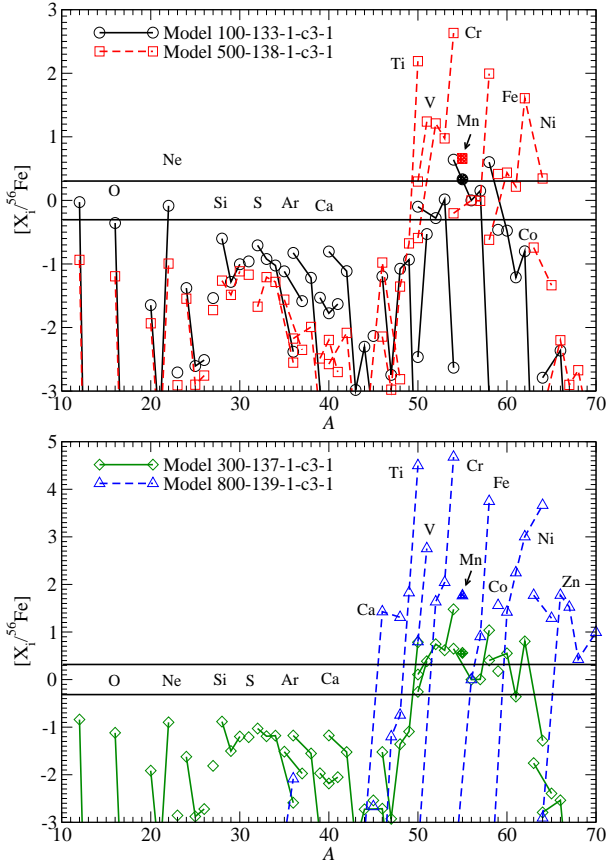

\centering
\includegraphics*[width=8cm,height=5.7cm]{fig12a.eps}
\includegraphics*[width=8cm,height=5.7cm]{fig12b.eps}
\caption{
$[X_i/^{56}$Fe] against
mass number for remnants of C+O WD models for Models
100-130-1-c3-1 ($\rho_c = 1 \times 10^9$ g cm$^{-3}$) and 
500-138-1-c3-1 ($\rho_c = 5 \times 10^9$ g cm$^{-3}$) in the top panel,
300-137-1-c3-1 ($\rho_c = 3 \times 10^9$ g cm$^{-3}$) and 
800-139-1-c3-1 ($\rho_c = 8 \times 10^9$ g cm$^{-3}$) in the bottom panel.
All models assume no hybrid O+Ne+Mg-rich matter
$X(^{22}$Ne) $= 0.025$, $c3$ initial flame and C/O ratio $= 1$.}
\label{fig:final_bnd_M_PTD_plot}
\end{figure}

In Figure \ref{fig:final_bnd_M_PTD_plot} we 
study the effects of 
initial $\rho_c$ of the WD
on the nucleosynthesis yields in the bounded remnants WDs. 
Similar to the 
characteristic model, the bounded remnant is defined
by the tracer particles which have a negative total energy
at the end of simulations (10 s after the flame starts).
The effects of $\rho_c$ are consistent with our earlier
work \citep{Leung2018Chand}.

In the top panel, we plot the Models 100-000-1-c3-1 and
Model 500-000-1-c3-1 of the remnant part and in the
bottom panel the other two models. Similar to the ejecta,
with increasing $\rho_c$, masses of IMEs decrease. For the very high
$\rho_c$, the remnant does not contain any C+O+Ne-rich matter
or IMEs. This is because those matter, which are synthesized
at the outer region of a lower density ($10^8$ g cm$^{-3}$ or below),
are also ejected without fallback by the stronger deflagration.
In all models, IMEs are always underproduced. The IMEs decrease
with increasing $\rho_c$ because
more matter can be burnt at a lower density before the
flame quenches.


The IPEs show more interesting features.
Isotopes with a high $Y_{\rm e}$ gradually decrease
in its abundance when $\rho_c$ increases. 
Neutron-rich isotopes are robustly over-produced
in the remnant. At a central density $3 \times 10^9$ g cm$^{-3}$ 
isotopes like $^{54}$Cr and $^{58}$Fe are 
over-produced, especially 
for Models 500-138-1-c3-1 and 800-139-1-c3-1. 
At a higher $\rho_c$, IPEs from Ca to Zn are
found in the remnant.
Their abundances can exceed the solar values by a factor 
of $10^{2-5}$ in Model 800-139-1-c3-1.

We remark that in interpreting the remnant composition,
it is also useful to examine the element abundances
instead of isotope abundances because the spectra 
from WD remnants do not distinguish isotopes. 
We also show the element distribution in
Section \ref{sec:discussion}.

\subsubsection{Dependence on Initial Flame Structure}

\begin{figure}
\centering
\includegraphics*[width=8cm,height=5.7cm]{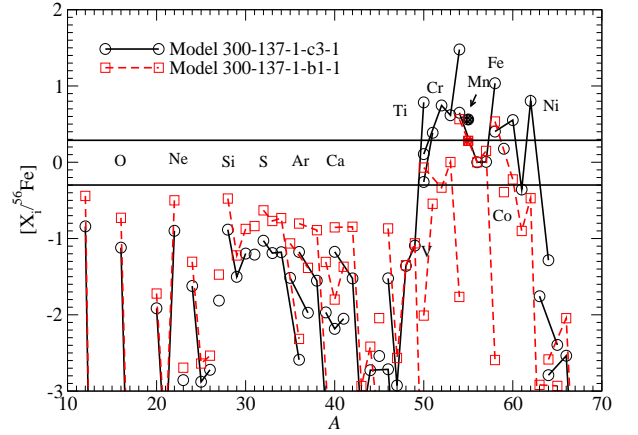}
\caption{$[X_i/^{56}$Fe] against
mass number for the bounded remnant
of C+O WD models for Models
300-137-1-c3-1 ($c3$ flame) and 
300-137-1-b1-1 ($b1$ flame).
All models assume $\rho_c = 3 \times 10^9$ g cm$^{-3}$, no hybrid O+Ne+Mg-rich matter,
$X(^{22}$Ne) $= 0.025$ and C/O ratio $= 1$.}
\label{fig:final_flame_bnd_PTD_plot}
\end{figure}

In Figure \ref{fig:final_flame_bnd_PTD_plot} we study the 
effects of initial flame on the remnant nucleosynthesis pattern.
We plot the abundance pattern of Models 300-137-1-c3-1 and 
300-137-1-b1-1. They differ only by the position of the initial
flame (center versus. off-center).
Consistent with the ejecta composition, the remnant consists
of high abundances of IMEs when the flame is off-center.
When the initial flame is farther from
the center, the over-production
of some neutron-rich isotopes is less severe.
This can be seen as a systematic drop in isotopes
like $^{52-54}$Cr, $^{55}$Mn, $^{58}$Fe and 
$^{60-64}$Ni. Higher $Y_{\rm e}$ isotopes of 
Fe and Ni remain similar. 
These features are
in general consistent with the picture that the off-center
flame can push matter more easily outward as the momentum 
transport does not necessarily start in the center.

\subsubsection{Dependence on C/O ratio}

\begin{figure}
\centering
\includegraphics*[width=8cm,height=5.7cm]{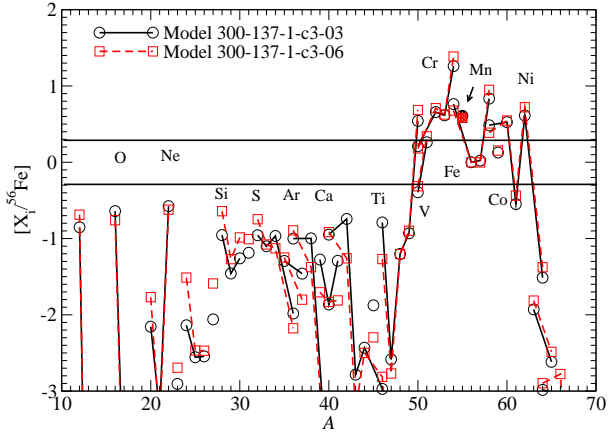}
\caption{$[X_i/^{56}$Fe] against
mass number for the bounded remnant 
of C+O WD models for Models
300-137-1-c3-06 (C/O = 0.6) and 
300-137-1-c3-03 (C/O = 0.3).
All models assume $\rho_c = 3 \times 10^9$ g cm$^{-3}$, no hybrid O+Ne+Mg-rich matter,
$X(^{22}$Ne) $= 0.025$ and $c3$ initial flame.}
\label{fig:final_CO_bnd_PTD_plot}
\end{figure}

In Figure \ref{fig:final_CO_bnd_PTD_plot} we plot the 
abundance pattern for the remnants of Models 
300-137-1-c3-06 and 300-137-1-c3-03. The two
models differ by the C/O ratio, which changes
not only the initial chemical composition, 
but also the energy production by the deflagration
and the laminar flame speed. 
The differences between the two models are 
slightly stronger than the ejecta. Enhanced 
isotopes of IMEs such as $^{38}$Ar and $^{42}$Ca 
can be observed. The IPEs are shifted upward
systematically when C/O ratio decreases,
again suggesting the changes of $^{56}$Fe.

\subsubsection{Dependence turbulent flame speed}

\begin{figure}
\centering
\includegraphics*[width=8cm,height=5.7cm]{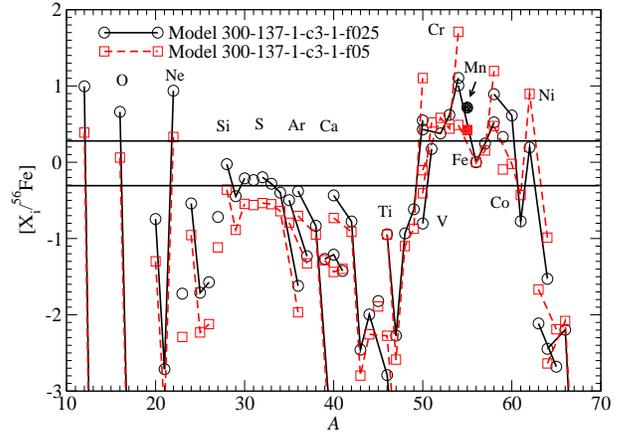}
\caption{$[X_i/^{56}$Fe] against
mass number for the bounded remnant 
of C+O WD models for Models
300-137-1-c3-1-f025 ($v_{{\rm flame}} = 0.25 ~v_{{\rm flame,0}}$) and 
300-137-1-c3-1-f05 ($v_{{\rm flame}} = 0.5 ~v_{{\rm flame,0}}$).
All models assume $\rho_c = 3 \times 10^9$ g cm$^{-3}$, no hybrid O+Ne+Mg-rich matter,
$X(^{22}$Ne) $= 0.025$, $c3$ initial flame and C/O ratio $= 1$.}
\label{fig:final_vflame_bnd_PTD_plot}
\end{figure}

At last in Figure \ref{fig:final_vflame_bnd_PTD_plot} we compare
the effects of asymptotic turbulent flame speed on the remnant
composition for Models 300-137-1-c3-1-f025 and 300-137-1-c3-1-f05. 
Similar to the ejecta, the difference of flame speed
we chose does not affect the abundance pattern in an observable 
level. A systematic decrease for most isotopes can be observed when 
$v_{\rm flame}$ increases, suggesting that the flame produces more
$^{56}$Fe as a result. When a slower flame model is used,
a systematic enhancement of C+O-rich matter and IMEs can be 
found in the remnant. The abundances of neutron-rich IPEs increase when 
the flame speed increases too. Meanwhile, there is no significant
enhancement for IPEs with a higher $Y_{\rm e}$.

\section{Hybrid C+O+Ne WD}

In this section we study how the model parameters affect the explosive
nucleosynthesis of the hybrid C+O+Ne WD.
In Table \ref{table:Isotope2}, \ref{table:Isotope2b} and 
\ref{table:Decay2} we tabulate the nucleosynthesis yields and
the masses of radioactive isotopes in the ejecta.

\subsection{Ejecta of Hybrid C+O+Ne WD}
\subsubsection{Dependence on Central Density}

\begin{figure}
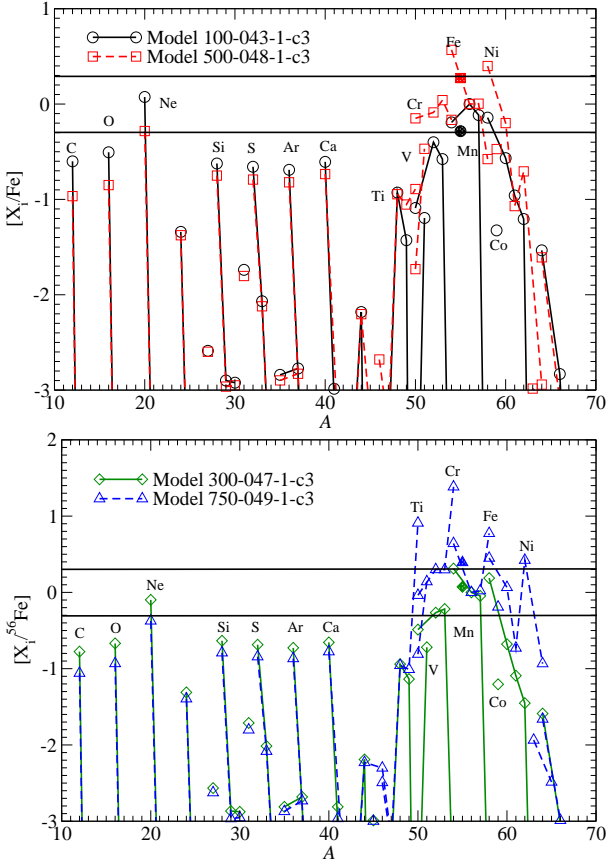

\centering
\includegraphics*[width=8cm,height=5.7cm]{fig16a.eps}
\includegraphics*[width=8cm,height=5.7cm]{fig16b.eps}
\caption{$[X_i/^{56}$Fe] against
mass number for the ejecta of C+O WD models for Models
100-043-1-c3 ($\rho_c = 1.0 \times 10^9$ g cm$^{-3}$) and 
500-048-1-c3 ($\rho_c = 5.0 \times 10^9$ g cm$^{-3}$) in the top panel, Models
300-047-1-c3 ($\rho_c = 3.0 \times 10^9$ g cm$^{-3}$) and 
750-049-1-c3 ($\rho_c = 7.5 \times 10^9$ g cm$^{-3}$) in the bottom panel.
All models assume $X(^{22}$Ne) = 0.025 and $c3$ initial flame.}
\label{fig:final_esc_M_plot}
\end{figure}

In Figure \ref{fig:final_esc_M_plot} we plot 
$[X_i/^{56}$Fe] in
the ejecta 
of the hybrid C+O+Ne WDs for
Models 100-043-1-c3, 300-047-1-c3,
500-048-1-c3 and 700-049-1-c3. 

The abundance pattern of the ejecta is
similar to the typical SN Ia.  (1) IMEs are
underproduced. (2) The V, Cr, Mn and Co isotopes are much higher
in the higher central density models than the lower density models.
(3) On the contrary, Fe and Ni isotopes are not sensitive to the central density
except that $^{54}$Fe and $^{58}$Ni are higher for the higher central density.
(4) Only in extreme cases such as
Model 750-049-1-c3 we observe the severe
overproduction of the neutron-rich isotopes
$^{50}$Ti, $^{54}$Cr, $^{58}$Fe and $^{62}$Ni.
(5) A growth of Mn with increasing central density is still seen but it saturates
at a value [Mn/Fe]$\sim 2$, while [Co/Fe] becomes
compatible with the solar value.


The overall trend is similar to C+O WD models.
The C+O+Ne composition 
only provides a lower energy
release due to the lower abundance of $^{12}$C and a higher
abundance of $^{20}$Ne. 
The C+O+Ne composition also makes the 
flame propagation slower. In 
general the turbulent flame dominates the flame propagation,
which is independent on the composition. But the composition
affects the turbulent flame indirectly by its energy
feedback, which changes the turbulent motion inside
the star, and hence the production and decay of turbulent motion.

In general, the CO-deflagration does not 
differ much from CONe-deflagration at high density
because in both cases, matter is burnt into NSE. 
By comparing
Figure \ref{fig:final_esc_M_plot} with Figure \ref{fig:final_esc_M_PTD_plot},
the overall patterns suggest that, indeed, the property
of CO-deflagration and CONe-deflagration does not differ much
except for the minor details, such as the minor isotopes in 
IMEs and lower mass IPEs.
But we remind that a hybrid C+O+Ne WD,
takes a longer time for the deflagration 
wave to spread and burn to release the necessary energy
for the expansion. This also means the momentum transport
from the hot ash to the cold fuel is slower. 

\subsubsection{Dependence on Initial Flame Structure}

\begin{figure}
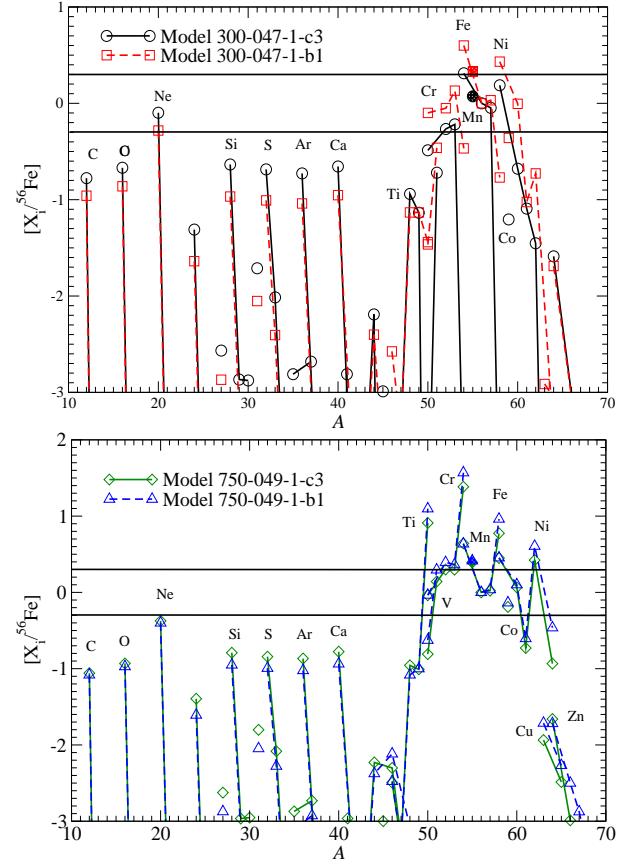

\centering
\includegraphics*[width=8cm,height=5.7cm]{fig17a.eps}
\includegraphics*[width=8cm,height=5.7cm]{fig17b.eps}
\caption{(top panel) $[X_i/^{56}$Fe] against
mass number for the ejecta in hybrid C+O+Ne WD models for
300-047-1-c3 ($c3$ flame) and 
300-047-1-b1 ($b1$ flame).
(bottom panel) Same as the upper panel, but for
Models 750-049-1-c3 and 750-049-1-b1.
All models assume $X(^{22}$Ne) $= 0.025$ and $c3$ initial flame.}
\label{fig:final_flame_plot}
\end{figure}

We examine the effects of initial flame structure 
on the nucleosynthesis yield for the hybrid C+O+Ne WDs.
In Figure \ref{fig:final_flame_plot} we plot 
[$X_i/^{56}$Fe] of the ejecta 
for Models 300-047-1-c3 and 300-047-1-b1. The two
WD models share the same configuration but with different
initial flame geometry (centered flame $c3$ vs. off-center flame $b1$).

As a demonstration we compare the final nucleosynthesis yields
of the ejecta in Models 300-047-1-c3 and 300-047-1-b1 in the 
upper panel and Models 750-049-1-c3 and 750-049-1-b1 in the 
lower panel. The
two models differ by the initial flame where one is the central
$c3$ flame and the other is off-centered $b1$ flame. 
The ejecta composition is very similar in both cases 
in the light elements of O, Ne, and IMEs such as Si, S, Ar and Ca. 
Off-center flame gives a slightly lower or a comparable 
amount of C, O, Ne and 
IMEs.
The difference of the flame mostly affects IPEs. 
In the lower density models, 
the off-center flame tends to produce more Cr, Fe, Mn and Ni
with very significant overproduction. Strong enhancement of $^{59}$Co 
for the $b1$-flame model can be seen.
This is because
the IPEs, which are produced in the core region,
can be transported more easily by the flame bubble which flows
with buoyancy. On the other hand, IPEs produced in the central region
tend to fallback when they transport their momentum to outer
fluid elements. 
In the high-density models, 
not much difference in the pattern can be found. This is because
the propagation becomes so rapid and energetic that the 
flame burn similarly and ejects most of the WD.

\subsection{Remnant of Hybrid C+O+Ne WD}

\subsubsection{Dependence on Central Density}

\begin{figure}
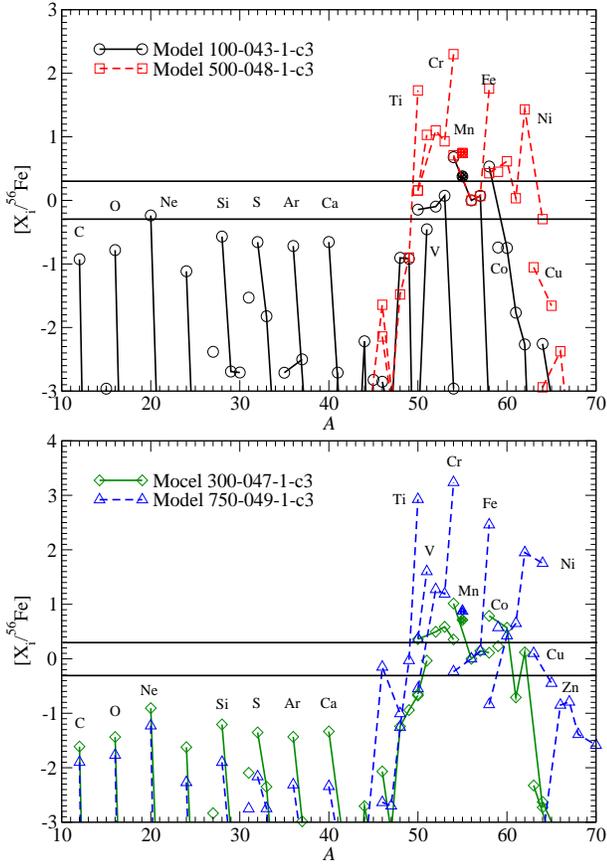

\centering
\includegraphics*[width=8cm,height=5.7cm]{fig19a.eps}
\includegraphics*[width=8cm,height=5.7cm]{fig19b.eps}
\caption{$[X_i/^{56}$Fe] against
mass number for the remnants of hybrid C+O+Ne WD models for Models
100-043-1-c3 ($\rho_c = 1.0 \times 10^9$ g cm$^{-3}$) and 
500-048-1-c3 ($\rho_c = 5.0 \times 10^9$ g cm$^{-3}$) in the top panel, Models
300-047-1-c3 ($\rho_c = 3.0 \times 10^9$ g cm$^{-3}$) and 
750-049-1-c3 ($\rho_c = 7.5 \times 10^9$ g cm$^{-3}$) in the bottom panel.
All models assume $X(^{22}$Ne) $= 0.025$ and $c3$ initial flame.}
\label{fig:final_bnd_M_plot}
\end{figure}

In Figure \ref{fig:final_bnd_M_plot} we plot the abundance patterns of the bound remnants
in Models 100-043-1-c3, 300-047-1-c3, 500-048-1-c3
and 750-049-1-c3. This series of models studies the 
effects of the central density of the WD
on the chemical composition of the remnant.

From both panels it shows that the initial central density strongly
influences the IPEs in the remnant. 
A higher central density leads to a stronger 
enhancement of neutron-rich isotopes like 
$^{50}$Ti, $^{54}$Cr, $^{58}$Fe and 
$^{62}$Ni. Their ratios to $^{56}$Fe can 
be from 10 to 10000 times of the solar ratio.
The trend can be seen already in the C+O WDs.
Isotopes, with a $Y_{\rm e} \approx$ 0.5, for example,
$^{46}$Ti, $^{50}$Cr, $^{54}$Fe and $^{58}$Ni,
share similar ratios among all models. This 
is consistent with our previous results that 
matter with a higher density has a faster
electron capture rate, 
which strongly favours the production of 
neutron-rich isotopes. 

The general abundances of the IMEs (Si, S, Ar, Ca)
drop when the progenitor mass increases. It is because
as the mass increases, the star becomes more compact,
and the density drop in the outer part becomes
steeper. Thus there is a lower mass of matter with 
a low density ($\sim 10^7$ g cm$^{-3}$). 
One exception is in Model 500-048-1-c3.
In that model no C+O+Ne-rich matter and no IMEs can be
found. This feature is comparable with Model 800-139-1-c3.
The non-monotonic trend 
suggests that the fallback event is sensitive to 
how outer matter is ejected. However, we remind that 
even the IME abundances are higher in Model 800-049-1-c3,
it is only about a few \% of the solar values.

Despite that some isotopes are extremely over-produced
compared to the solar ratio, we remind that their 
corresponding masses as a part of the element can still
be smaller or only comparable with the major isotopes
of their corresponding elements. In particular, in 
spectrography, the atomic lines are sensitive to the 
elements, but not individual isotopes.

\subsubsection{Dependence on Initial Flame Structure}

\begin{figure}
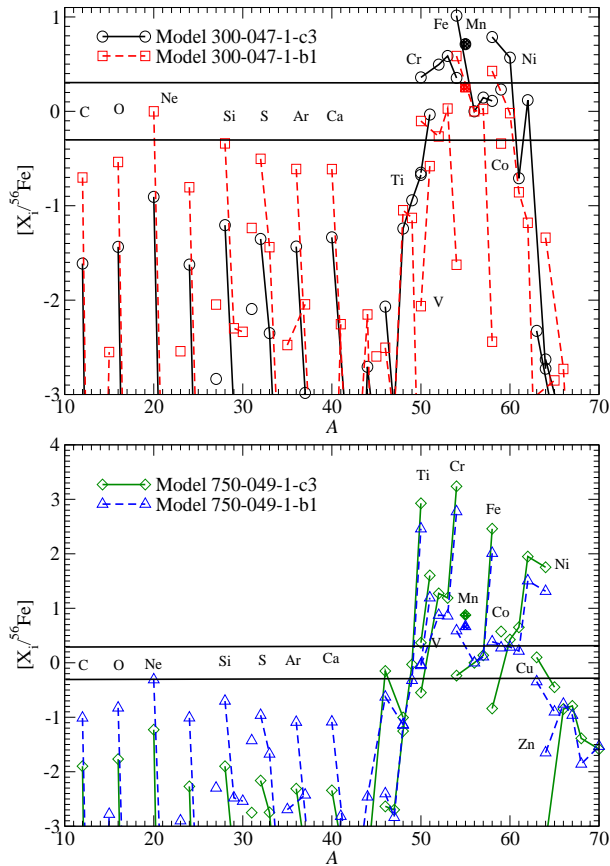

\centering

\includegraphics*[width=8cm,height=5.7cm]{fig20a.eps}
\includegraphics*[width=8cm,height=5.7cm]{fig20b.eps}
\caption{(top panel)$[X_i/^{56}$Fe] against
mass number for the remnants 
of hybrid C+O+Ne WD models for Models
300-047-1-c3 ($c3$ flame) and 
300-047-1-b1 ($b1$ flame). (bottom panel) Similar to the 
top panel but for 750-049-1-c3 ($c3$ flame) and 
750-049-1-b1 ($b1$ flame).
All models assume $X(^{22}$Ne) $= 0.025$ and C/O ratio $= 1$.}
\label{fig:final_flame_bnd_plot}
\end{figure}

In Figure \ref{fig:final_flame_bnd_plot} we compare the chemical
composition of the remnant in Models 300-047-1-c3 and 300-047-1-b1
in the upper panel and Models 750-049-1-c3 and 750-049-1-b1
in the lower panel.
The two models differ by the initial flame. 

In the low density models,
the remnant composition is characterized by a significant overproduction
of IPEs. 
Larger amounts of C+O+Ne-rich matter and IMEs
are observed, again suggesting that 
an off-center flame help to eject matter more easily due to weaker
gravitational attraction and less matter on top 
of burnt matter.
On the other hand, a farther flame from the center
produces ash which has experienced less
electron capture before the ash cools down. The difference
can be seen by the abundance of neutron-rich isotopes 
$^{54}$Cr, $^{58}$Fe and $^{59}$Co, $^{62}$Ni. Apart from that 
the pattern of IPEs is similar. 

Then we also compare the abundance yields using different initial
flame for the high density models. The abundance pattern is less sensitive
to the initial flame structures. C+O+Ne-rich matter and IMEs are higher
in the off-center flame model (750-049-1-b1) while there is a higher
abundance of IPEs in the centered flame model (750-049-1-c3).

\section{Discussion}
\label{sec:discussion}

\subsection{Ejecta and its Observable}

\subsubsection{Ejecta mass - Ejecta energy relation}

In observing SNe Iax, the parameter space 
$(M_{{\rm ej}}, E_{{\rm ej}})$, i.e. the pair of ejecta
mass and the ejecta energy, is important 
owing to the presence of bounded remnant. 

\begin{figure}
\centering
\includegraphics*[width=8cm,height=5.7cm]{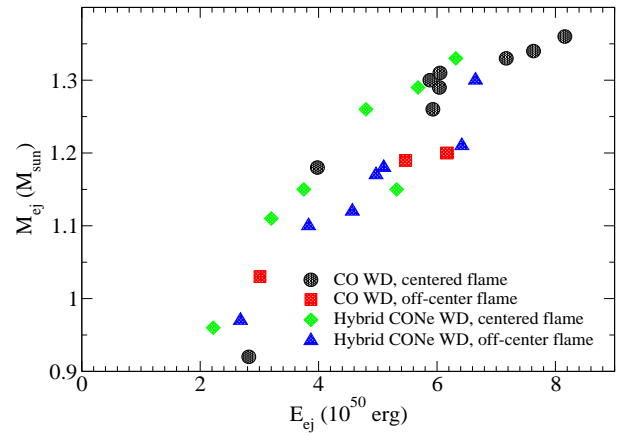}
\caption{Ejecta mass against final WD ejecta energy
for all models in this work.}
\label{fig:Mej_Etot_plot}
\end{figure}

To derive this relation, we collect the final 
kinetic energy of the ejecta, together with 
their total mass. 
In Figure \ref{fig:Mej_Etot_plot} we plot the 
ejecta mass against the WD final energy for
the models presented in this work. 
We can see a clear monotonic trend 
where the higher final energy corresponds to a 
higher ejecta mass. A quasi-linear relation can be seen
for $E_{{\rm ej}} < 6 \times 10^{50}$ erg. 
For $E_{{\rm ej}}$ greater than that, the data approaches
the asymptotic value of $\sim 1.4~M_{\odot}$
and levels off.

We further fit the data points with a linear relation
at low $E_{{\rm ej}}$. We observe a linear relation
for $E_{\rm ej}$ between 2 -- 6 $\times 10^{50}$ erg
as $M_{{\rm ej}} = 0.9 + 0.1 E_{{\rm ej}}$ 
with a chi-square about 0.28. 
However we expect that the linear relation 
will break down in the low $E_{\rm ej}$ limit
as it corresponds to the limit where the ejecta has no energy, meaning that
no ejecta exists. In that case, the relation
should fall steeply towards the origin.

\subsubsection{Ejecta mass -- $^{56}$Ni Mass Relation}

\begin{figure}
\centering
\includegraphics*[width=8cm,height=5.7cm]{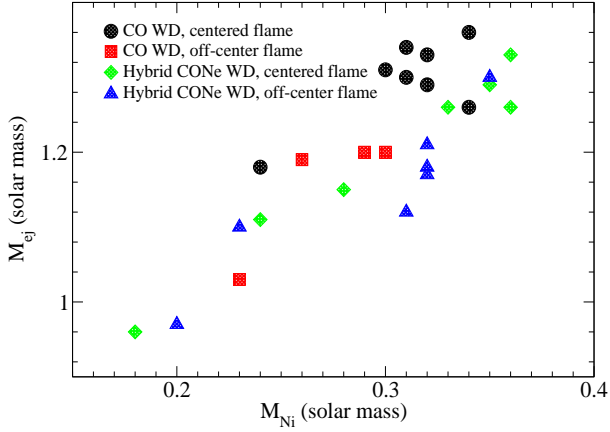}
\caption{Ejecta mass against ejected $^{56}$Ni mass
for all models in this work.}
\label{fig:Mej_Mni_plot}
\end{figure}

The ejecta mass - $^{56}$Ni mass relation provides a test on 
how explosion strength is connected to the light curve feature. 
In general the ejecta mass is related to the width
and the $^{56}$Ni mass are related to the width and 
peak luminosity of the light curve respectively. 
We plot in Figure \ref{fig:Mej_Mni_plot}
the ejecta mass against $^{56}$Ni for all models we presented 
in this work. Each group of SN Iax models has
their own slope due to their different ways to 
explode the star and their corresponding energy 
production. For example, the data for the C+O WD models
with centered flame is more clustered near $M_{\rm Ni} = 0.3$.
This is because those data points correspond to where the 
star is almost completely disrupted, while the lack of 
the detonation bounds the possible $^{56}$Ni production.

However, globally, the models show a 
general linear trend with dispersion. We again see at low $^{56}$Ni, the 
pair has a linear relation to a good approximation, which can be fitted by
$M_{\rm ej} \approx 4 M_{\rm Ni}$ 
where we require the fitting line to pass through the origin.

Accompanying with the partial ejection of the WD materials, 
a small mass WD, which is 
bounded gravitationally, survives. The typical bounded mass
ranges from 0.2 to 1.1 $M_{\odot}$, depending on the explosion strength.
Such white dwarf remnant is observable. 
WDs of mass $\sim$ 0.3 $M_{\odot}$ are observed (See e.g. \cite{Brown2010a}).

\subsubsection{$^{56}$Ni Mass -- Ejecta Velocity Relation}

\begin{figure}
\centering
\includegraphics*[width=8cm,height=5.7cm]{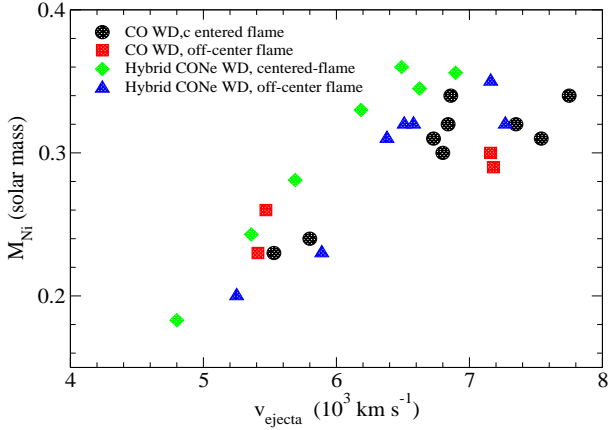}
\caption{Ejecta $^{56}$Ni mass against 
ejecta characteristic velocity for the SN Iax models
presented in this work.}
\label{fig:vej_ni_logplot}
\end{figure}

At last we examine the ejecta velocity relation with $^{56}$Ni production $(M_{\rm Ni})$.
These two quantities are associated with the observable pair
Si II Velocity against absolute magnitude at peak M$_V$.
In Figure \ref{fig:vej_ni_logplot} we plot the ejecta
$^{56}$Ni against ejecta characteristic velocity. 
The velocity of the ejecta $v_{\rm ejecta}$ is computed by first summing
up the kinetic energy $E_k$ of all the tracer particles,
and then we obtain $v_{\rm ejecta} = \sqrt{2 E_k/M_{\rm ej}}$.

We observe in this pair of observables that the ejecta velocity 
increases with $^{56}$Ni mass linearly. The typical 
ejecta velocity ranges between 5000 - 8000 km s$^{-1}$.
This corresponds to models with a strong explosion when $M$ is large,
where the energy production increases more rapidly than
the growth of mass.
For a given ejecta velocity, it can correspond
a dispersion of $^{56}$Ni mass of $\sim 0.1~M_{\odot}$ in
the high velocity limit and the dispersion is smaller
otherwise. This "fan"-shape pattern allows more
diversity of SNe Iax when they are more
luminous.

However, in this figure we do not attempt to directly compare
our theoretical models with the observational data such as 
those in \cite{Foley2013}. It is because to extract the 
reliable estimation of Si II velocity, the photosphere 
position and the corresponding velocity of the Si shell are essential.
The exact value can be obtained by carrying out radiative 
transfer directly. To further relate the expected peak luminosity 
with the $^{56}$Ni production, radiative transfer with 
gamma-ray energy deposition is necessary for a consistent prediction.

\subsubsection{Ejecta Elemental Abundance}

\begin{figure}
\centering
\includegraphics*[width=8cm,height=5.7cm]{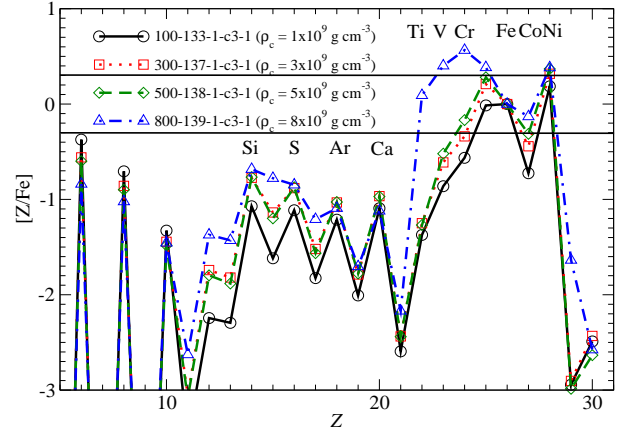}
\caption{Ejecta chemical abundance [Z/Fe] against $Z$ for 
Models 100-133-1-c3-1 ($\rho_c = 1 \times 10^9$ g cm$^{-3}$), 
300-137-1-c3-1 ($\rho_c = 3 \times 10^9$ g cm$^{-3}$), 500-138-1-c3-1 ($\rho_c = 5 \times 10^9$ g cm$^{-3}$),
and 800-139-1-c3-1 ($\rho_c = 8 \times 10^9$ g cm$^{-3}$).
All models assume C+O WD, $c3$-flame,
$X(^{22}$Ne) $= 0.025$ and C/O mass fraction ratio $= 1$.}
\label{fig:ejecta_elements_plot}
\end{figure}

In typical spectral observations of SNe Iax and their remnants,
the elemental abundances instead of the 
isotopic abundances are measured. 
Here we examine how the chemical abundance of SN Iax
depends on the progenitor mass. By using the post-decay (i.e. after $10^6$ years)
stable isotopic contributions in the ejecta, 
we sum the mass of the isotopes of each elements from C to Zn,
then we compute
the corresponding [Z/Fe] for Models 100-133-1-c3-1 ($\rho_c = 1 \times 10^9$ g cm$^{-3}$), 
300-137-1-c3-1 ($\rho_c = 3 \times 10^9$ g cm$^{-3}$), 500-138-1-c3-1 ($\rho_c = 5 \times 10^9$ g cm$^{-3}$),
and 800-139-1-c3-1 ($\rho_c = 8 \times 10^9$ g cm$^{-3}$).
We remind that minor long-lived radioactive isotopes
still contribute but their amounts are much smaller than
the major stable isotopes.
These models correspond to models with different central density
and hence different progenitor mass.

In Figure \ref{fig:ejecta_elements_plot} we plot
the element abundances of these models. 
The ejecta composition, after fallback when the hot ash pushes the 
external envelope and atmosphere away from the WD, appear 
to have a similar abundance pattern. Except Model 800-139-1-c3-1,
other models share features including sub-solar IMEs with similar
ratios of [Ca/S]. Light IMEs are also under-produced. 
It is because the turbulent flame is sub-sonic, which cannot
follow the pace of expanding matter for creating an adequate amount
of IMEs.
The ejecta features an abundant amount of Mn, Fe and Ni
with respect to the solar composition. For the contrasting model
800-139-1-c3-1, it shows a much higher Ti, V and Cr. 
Such difference can be the key to identify the difference in 
progenitor from future SNR observations. 

\subsection{Application to Observed Low-Mass WD}

\begin{figure}
\centering
\includegraphics*[width=8cm,height=5.7cm]{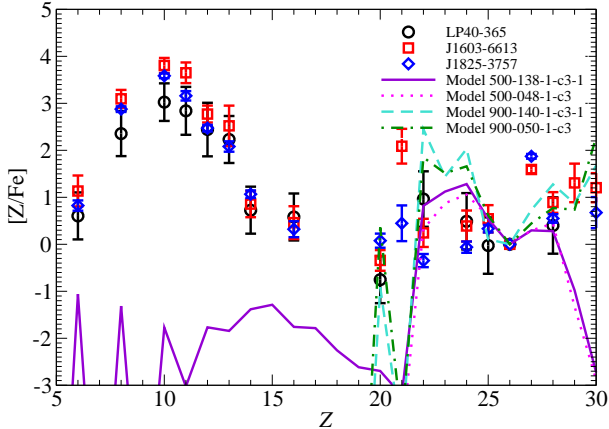}
\caption{The abundance patterns [$Z/$Fe] of low-mass WDs including 
LP40-365, J1603-6613 and J1825-3757. Representative SN Iax
models including 500-138-1-c3-1, 500-047-1-c3-1, 900-140-1-c3-1
and 900-050-1-c3-1 are included for comparison.}
\label{fig:WD_abundance_plot}
\end{figure}

In \cite{Raddi2018,Raddi2019} the abundance patterns are extracted
from the spectra of some low-mass WDs with unusually high metal 
fractions. The are also traveling with a high velocity.
They include LP40-365, J1603-6613 and J1825-3757.

In Figure \ref{fig:WD_abundance_plot} we plot the abundance
patterns of these objects together with some of our representative
models presented in this work. To obtain the final chemical
composition, after the remnant chemical composition is obtained
by post-processing, the composition in the remnant is assumed to be completely
mixed. We allow the remnant to pass $\sim 10^6$ 
years until most typical radioactive elements have completely
decayed. Exceptions include very long-lived radioactive elements
such as $^{27}$Al and $^{60}$Fe. In this work
when the radioactive decay injects energy to the remnant,
the remnant is assumed to be static. 
No mass loss is assumed in this process.

The models include 
500-138-1-c3-1, 500-050-1-c3-1, 900-140-1-c3-1
and 900-050-1-c3-1. The first two models correspond to the 
the C+O WD and hybrid C+O+Ne WD models with $\rho_c = 5 \times 10^9$ g cm$^{-3}$.
The last two models correspond to the C+O and hybrid C+O+Ne WDs
with $\rho_c = 9 \times 10^9$ g cm$^{-3}$.

These observed WDs feature extremely abundant of light elements
including C, O, Ne, Na, Mg and Al. These are as high as $10^4$
times of the solar abundances with respect to Fe. 
The observed abundances of IMEs (Si, S and Ca) in these WDs are comparable with 
the solar abundances.  
It would be interesting to note that these WDs have super-solar
IPEs especially Mn, Co and Ni. 

Our models all show a systematic underproduction of light elements, 
none of which is super-solar. The IMEs are also too low compared to the 
objects. However, these features could originate from 
other reasons, e.g., the progenitors of the WDs.  Possible later energy deposition
due to the $^{56}$Ni-decay in the remnant can also trigger further mass loss,
which may further lower the Fe contribution. 

The IPEs of our models 
appear to be similar to the observations.
The values for Ti, Cr and
Ni are particularly close to the observed values. We remark that 
the ejecta of this model
shows a much higher V, but not in the remnant WD. 
However, none of our models can reproduce the
$\sim 100$ times production of Co/Fe as seen in the two 
WDs shown here. 

We note that there are 
SN Iax models without leaving a WD remnant.
For example,  \cite{Sahu2008} showed that 
the early and late time light curves and spectra of SN 2005hk are well-reproduced
with the pure deflagration model which has no WD remnant.

Such weak explosion models can be found in PTD models with pulsation
\cite[e.g.,][]{Nomoto1976} and
relatively fast flame propagation \cite[e.g.,][]{Fink2014}.  The exact
ejection mass depends strongly on the initial flame location. An
off-center flame tends to suppress the ejected mass.

\subsection{Application to Supernova Remnant}

\begin{figure}
\centering
\includegraphics*[width=8cm,height=5.7cm]{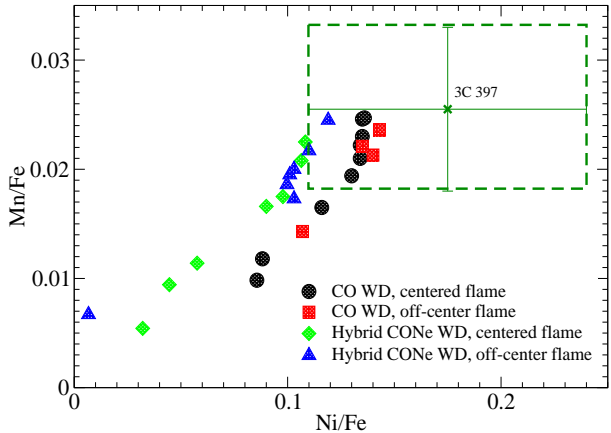}
\caption{The mass ratios of Mn/Fe against Ni/Fe for the ejecta from C+O and hybrid C+O+Ne WD
models in this work. The data point corresponds to the 
SN remnant 3C 397.}
\label{fig:ejecta_MnNi_plot}
\end{figure}

We further apply the explosion yield to some supernova remnants (SNRs). 
In \cite{Yamaguchi2015} the chemical abundance pattern of the 
SNR 3C 397 is discussed. Similar analysis has been done 
for various SNRs in the galaxy and in the Large Magellanic Cloud
as reported in \cite{MartinezRodriguez2017}.
From the X-ray spectra, it is found that this galactic 
SN remnant (3C 397) contains extremely high mass ratios Mn/Fe and Ni/Fe,
which hints on the possibility of super-solar metallicity of the progenitor star.

However, in \cite{Dave2017} another possibility of 
using the Chandrasekhar mass model in the high mass end ($\rho_c \sim 6 \times 10^9$ g cm$^{-3}$)
is proposed to explain the high Mn/Ni mass ratio. Here we further investigate if 
SN Iax models can approach this data point without 
invoking high metallicity. Also, in \cite{MartinezRodriguez2017}
the Ca/S mass ratio can be the hint to understand the 
diversity of observed SNRs with the parameters other than 
metallicity. Here we examine how the central density of 
the WD contributes to the diversity.  

In Figure \ref{fig:ejecta_MnNi_plot} we plot the mass ratios Mn/Fe
against Ni/Fe for models in Table \ref{table:Isotope1b}.
The data point corresponds to the SNR 3C 397 with the one sigma error box is shown. 
C+O WD models with $\rho_c > 6 \times 10^9$ g cm$^{-3}$
are found in the observational error box.
We remark that in PTD models, neutron-rich IPEs are synthesized
by the deflagration in the central region where $Y_{\rm e}$ is determined
by electron capture in NSE.  Thus Mn/Fe and Ni/Fe are sensitive
to the central $Y_{\rm e}$ and thus the central density, but not the initial 
metallicity.
In DDT models, on the other hand, IPEs are produced (in addition to 
the central deflagration) in the detonation at the low density outer
region, so that Mn/Fe and Ni/Fe are more sensitive to the initial $Y_{\rm e}$
and thus metallicity.

In \cite{Yamaguchi2015} the Cr/Fe mass ratio is also
measured as Cr/Fe = $0.027 \pm^{0.007}_{0.006}$. By comparing 

with our chemical yields listed in Tables \ref{table:Isotope1b}
and \ref{table:Isotope2b}, Cr/Fe is sensitive to the
central density as well. It sharply increases from 
$\sim 0.003$ in Model 100-133-1-c3-1 to
$\sim 0.064$ in Model 900-140-1-c3-1. The best
value lies around 0.029 of Model 750-139-1-c3-1 with $\rho_c = 7.5 \times 10^9$ g cm$^{-3}$.
Similar results appear for the hybrid
C+O+Ne WD which has Cr/Fe $\sim 0.005 - 0.045$
in our sampled density range. Model 750-050-1-c3-1 with $\rho_c = 7.5 \times 10^9$ g cm$^{-3}$
has the mass ratio Cr/Fe = 0.026 being closest to the observational
data. The sharp dependence on the central density
provides another precise indicator to identify the 
required numerical model. It suggests that by treating
Cr/Fe, Mn/Fe and Ni/Fe as a three-dimensional parameter
space can largely confine the potential SN Iax model
as a physical origin of SNR 3C397.

In order to judge if PTD is the origin of 3C397, 
further constraints apply to this object, 
such as the mass ratios of Ca/S and Ar/S.  The observed
mass ratios are Ca/S = 0.12 -- 0.16 and Ar/S = 0.17 -- 0.18 
\citep{MartinezRodriguez2017}. Our models with 
$\rho_c \geq 5.5 \times 10^9$ g cm$^{-3}$ give 
the mass ratios of Ca/S = 0.08 -- 0.16 and Ar/S = 0.14 -- 0.17
for C+O WDs and Ca/S = 0.18 -- 0.19 and Ar/S $\sim$0.20
for hybrid C+O+Ne WDs. The tight constraint by
Ar/S is challenging to the claim that 
3C 397 is an SN Iax origin. Future detection or 
no-detection of a low mass WD remnant will provide a 
definite indication to this physical picture.

Besides SNRs, the nucleosynthesis results presented in this
work can be further applied to the galactic chemical evolution
\citep[e.g.][]{Matteucci1986,Pagel1997,Kobayashi2019}
to identify the role of individual types of SNe
to the trend formation of specific elements as
a function of metallicity. In particular, SNe Iax can be 
important in dSph where the star formation history 
is largely different from ordinary galaxies. The unique 
abundance pattern of SNe Iax yields leaves 
observable effects. For example, in models for dSph
\citep{Kobayashi2015,Kobayashi2019,Cescutti2017},
SNe Iax contribute to form the evolutionary trends of 
[Mn/Fe] and [O/Fe] consistently.

\section{Conclusion}

In this work we have carried out the parameter survey for the explosive
nucleosynthesis in SNe Iax models using the pure turbulent deflagration model
as the explosion mechanism. We explored near-Chandrasekhar mass                   
C+O WDs and hybrid C+O+Ne WDs as the progenitors. 
We studied nucleosynthesis products in both the ejecta and bound remnants and their
parameter dependence, including the central density of the progenitor WD,  initial
flame structure, initial C/O ratio, and
turbulent flame speed formula for the two types of WDs.
Our results are summarized as follows:

\noindent (1) The ejecta is a mixture of burnt ash with iron-peak elements (IPEs) and
unburnt C+O-rich or C+ON+Ne-rich matter.

\noindent (2) The central density of the progenitor WD is the most important parameter
for chemical compositions of the ejecta and the remnant WD. 
The explosion models with higher central densities enhance production of V, Cr, Fe and Ni relative to Fe
by up to $\sim 100$ times the solar abundances  
in the ejecta and close to $10^3$ times in the remnant WDs.
The abundance in the ejecta is also sensitive to the 
initial flame structure. It is less sensitive to the 
C/O ratio and turbulent flame speed formula.

\noindent (3) The hybrid C+O+Ne WD shows similar features
to the C+O WD except that the ejected mass is smaller
due to the smaller nuclear energy release.
The ejecta includes IPEs with similar abundance patterns to the solar, 
while the remnant WD contains overproduced 
Ti, V and Cr. IMEs such as Si, S, Ar and Ca are underproduced
in both ejecta and the remnant WD.

\noindent (4) We compare the relations between the ejecta mass, explosion energy, 
and $^{56}$Ni mass in our models. We find 
a linear trend for the low energy explosion up to 
$\sim 4 \times 10^{50}$ erg and $^{56}$Ni up to 
$\sim 0.3 ~M_{\odot}$. We also examine the $^{56}$Ni mass
against the ejecta velocity to compare with 
the observational trends of SNe Iax. A clear
linear trend can be observed for the ejecta mass against
ejecta energy and $^{56}$Ni mass in the ejecta. 

\noindent (5) We further compare the model abundances with
the recently observed low mass WDs: LP40-365, J1603-6613 and J1825-3757.
Our models reproduce some aspects of the observed abundance pattern, 
especially the enhanced IPEs.
The observed high abundances of Cr, Mn, and Ni are consistent with 
the explosion at the central density as high as $\sim 5 \times 10^9$ g cm$^{-3}$.
However the high ratio [Co/Fe] cannot be reproduced.

\noindent (6) We compare our yields with the recently observed
supernova remnant 3C 397, which has super-solar Mn/Fe 
and Ni/Fe ratios. Our SN Iax models with the central density
higher than $\sim 5 \times 10^9$ g cm$^{-3}$ reproduce the measured high
Mn/Fe and Ni/Fe ratios. 
Thus SN Iax from a near Chandrasekhar mass WD with a high enough
central density (and solar metallicity) is a possible alternative to
the SN Ia with very high metallicity ($5 ~Z_{\odot}$)
\citep[e.g.,][]{Leung2018Chand, Leung2019subChand} for the model of 3C 397.
However, further accurate determination of the observed abundances
may be necessary to distinguish between models for 3C 397.

\section{Acknowledgments}
This work has been supported by the World Premier International
Research Center Initiative (WPI Initiative), MEXT, Japan, and JSPS
KAKENHI Grant Number JP17K05382 and JP20K04024.
S.C.L. also acknowledges the support by the funding HST-AR-15021.001-A
and 80NSSC18K1017.

We thank Francis Xavier Timmes for the open-source subroutines including the helmholtz equation of state, seven-isotope network and the torch nuclear reaction network.
S.C.L. thanks Friedrich Roepke and Florian Lach for the insightful discussion in the topic of deflagration and flame geometry. 
We thank Chiaki Kobayashi for the motivation
of this project from the galactic chemical evolution perspective. 
We thank Samuel Jones for useful discussion in nucleosynthesis.
We thank Hiroya Yamaguchi about the detailed modeling
of the supernova remnant 3C 397.

\appendix

\section{Overview and Input Physics of the Simulation Code}

In this section we briefly review the input physics we use 
for modeling the SNe Iax in this article. 

We use our own supernova hydrodynamics code 
for all the hydrodynamics simulation here. The code solves 
the two-dimensional Euler equations with shock-capturing
scheme. The spatial discretization is obtained by the 
$5^{\rm th}$-order Weighted Essentially Non-Oscillatory Scheme \citep{Barth1999}
and the time discretization is obtained by the 
5-step $3^{rd}$-order non-strong Stability Preserving
Runge-Kutta scheme \citep{Wang2007}.

We use the Helmholtz equation of state for modeling 
the microphysics \citep{Timmes1999a}.
The equation of state describes the properties
of non-interacting electron gas of arbitrarily 
relativistic and degenerate levels. It also 
contains contribution from nuclei as a classical
ideal gas, photon gas in Planck distribution
and electron-positron pair. The Coulomb correction
is also included for the screening effects
between electron gas and nuclei.

To describe the chemistry, the use the 
7-isotope network identical to \cite{Timmes1999b}. 
This network contains $^{4}$He, $^{12}$C, $^{16}$O,
$^{20}$Ne, $^{24}$Mg, $^{28}$Si and $^{56}$Ni.
We note that this is the most simplified network
one can use to describe the nuclear reaction
of CO matter and ONe matter. However, as all 
isotopes in this network are along the $\alpha$-chain
network, to accommodate the electron capture physics,
we treat the mean electron mole number $Y_e$ (also
known as the electron fraction) as an independent
quantity. It follows the fluid motion as a scalar
quantity, but it can be modified by 
including the electron capture rate (see below
for further discussion). With $Y_e$ is an
extra quantity, in the code, we treat the 
mean atomic number $\bar{Z}$ is the implied value 
from the mean mass number $\bar{A}$ and $Y_e$
that $\bar{Z} = \bar{A} Y_e$. 


\section{Comparison with Literature}

\begin{table*}
\begin{center}
\caption{Comparison of the input physics with the works in the literature.
"Hydrodynamics" is the hydrodynamics solver used for solving the Euler equations.
"Dimensionality" is the number of dimension used in the simulations. 
"Microphysics" is the equation of state used.
"E-cap scheme" is the electron capture scheme.
"PP isotope no." is the number of isotopes used in the post-processing
with "tracer no." is the number of passive tracers for recording the thermodynamical history.
"Flame Cap" is the flame capturing scheme used for tracking the deflagration front. 
"Nuc. network" is the simplified network used in the hydrodynamics simulations
with "3-step" for the three-step nuclear reaction scheme described in \cite{Townsley2007}
and "table" for the pre-built nuclear reaction table.
"Hyd isotope no." is the number of isotopes used in the hydrodynamics simulations.
"SSG" is the sub-grid scale turbulence scheme used to model the development of eddy motion.
"n/a" in the table means that no exact implementation details can be found. 
}
\begin{tabular}{|c|c|c|c|c|c|}
\hline
physics & this work & \cite{Reinecke2002a} & \cite{Jones2016} & \cite{Long2014} & \cite{Fink2014} \\ \hline
hydrodynamics 	& WENO & prometheus & prometheus & FLASH & prometheus \\ \hline
dimensionality 	& 2 & 2 & 3 & 3 & 3 \\ \hline
microphysics 	& helmholtz & private & private & helmholtz & private \\ \hline
e-cap scheme 	& extended & n/a & extended & n/a & \cite{Seitenzahl2009} \\
pp isotope no. 	& 495 & n/a & 384 & n/a & 384 \\
tracer no. 		& 160$^2$ & n/a &  & $10^7$ & 200$^3$ \\ 
flame cap 		& level-set & level-set & level-set & 3-step & level-set \\
nuc. network 	& 3-step & table & table & 3-step & table \\
Hyd isotope	no.	& 7 & 5 & 5 & 3 & 5 \\ \hline
SSG 			& one-eq. model & one-eq. model & one-eq. model & n/a & one-eq. model \\ \hline

\end{tabular}

\end{center}
\label{table:compare}
\end{table*}

\begin{figure*}
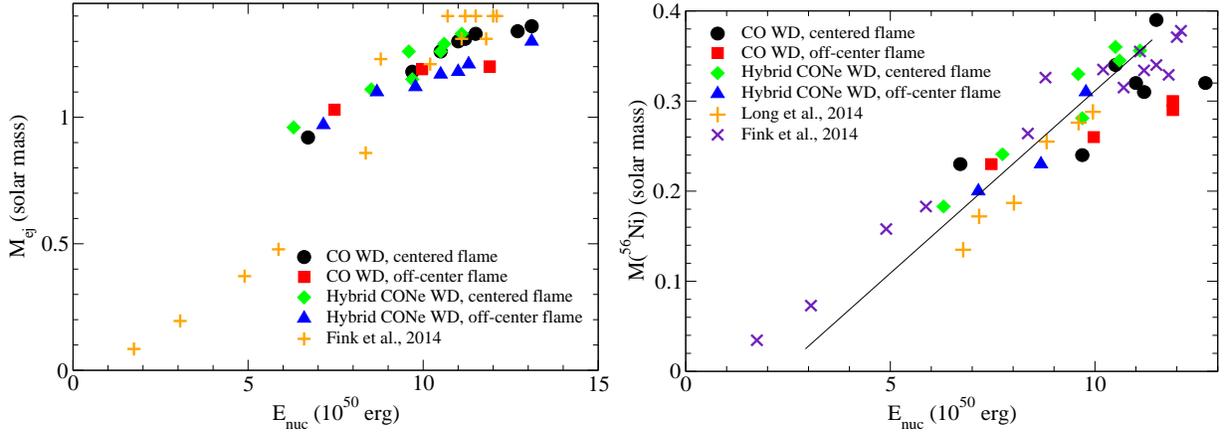

\centering
\includegraphics*[width=8cm,height=5.7cm]{fig32a.eps}
\includegraphics*[width=8cm,height=5.7cm]{fig32b.eps}
\caption{(left panel) The ejecta mass against the total energy released 
by nuclear reactions computed by our models, and in those
reported in \cite{Fink2014}.  (right panel) Similar to the left
panel but for the $^{56}$Ni-mass against the total energy
released by nuclear reactions and also in those reported
in \cite{Fink2014,Long2014}. A fitted straight line is shown
to demonstrate the trend of the models.}
\label{fig:Mej_Enuc_comp_plot}
\end{figure*}

\subsection{\cite{Reinecke2002a} and related works}

Our code has a similar structure with their works but with 
some distinctive differences. 
In Table \ref{table:compare} we tabulate the similarities and 
differences of our code compared with theirs. In general our
codes are similar as we make detailed references while
validating our code performance. 

Our model 300-000-1-c3-1 is similar to their model c3\_2d\_256 
in terms of resolution and initial flame structure. 
They have a total nuclear reaction energy of $7.19 \times 10^{50}$ erg
and "Ni"-production of 0.40 $M_{\odot}$. $\sim 0.6~M_{\odot}$
of matter is burnt. In our model, we have a stronger explosion of
$9.69 \times 10^{50}$ erg
released by nuclear reaction and 0.35 $M_{\odot}$ of $^{56}$Ni.
0.625 $M_{\odot}$ of matter is burnt by the deflagration. 
The difference in the choice of equation of state, 
electron capture rate and also the detailed implementation
of sub-grid scale turbulence can contribute to the observed
differences. Despite that, we obtain qualitative very similar
flame structure as seen from their Figure 2. 

\subsection{\cite{Jones2016}}

This work is based on the extension from \cite{Reinecke2002a} and
the later extension in the microphysics, in particular the electron 
capture table for the NSE matter by including rates from e.g. 
\cite{Nabi2004}. They focus on the deflagration phase of the 
ONe core in the context of electron capture supernova evolved
super-AGB stars \citep{Jones2013}.  

Their three-dimensional code allows them to explore complex
off-center flame structure. Their model G15 is similar to 
our Model 900-140-1-c3-1 (in Appendix) but differ by flame structure owing to
their three-dimensional freedom in flame placement. 
They observe ejecta and remnant masses of 0.177 and 1.212 $M_{\odot}$.
On the other hand we observe a larger ejecta mass by 0.567 $M_{\odot}$.
The difference might be originated from the initial flame we use (c3 flame), which is extended in size $\sim 100$ km. This also
enhances the energy release by deflagration. Also, its 
aspherical shape facilitates the turbulence production and
hence the amount of matter burnt at early time. This suppresses
the initial electron capture effect.

\subsection{\cite{Long2014}}

This code uses another code FLASH for modeling the deflagration
phase of SNe Ia. Again, we tabulate their input physics in Table \ref{table:compare}.
Major differences of this code is that the flame propagation does
not depend on sub-grid scale turbulence, but only Rayleigh-Taylor
instabilities. Notice that such instabilities depend on 
local gravity $g$. Near the core, the local gravity
scales as $g(r) \sim m(r)/r^2 \sim \rho r \rightarrow 0$. 
This means the flame is almost like laminar near the center.
On the other hand, there is no such restriction in 
sub-grid scale turbulence model. As long as the shear-stress is 
strong, eddy motion can be generated anywhere inside the star
which accelerates the flame propagation.

\cite{Long2014} construct a WD of mass 1.365 $M_{\odot}$ with equal mass fractions
of $^{12}$C and $^{16}$O. Again, the use of three-dimensional hydrodynamics
offers the possibility to explore off-center bubble flame structure. 

Due to the weaker flame propagation, they observe a higher bound
mass for the same amount of bubbles used, when compared with \cite{Fink2014}.
We notice the drastic difference in the flame structure. 
While flame structure
affects strongly the explosion energetics in three-dimensional simulation,
we only compare with the global trends of their models. 
We do not compare the ejecta mass because we cannot find the 
corresponding numbers. 

In the right panel of Figure \ref{fig:Mej_Enuc_comp_plot} we plot the 
$^{56}$Ni-mass in the ejecta against nuclear energy release 
of all models. We can see a very consistent trend
between our work and their work. 
This suggests that our code 
can capture a consistent result in following how 
the energy input from deflagration contributes to the final 
ejecta mass. Our models show also a narrow band
as theirs which can teach as high as $M_{\rm Ni}$. This means that
in the high-mass regime, the faster and stronger nuclear
flame is balanced by other effects, such as the electron
capture effects. The production of $^{56}$Ni therefore 
becomes saturated.
However, our models does not show that $E_{\rm nuc}$ reaches as 
low as theirs. Such models will need further exploration of 
three-dimensional flame structure where much smaller initial
flame in the form of a few bubbles is used.

\subsection{\cite{Fink2014}}

Similar to \cite{Seitenzahl2013}, \cite{Fink2014} carried out a parameter
survey for the pure turbulent deflagration for the near-Chandrasekhar mass
Model in three-dimensions. The code structure inherits from previous 
version such as \cite{Reinecke2002a} but with updated turbulence
calculation described in \citep{Schmidt2006b}.
A wide range of flame structure from 1 bubble to 1600 
bubbles are used to study their effects on the chemical abundance pattern and its observable. The progenitor has a central density $2.9 \times 10^9$ g cm$^{-3}$
and a composition of $X(^{12}$C)$ = 0.475$, $X(^{16}$O)$ = 0.5$
and $X(^{22}$Ne)$ = 0.025$. The ejecta mass increases from $\sim 0.08 ~M_{\odot}$
to complete disruption $(= 1.4 ~M_{\odot})$ when bubble count exceeds 150. 

Since it is difficult to compare models with a different flame 
structure, we compare the global trends of their models with 
ours. In their models they show that the ejecta mass increases
almost linearly with the explosion energy. We also plot the explosion
energetics of their models in both panels of Figure \ref{fig:Mej_Enuc_comp_plot}.

In the left panel, we plot the relation between the ejecta mass and 
the total nuclear energy released. 
We see a very clear linear trend in our models. 
Their models also show a similar trend but with a lower
end and higher end of the ejecta mass.
Despite the difference in the range, the slope, which characterizes
how the deflagration contributes to the mass ejection, agrees
with each other in the high $E_{\rm nuc}$ limit. 
Furthermore, our models are 
confined in a narrow bin from the fitted straight line.
But we also notice for model with $E_{\rm nuc} < 10^{51]}$ erg,
their model has a steeper slope than ours. We notice
that the diversity of their model depends on the initial
flame instead of the progenitor mass. A centered flame like ours
provides a sufficiently large surface area to maintain 
burning before the expansion quenches the deflagration wave.
Therefore our models tend to cluster in the strong explosion side
while in their work the models span from weak explosions to 
strong explosions.

In the right panel, we make a similar plot to the left panel but
for the $^{56}$Ni-mass production. Again the trends among 
all works show a promising similarity. The models in their work
show again the lower $^{56}$Ni-mass in the lower end. Their models
show a similar                            
$^{56}$Ni-mass in the upper end. The consistency in the 
slope, as well as the thin band formed by all the models,
show that our code agree well with their hydrodynamics
and post-process results.

\section{O+Ne+Mg WD}

In the main text we have presented a detailed description about
how the nucleosynthesis varies with the model parameters of 
a C+O WD and a hybrid C+O+Ne
WD. In fact, super-AGB star
with a mass $8 - 10M_{\odot}$ can also form a massive O+Ne+Mg WD,
where the core or shell O-burning is ignited by by electron capture
and
can trigger oxygen deflagration. 

The outcome of the oxygen deflagration is most sensitive to the
central density of the WD when the deflagration is initiated, which is
called as the deflagration density $\rho_{\rm c,def}$.  This is
because electron capture on NSE materials leads to the collapse of the
WD, while nuclear energy release by the oxygen deflagration leads to
the expansion.

\cite{Zha2019} investigated how $\rho_{\rm c,def}$ depends on the
electron capture rate, nuclear reaction rate, URCA process, and
convective criterion and concluded $\log_{10}(\rho_{\rm
  c,def}/\mathrm{g~cm^{-3}})>10.10$ is most likely.  Such high
$\rho_{\rm c,def}$ has been found to lead to collapse to form a
neutron star rather than thermonuclear explosion
\citep{Leung2019ECSN,Zha2019}.

Earlier \cite{Jones2016,Jones2019} have found the the outcome 
for $\log_{10}(\rho_{\rm c,def}/\mathrm{g~cm^{-3}})=9.95$ is
thermonuclear explosion that leaves a small mass WD behind, being
similar to PTD models presented in the main text.

Although we think the collapse is the likely outcome of oxygen
deflagration in the actually O+Ne+Mg WDs with$\log_{10}(\rho_{\rm
  c,def}/\mathrm{g~cm^{-3}})>10.10$, we perform numerical experiments
of the pure oxygen deflagration at lower $\rho_{\rm c,def}$, i.e.,
$\log_{10}(\rho_{\rm c,def}/\mathrm{g~cm^{-3}})=9.95$ and 9.90 for the
purpose of comparisons with \cite{Jones2016} and with the similar
$\rho_{\rm c,def}$ as C+O and C+O+Ne WD models.


In this section, we examine the explosive nucleosynthesis of the
O+Ne+Mg WD. In Table \ref{table:Isotope3}, \ref{table:Isotope3b} and
\ref{table:Decay3} we tabulate the nucleosynthesis yield and the
radioactive isotopes in the ejecta.


\subsection{Ejecta of O+Ne+Mg WD}

\subsubsection{Dependence on Central Density}

\begin{figure}
\centering
\includegraphics*[width=8cm,height=5.7cm]{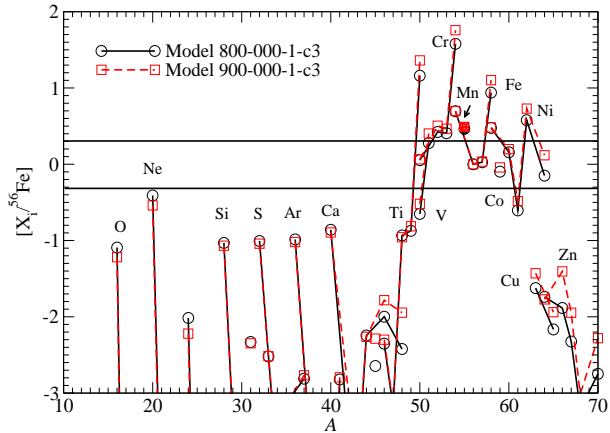}
\caption{$[X_i/^{56}$Fe] against
mass number for the ejecta 
of O+Ne+Mg WD models for
800-000-1-c3-1 ($\rho_c \approx 10^{9.9}$ g cm$^{-3}$) and 
900-000-1-c3-1 ($\rho_c \approx 10^{9.95}$ g cm$^{-3}$).
All models assume O+Ne+Mg composition with,
$X(^{22}$Ne) = 0.025 and $c3$ initial flame.}
\label{fig:final_esc_M_ONeMg_plot}
\end{figure}

Here we compare the nucleosynthesis pattern of O+Ne+Mg WD models
of different central densities. 
For our numerical experiments, we adopt the $\rho_c$ range
from $10^{9.90-9.95}$ g cm$^{-3}$
(i.e. $8 - 9 \times 10^9$ g cm$^{-3}$). Higher $\rho_c$ 
in general result in a collapse to form a neutron star.

We compare in Figure \ref{fig:final_esc_M_ONeMg_plot}
the ejecta chemical composition of Models 800-000-1-c3-1
and 900-000-1-c3-1. 
Notice that the typical explosion of the O+Ne+Mg WD is weaker than the
C+O counterpart because the nuclear energy release from O+Ne+Mg matter to NSE is smaller than
the energy release from C+O matter to NSE.

The central density plays an important role 
in the ejecta composition for the O+Ne+Mg WD case. 
IMEs are underproduced and most IPEs 
are significantly produced. The models show to have a 
similar composition except that the lower mass 
one shows a higher yield of $^{50}$Ti, $^{50-51}$V,
$^{54}$Cr, $^{58}$Fe and $^{60,62}$Ni. 
During the fallback 
of matter in the inner layer, most IPEs
including neutron-rich isotopes are trapped in the 
remnant. Also, the higher density WD is more compact,
thus the inner layer requires a higher escape velocity
for mass ejection. As a result, a larger amount of the inner part of
matter is trapped, which suppresses the increase in neutron-rich isotopes.

\subsubsection{Dependence on Initial Flame Structure}

\begin{figure}
\centering
\includegraphics*[width=8cm,height=5.7cm]{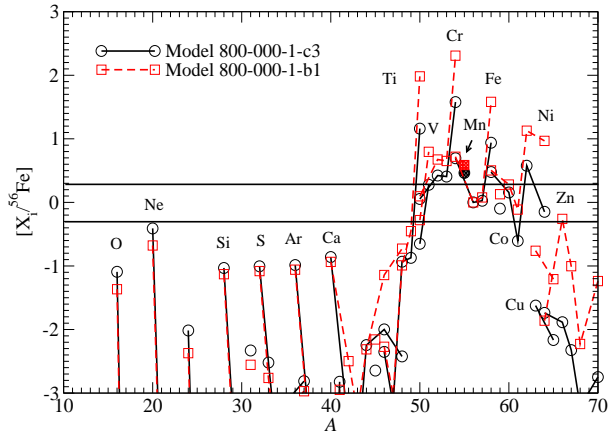}
\caption{$[X_i/^{56}$Fe] against
mass number for the ejecta 
of pure ONe WD models for
800-000-1-c3-1 ($c3$ flame) and 
800-000-1-b1-1 ($b1$ flame).
All models assume $\rho_c = 10^{9.90} \times 10^9$ g cm$^{-3}$, O+Ne+Mg composition and
$X(^{22}$Ne) = 0.025.}
\label{fig:final_esc_flame_ONeMg_plot}
\end{figure}

Here we compare the nucleosynthesis yield of the O+Ne+Mg WD
models with different initial flame. In Figure \ref{fig:final_esc_flame_ONeMg_plot}
we plot $[X_i/^{56}$Fe] for Models 800-000-1-c3-1 and 800-000-1-b1-1. 

The ejecta properties are significantly different
when the flame starts from off-center. 
Major isotopes of IPEs are similar. But the 
amount of neutron-rich isotopes can be 1 -- 2 orders of magnitudes
higher for the off-center flame than the centered flame. This is because
the outer flame can channel the ash outwards 
at an earlier time, which allows more matter containing neutron-rich isotopes
to escape from the gravitational pull of the star. 
On the other hand, IMEs are only mildly changed. 

\subsection{Remnant White Dwarf}

\subsubsection{Dependence on Central Density}

\begin{figure}
\centering
\includegraphics*[width=8cm,height=5.7cm]{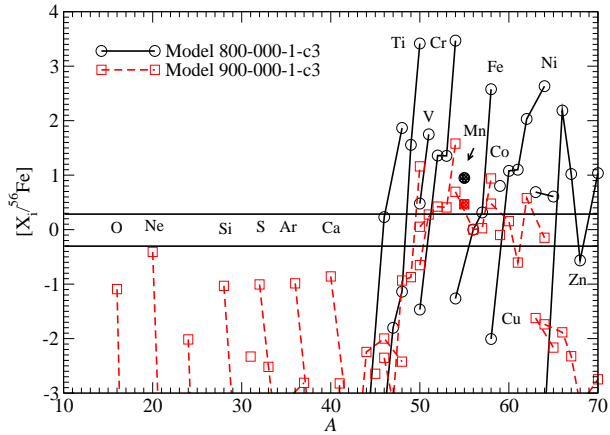}
\caption{$[X_i/^{56}$Fe] against
mass number for the bounded remnant
of O+Ne+Mg WD models for Models
800-000-1-c3-1 ($\rho_c \approx 10^{9.9}$ g cm$^{-3}$) and 
900-000-1-c3-1 ($\rho_c \approx 10^{9.95}$ g cm$^{-3}$).
All models assume O+Ne+Mg composition with
$X(^{22}$Ne) = 0.025 and $c3$ initial flame.}
\label{fig:final_bnd_M_ONeMg_plot}
\end{figure}

We compare the remnant composition for O+Ne+Mg WDs with different
$\rho_c$ by comparing Models 800-000-1-c3-1 and 
900-000-1-c3-1. 
In Figure \ref{fig:final_bnd_M_ONeMg_plot}, 
we plot [$X_i/^{56}$Fe] for these two models.
The remnant WD exhibits very peculiar abundance
as also seen in the high density models of C+O and C+O+Ne WDs,
where IPEs can be $10^3$ times higher than the solar value. 
The different central densities provides a very different abundance pattern
in the remnant WD.
The lower density model (800-000-1-c3) has a much higher 
neutron-rich isotopes of IPEs than the higher density model
(900-000-1-c3) in the remnant WD. They can be $\sim 10^{3-4}$ times higher than 
the solar ratios. Also, there is no observable amount of IMEs.
Meanwhile, Model 900-000-1-c3 shows the abundance pattern closer
to the solar composition.
The IMEs are underproduced at a level of $\sim 10 \%$ of the solar
values. The most overproduced isotope is $^{54}$Cr and it is only
$\sim 100$ times higher, which is almost two orders of magnitude
lower than that in Model 800-000-1-c3.

\subsubsection{Dependence on Initial Flame Structure}

\begin{figure}
\centering
\includegraphics*[width=8cm,height=5.7cm]{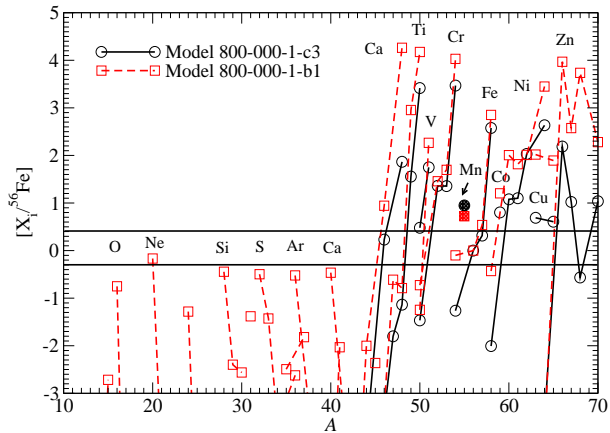}
\caption{$[X_i/^{56}$Fe] against
atomic mass for the remnant of O+Ne+Mg WD models for
800-000-1-c3-1 ($c3$ flame) and 
800-000-1-b1-1 ($b1$ flame).
All models assume $\rho_c = 10^{9.90}$ g cm$^{-3}$, O+Ne+Mg composition with $X(^{22}$Ne) = 0.025.}
\label{fig:final_bnd_flame_ONeMg_plot}
\end{figure}

In Figure \ref{fig:final_bnd_flame_ONeMg_plot} we compare
the abundance pattern for the remnant part of 
Models 800-000-1-c3-1 and 800-000-1-b1-1. 
The remnant in the centered flame model has no observable O+Ne+Mg-rich
matter or IMEs. The overproduction of IPEs, especially 
neutron-rich isotopes, are similar to the high density 
C+O and hybrid C+O+Ne models.
However, the off-center model has more neutron-rich isotopes
than the centered flame model. Again, in such a high
density, electron capture begins to be important to the global
dynamics as it suppresses the pressure jump after the 
matter is burnt. The ejecta in the centered flame is
more suppressed than the ejecta in the off-center flame.

\bibliographystyle{apj}
\pagestyle{plain}
\bibliography{biblio}

\newpage

\begin{table*}

\begin{center}
\caption{Mass of major isotopes in the ejecta after all 
short-lived radioactive isotopes have decayed. The 
isotope masses are in units of solar mass.}
\label{table:Isotope1}
\begin{tabular}{|c| c c c c c c c|}
\hline
Isotopes &  100-133-1-c3-1 & 200-135-1-c3-1 & 300-137-1-c3-1 & 500-138-1-c3-1 & 550-138-1-c3-1 & 750-139-1-c3-1 & 900-140-1-c3-1  \\ \hline

$^{12}$C & $2.96 \times 10^{-1}$ & $3.81 \times 10^{-1}$ & $3.21 \times 10^{-1}$ & $3.14 \times 10^{-1}$ & $3.14 \times 10^{-1}$ & $2.43 \times 10^{-1}$ & $1.82 \times 10^{-1}$ \\
 $^{13}$C & $6.56 \times 10^{-12}$ & $4.44 \times 10^{-11}$ & $6.50 \times 10^{-11}$ & $6.55 \times 10^{-11}$ & $5.31 \times 10^{-11}$ & $1.42 \times 10^{-10}$ & $2.9 \times 10^{-10}$ \\
 $^{14}$N & $9.12 \times 10^{-10}$ & $3.85 \times 10^{-9}$ & $5.85 \times 10^{-9}$ & $5.37 \times 10^{-9}$ & $4.72 \times 10^{-9}$ & $1.23 \times 10^{-8}$ & $1.87 \times 10^{-8}$ \\
 $^{15}$N & $2.77 \times 10^{-10}$ & $1.44 \times 10^{-9}$ & $1.44 \times 10^{-9}$ & $1.29 \times 10^{-9}$ & $1.43 \times 10^{-9}$ & $3.17 \times 10^{-9}$ & $4.53 \times 10^{-9}$ \\
 $^{16}$O & $3.1 \times 10^{-1}$ & $4.8 \times 10^{-1}$ & $3.54 \times 10^{-1}$ & $3.44 \times 10^{-1}$ & $3.40 \times 10^{-1}$ & $3.17 \times 10^{-1}$ & $2.82 \times 10^{-1}$ \\
 $^{17}$O & $2.96 \times 10^{-10}$ & $1.33 \times 10^{-9}$ & $2.9 \times 10^{-9}$ & $1.93 \times 10^{-9}$ & $1.70 \times 10^{-9}$ & $4.41 \times 10^{-9}$ & $6.78 \times 10^{-9}$ \\
 $^{18}$O & $9.15 \times 10^{-12}$ & $3.86 \times 10^{-11}$ & $6.27 \times 10^{-11}$ & $5.59 \times 10^{-11}$ & $5.2 \times 10^{-11}$ & $1.29 \times 10^{-10}$ & $1.99 \times 10^{-10}$ \\
 $^{19}$F & $3.57 \times 10^{-12}$ & $1.52 \times 10^{-11}$ & $1.81 \times 10^{-11}$ & $1.75 \times 10^{-11}$ & $1.65 \times 10^{-11}$ & $4.23 \times 10^{-11}$ & $6.14 \times 10^{-11}$ \\
 $^{20}$Ne & $7.40 \times 10^{-4}$ & $3.10 \times 10^{-3}$ & $3.23 \times 10^{-3}$ & $3.17 \times 10^{-3}$ & $3.17 \times 10^{-3}$ & $8.1 \times 10^{-3}$ & $1.11 \times 10^{-2}$ \\
 $^{21}$Ne & $2.46 \times 10^{-8}$ & $1.3 \times 10^{-7}$ & $1.8 \times 10^{-7}$ & $1.12 \times 10^{-7}$ & $1.3 \times 10^{-7}$ & $2.80 \times 10^{-7}$ & $3.88 \times 10^{-7}$ \\
 $^{22}$Ne & $1.20 \times 10^{-2}$ & $1.55 \times 10^{-2}$ & $1.31 \times 10^{-2}$ & $1.28 \times 10^{-2}$ & $1.28 \times 10^{-2}$ & $9.90 \times 10^{-3}$ & $7.43 \times 10^{-3}$ \\
 $^{23}$Na & $2.20 \times 10^{-6}$ & $9.76 \times 10^{-6}$ & $1.5 \times 10^{-5}$ & $1.5 \times 10^{-5}$ & $1.2 \times 10^{-5}$ & $2.49 \times 10^{-5}$ & $3.62 \times 10^{-5}$ \\
 $^{24}$Mg & $7.48 \times 10^{-4}$ & $3.58 \times 10^{-3}$ & $3.99 \times 10^{-3}$ & $3.75 \times 10^{-3}$ & $3.44 \times 10^{-3}$ & $9.1 \times 10^{-3}$ & $1.18 \times 10^{-2}$ \\
 $^{25}$Mg & $5.9 \times 10^{-6}$ & $1.98 \times 10^{-5}$ & $2.20 \times 10^{-5}$ & $2.13 \times 10^{-5}$ & $2.6 \times 10^{-5}$ & $5.39 \times 10^{-5}$ & $7.45 \times 10^{-5}$ \\
 $^{26}$Mg & $7.24 \times 10^{-6}$ & $3.14 \times 10^{-5}$ & $3.35 \times 10^{-5}$ & $3.28 \times 10^{-5}$ & $3.20 \times 10^{-5}$ & $8.3 \times 10^{-5}$ & $1.14 \times 10^{-4}$ \\
 $^{26}$Al & $2.39 \times 10^{-29}$ & $3.6 \times 10^{-29}$ & $3.27 \times 10^{-29}$ & $9.62 \times 10^{-10}$ & $9.33 \times 10^{-10}$ & $2.54 \times 10^{-9}$ & $3.53 \times 10^{-29}$ \\
 $^{27}$Al & $6.4 \times 10^{-5}$ & $2.67 \times 10^{-4}$ & $3.1 \times 10^{-4}$ & $2.78 \times 10^{-4}$ & $2.59 \times 10^{-4}$ & $7.23 \times 10^{-4}$ & $9.40 \times 10^{-4}$ \\
 $^{28}$Si & $1.21 \times 10^{-2}$ & $3.64 \times 10^{-2}$ & $4.0 \times 10^{-2}$ & $4.36 \times 10^{-2}$ & $3.61 \times 10^{-2}$ & $5.5 \times 10^{-2}$ & $6.3 \times 10^{-2}$ \\
 $^{29}$Si & $6.71 \times 10^{-5}$ & $3.9 \times 10^{-4}$ & $3.51 \times 10^{-4}$ & $3.36 \times 10^{-4}$ & $2.97 \times 10^{-4}$ & $8.21 \times 10^{-4}$ & $1.10 \times 10^{-3}$ \\
 $^{30}$Si & $9.78 \times 10^{-5}$ & $4.89 \times 10^{-4}$ & $5.78 \times 10^{-4}$ & $5.34 \times 10^{-4}$ & $4.62 \times 10^{-4}$ & $1.36 \times 10^{-3}$ & $1.78 \times 10^{-3}$ \\
 $^{31}$P & $2.52 \times 10^{-5}$ & $1.6 \times 10^{-4}$ & $1.27 \times 10^{-4}$ & $1.17 \times 10^{-4}$ & $1.0 \times 10^{-4}$ & $2.98 \times 10^{-4}$ & $3.92 \times 10^{-4}$ \\
 $^{32}$S & $6.28 \times 10^{-3}$ & $1.61 \times 10^{-2}$ & $1.75 \times 10^{-2}$ & $1.92 \times 10^{-2}$ & $1.58 \times 10^{-2}$ & $1.97 \times 10^{-2}$ & $2.25 \times 10^{-2}$ \\
 $^{33}$S & $1.99 \times 10^{-5}$ & $7.75 \times 10^{-5}$ & $9.38 \times 10^{-5}$ & $8.88 \times 10^{-5}$ & $7.29 \times 10^{-5}$ & $2.23 \times 10^{-4}$ & $2.91 \times 10^{-4}$ \\
 $^{34}$S & $1.6 \times 10^{-4}$ & $5.7 \times 10^{-4}$ & $6.13 \times 10^{-4}$ & $6.2 \times 10^{-4}$ & $4.76 \times 10^{-4}$ & $1.8 \times 10^{-3}$ & $1.47 \times 10^{-3}$ \\
 $^{36}$S & $1.2 \times 10^{-8}$ & $4.88 \times 10^{-8}$ & $5.80 \times 10^{-8}$ & $5.46 \times 10^{-8}$ & $4.62 \times 10^{-8}$ & $1.52 \times 10^{-7}$ & $1.97 \times 10^{-7}$ \\
 $^{35}$Cl & $1.11 \times 10^{-5}$ & $3.30 \times 10^{-5}$ & $3.82 \times 10^{-5}$ & $3.62 \times 10^{-5}$ & $3.24 \times 10^{-5}$ & $7.73 \times 10^{-5}$ & $1.6 \times 10^{-4}$ \\
 $^{37}$Cl & $1.84 \times 10^{-6}$ & $5.6 \times 10^{-6}$ & $5.83 \times 10^{-6}$ & $6.19 \times 10^{-6}$ & $4.75 \times 10^{-6}$ & $8.54 \times 10^{-6}$ & $1.10 \times 10^{-5}$ \\
 $^{36}$Ar & $1.15 \times 10^{-3}$ & $2.60 \times 10^{-3}$ & $2.83 \times 10^{-3}$ & $3.6 \times 10^{-3}$ & $2.53 \times 10^{-3}$ & $2.69 \times 10^{-3}$ & $2.89 \times 10^{-3}$ \\
 $^{38}$Ar & $6.68 \times 10^{-5}$ & $2.33 \times 10^{-4}$ & $2.71 \times 10^{-4}$ & $2.83 \times 10^{-4}$ & $2.21 \times 10^{-4}$ & $2.87 \times 10^{-4}$ & $3.46 \times 10^{-4}$ \\
 $^{40}$Ar & $1.71 \times 10^{-10}$ & $6.25 \times 10^{-10}$ & $7.50 \times 10^{-10}$ & $6.68 \times 10^{-10}$ & $5.94 \times 10^{-10}$ & $2.42 \times 10^{-9}$ & $3.11 \times 10^{-9}$ \\
 $^{39}$K & $6.49 \times 10^{-6}$ & $1.64 \times 10^{-5}$ & $1.82 \times 10^{-5}$ & $1.98 \times 10^{-5}$ & $1.49 \times 10^{-5}$ & $1.96 \times 10^{-5}$ & $2.61 \times 10^{-5}$ \\
 $^{40}$K & $3.89 \times 10^{-9}$ & $9.94 \times 10^{-9}$ & $1.19 \times 10^{-8}$ & $1.3 \times 10^{-8}$ & $9.65 \times 10^{-9}$ & $3.31 \times 10^{-8}$ & $4.39 \times 10^{-8}$ \\
 $^{41}$K & $4.85 \times 10^{-7}$ & $1.9 \times 10^{-6}$ & $1.14 \times 10^{-6}$ & $1.27 \times 10^{-6}$ & $9.28 \times 10^{-7}$ & $1.17 \times 10^{-6}$ & $1.47 \times 10^{-6}$ \\
 $^{40}$Ca & $1.4 \times 10^{-3}$ & $2.12 \times 10^{-3}$ & $2.34 \times 10^{-3}$ & $2.47 \times 10^{-3}$ & $2.5 \times 10^{-3}$ & $1.98 \times 10^{-3}$ & $1.97 \times 10^{-3}$ \\
 $^{42}$Ca & $2.28 \times 10^{-6}$ & $6.57 \times 10^{-6}$ & $7.47 \times 10^{-6}$ & $7.98 \times 10^{-6}$ & $6.8 \times 10^{-6}$ & $8.44 \times 10^{-6}$ & $1.3 \times 10^{-5}$ \\
 $^{43}$Ca & $7.46 \times 10^{-9}$ & $1.55 \times 10^{-8}$ & $2.24 \times 10^{-8}$ & $1.92 \times 10^{-8}$ & $1.62 \times 10^{-8}$ & $5.71 \times 10^{-8}$ & $7.92 \times 10^{-8}$ \\
 $^{44}$Ca & $8.41 \times 10^{-7}$ & $1.40 \times 10^{-6}$ & $1.87 \times 10^{-6}$ & $1.84 \times 10^{-6}$ & $1.56 \times 10^{-6}$ & $1.84 \times 10^{-6}$ & $2.3 \times 10^{-6}$ \\
 $^{46}$Ca & $2.80 \times 10^{-12}$ & $9.47 \times 10^{-12}$ & $1.29 \times 10^{-11}$ & $1.31 \times 10^{-11}$ & $2.82 \times 10^{-10}$ & $6.17 \times 10^{-9}$ & $4.36 \times 10^{-8}$ \\
 $^{48}$Ca & $1.68 \times 10^{-17}$ & $5.34 \times 10^{-17}$ & $6.59 \times 10^{-17}$ & $2.66 \times 10^{-14}$ & $2.3 \times 10^{-11}$ & $3.11 \times 10^{-9}$ & $1.78 \times 10^{-6}$ \\
 $^{45}$Sc & $1.84 \times 10^{-8}$ & $3.64 \times 10^{-8}$ & $4.41 \times 10^{-8}$ & $4.68 \times 10^{-8}$ & $3.63 \times 10^{-8}$ & $7.75 \times 10^{-8}$ & $9.94 \times 10^{-8}$ \\
 $^{46}$Ti & $1.6 \times 10^{-6}$ & $2.86 \times 10^{-6}$ & $3.20 \times 10^{-6}$ & $3.64 \times 10^{-6}$ & $2.65 \times 10^{-6}$ & $3.27 \times 10^{-6}$ & $3.98 \times 10^{-6}$ \\
 $^{47}$Ti & $4.20 \times 10^{-8}$ & $7.71 \times 10^{-8}$ & $1.6 \times 10^{-7}$ & $1.10 \times 10^{-7}$ & $9.56 \times 10^{-8}$ & $1.96 \times 10^{-7}$ & $2.77 \times 10^{-7}$ \\
 $^{48}$Ti & $2.22 \times 10^{-5}$ & $3.68 \times 10^{-5}$ & $4.80 \times 10^{-5}$ & $4.75 \times 10^{-5}$ & $4.42 \times 10^{-5}$ & $4.97 \times 10^{-5}$ & $5.6 \times 10^{-5}$ \\
 $^{49}$Ti & $1.96 \times 10^{-6}$ & $3.40 \times 10^{-6}$ & $4.77 \times 10^{-6}$ & $5.27 \times 10^{-6}$ & $5.34 \times 10^{-6}$ & $6.74 \times 10^{-6}$ & $1.29 \times 10^{-5}$ \\
 $^{50}$Ti & $2.47 \times 10^{-11}$ & $6.76 \times 10^{-11}$ & $2.84 \times 10^{-10}$ & $8.4 \times 10^{-7}$ & $5.20 \times 10^{-5}$ & $6.77 \times 10^{-4}$ & $3.68 \times 10^{-3}$ \\
 $^{50}$V & $1.74 \times 10^{-10}$ & $4.0 \times 10^{-10}$ & $1.44 \times 10^{-9}$ & $6.51 \times 10^{-9}$ & $2.43 \times 10^{-8}$ & $7.21 \times 10^{-8}$ & $1.3 \times 10^{-7}$ \\
 $^{51}$V & $9.23 \times 10^{-6}$ & $1.45 \times 10^{-5}$ & $2.73 \times 10^{-5}$ & $3.60 \times 10^{-5}$ & $6.88 \times 10^{-5}$ & $2.36 \times 10^{-4}$ & $5.16 \times 10^{-4}$ \\
  
 \end{tabular}
\end{center}
\end{table*}
 
\begin{table*}

\begin{center}
\caption{(cont'd) Mass of major isotopes in the ejecta after all 
short-lived radioactive isotopes have decayed. The 
isotope masses are in units of solar mass.}
\label{table:Isotope1b}
\begin{tabular}{|c| c c c c c c c|}
\hline
Isotopes &  100-133-1-c3-1 & 200-135-1-c3-1 & 300-137-1-c3-1 & 500-138-1-c3-1 & 550-138-1-c3-1 & 750-139-1-c3-1 & 900-140-1-c3-1  \\ \hline

$^{50}$Cr & $3.53 \times 10^{-5}$ & $6.40 \times 10^{-5}$ & $1.46 \times 10^{-4}$ & $2.5 \times 10^{-4}$ & $2.19 \times 10^{-4}$ & $2.49 \times 10^{-4}$ & $2.74 \times 10^{-4}$ \\
 $^{52}$Cr & $7.60 \times 10^{-4}$ & $1.13 \times 10^{-3}$ & $2.1 \times 10^{-3}$ & $3.25 \times 10^{-3}$ & $6.4 \times 10^{-3}$ & $1.25 \times 10^{-2}$ & $1.63 \times 10^{-2}$ \\
 $^{53}$Cr & $1.36 \times 10^{-4}$ & $1.94 \times 10^{-4}$ & $4.59 \times 10^{-4}$ & $6.56 \times 10^{-4}$ & $8.40 \times 10^{-4}$ & $1.36 \times 10^{-3}$ & $1.93 \times 10^{-3}$ \\
 $^{54}$Cr & $1.14 \times 10^{-8}$ & $3.74 \times 10^{-8}$ & $1.47 \times 10^{-6}$ & $2.22 \times 10^{-5}$ & $6.20 \times 10^{-4}$ & $5.33 \times 10^{-3}$ & $1.99 \times 10^{-2}$ \\
 $^{55}$Mn & $2.56 \times 10^{-3}$ & $3.29 \times 10^{-3}$ & $7.18 \times 10^{-3}$ & $9.3 \times 10^{-3}$ & $1.1 \times 10^{-2}$ & $1.27 \times 10^{-2}$ & $1.53 \times 10^{-2}$ \\
 $^{54}$Fe & $2.5 \times 10^{-2}$ & $2.89 \times 10^{-2}$ & $7.40 \times 10^{-2}$ & $9.76 \times 10^{-2}$ & $1.5 \times 10^{-1}$ & $1.16 \times 10^{-1}$ & $1.23 \times 10^{-1}$ \\
 $^{56}$Fe & $2.32 \times 10^{-1}$ & $2.42 \times 10^{-1}$ & $3.49 \times 10^{-1}$ & $3.55 \times 10^{-1}$ & $3.62 \times 10^{-1}$ & $4.8 \times 10^{-1}$ & $4.46 \times 10^{-1}$ \\
 $^{57}$Fe & $7.41 \times 10^{-3}$ & $7.65 \times 10^{-3}$ & $1.20 \times 10^{-2}$ & $1.22 \times 10^{-2}$ & $1.24 \times 10^{-2}$ & $1.38 \times 10^{-2}$ & $1.57 \times 10^{-2}$ \\
 $^{58}$Fe & $3.53 \times 10^{-8}$ & $9.64 \times 10^{-8}$ & $2.73 \times 10^{-6}$ & $1.13 \times 10^{-4}$ & $2.34 \times 10^{-3}$ & $1.42 \times 10^{-2}$ & $3.86 \times 10^{-2}$ \\
 $^{60}$Fe & $1.60 \times 10^{-20}$ & $5.82 \times 10^{-20}$ & $4.49 \times 10^{-17}$ & $4.39 \times 10^{-11}$ & $4.54 \times 10^{-9}$ & $1.14 \times 10^{-7}$ & $1.53 \times 10^{-5}$ \\
 $^{59}$Co & $1.26 \times 10^{-4}$ & $1.19 \times 10^{-4}$ & $4.5 \times 10^{-4}$ & $5.85 \times 10^{-4}$ & $7.6 \times 10^{-4}$ & $1.1 \times 10^{-3}$ & $1.26 \times 10^{-3}$ \\
 $^{58}$Ni & $2.13 \times 10^{-2}$ & $2.36 \times 10^{-2}$ & $4.69 \times 10^{-2}$ & $5.47 \times 10^{-2}$ & $5.69 \times 10^{-2}$ & $6.11 \times 10^{-2}$ & $6.44 \times 10^{-2}$ \\
 $^{60}$Ni & $7.6 \times 10^{-4}$ & $6.97 \times 10^{-4}$ & $3.25 \times 10^{-3}$ & $5.45 \times 10^{-3}$ & $6.76 \times 10^{-3}$ & $8.81 \times 10^{-3}$ & $1.5 \times 10^{-2}$ \\
 $^{61}$Ni & $2.45 \times 10^{-5}$ & $2.11 \times 10^{-5}$ & $3.93 \times 10^{-5}$ & $3.31 \times 10^{-5}$ & $4.75 \times 10^{-5}$ & $8.95 \times 10^{-5}$ & $1.72 \times 10^{-4}$ \\
 $^{62}$Ni & $2.11 \times 10^{-4}$ & $1.74 \times 10^{-4}$ & $2.79 \times 10^{-4}$ & $2.94 \times 10^{-4}$ & $1.10 \times 10^{-3}$ & $4.26 \times 10^{-3}$ & $8.59 \times 10^{-3}$ \\
 $^{64}$Ni & $8.88 \times 10^{-15}$ & $1.84 \times 10^{-12}$ & $3.4 \times 10^{-12}$ & $3.60 \times 10^{-8}$ & $2.36 \times 10^{-6}$ & $4.16 \times 10^{-5}$ & $1.34 \times 10^{-3}$ \\
 $^{63}$Cu & $1.47 \times 10^{-7}$ & $1.8 \times 10^{-7}$ & $2.31 \times 10^{-7}$ & $2.53 \times 10^{-7}$ & $7.61 \times 10^{-7}$ & $3.76 \times 10^{-6}$ & $1.95 \times 10^{-5}$ \\
 $^{65}$Cu & $5.51 \times 10^{-8}$ & $4.91 \times 10^{-8}$ & $1.40 \times 10^{-7}$ & $7.48 \times 10^{-8}$ & $1.17 \times 10^{-7}$ & $4.96 \times 10^{-7}$ & $3.0 \times 10^{-6}$ \\
 $^{64}$Zn & $4.29 \times 10^{-7}$ & $3.68 \times 10^{-7}$ & $9.7 \times 10^{-7}$ & $6.43 \times 10^{-7}$ & $5.93 \times 10^{-7}$ & $8.88 \times 10^{-7}$ & $1.25 \times 10^{-6}$ \\
 $^{66}$Zn & $9.51 \times 10^{-7}$ & $8.1 \times 10^{-7}$ & $1.72 \times 10^{-6}$ & $1.12 \times 10^{-6}$ & $1.1 \times 10^{-6}$ & $1.67 \times 10^{-6}$ & $2.50 \times 10^{-5}$ \\
 $^{67}$Zn & $5.57 \times 10^{-10}$ & $4.61 \times 10^{-10}$ & $1.37 \times 10^{-9}$ & $6.64 \times 10^{-10}$ & $1.88 \times 10^{-9}$ & $3.4 \times 10^{-9}$ & $1.15 \times 10^{-6}$ \\
 $^{68}$Zn & $2.14 \times 10^{-10}$ & $4.97 \times 10^{-10}$ & $5.35 \times 10^{-10}$ & $3.13 \times 10^{-10}$ & $6.52 \times 10^{-9}$ & $3.70 \times 10^{-8}$ & $4.43 \times 10^{-7}$ \\
 $^{70}$Zn & $1.77 \times 10^{-22}$ & $1.51 \times 10^{-12}$ & $5.64 \times 10^{-22}$ & $1.24 \times 10^{-15}$ & $1.41 \times 10^{-10}$ & $1.73 \times 10^{-11}$ & $7.49 \times 10^{-8}$ \\ \hline

\end{tabular}
\end{center}
\end{table*}

\begin{table*}

\begin{center}
\caption{Masses of the radioactive isotopes in the ejecta after
the explosion. The isotope masses are in units of solar mass.}
\label{table:Decay1}
\begin{tabular}{|c| c c c c c c c|}
\hline
Isotopes &  100-133-1-c3-1 & 200-135-1-c3-1 & 300-137-1-c3-1 & 500-138-1-c3-1 & 550-138-1-c3-1 & 750-139-1-c3-1 & 900-140-1-c3-1  \\ \hline

$^{22}$Na & $2.29 \times 10^{-9}$ & $9.80 \times 10^{-9}$ & $1.6 \times 10^{-8}$ & $1.0 \times 10^{-8}$ & $9.79 \times 10^{-9}$ & $2.62 \times 10^{-8}$ & $3.66 \times 10^{-8}$ \\
 $^{26}$Al & $1.11 \times 10^{-6}$ & $4.61 \times 10^{-6}$ & $4.82 \times 10^{-6}$ & $5.12 \times 10^{-6}$ & $4.85 \times 10^{-6}$ & $1.22 \times 10^{-5}$ & $1.69 \times 10^{-5}$ \\
 $^{39}$Ar & $7.40 \times 10^{-10}$ & $2.13 \times 10^{-9}$ & $2.2 \times 10^{-9}$ & $1.94 \times 10^{-9}$ & $2.1 \times 10^{-9}$ & $6.51 \times 10^{-9}$ & $9.2 \times 10^{-9}$ \\
 $^{40}$K & $3.91 \times 10^{-9}$ & $1.11 \times 10^{-8}$ & $1.4 \times 10^{-8}$ & $9.70 \times 10^{-9}$ & $1.5 \times 10^{-8}$ & $3.33 \times 10^{-8}$ & $4.42 \times 10^{-8}$ \\
 $^{41}$Ca & $4.29 \times 10^{-7}$ & $1.0 \times 10^{-6}$ & $1.23 \times 10^{-6}$ & $8.93 \times 10^{-7}$ & $1.11 \times 10^{-6}$ & $1.15 \times 10^{-6}$ & $1.46 \times 10^{-6}$ \\
 $^{44}$Ti & $8.12 \times 10^{-7}$ & $1.64 \times 10^{-6}$ & $1.74 \times 10^{-6}$ & $1.51 \times 10^{-6}$ & $1.52 \times 10^{-6}$ & $1.68 \times 10^{-6}$ & $1.81 \times 10^{-6}$ \\
 $^{48}$V & $3.94 \times 10^{-9}$ & $8.54 \times 10^{-9}$ & $1.5 \times 10^{-8}$ & $8.45 \times 10^{-9}$ & $1.2 \times 10^{-8}$ & $1.24 \times 10^{-8}$ & $1.63 \times 10^{-8}$ \\
 $^{49}$V & $8.9 \times 10^{-9}$ & $4.31 \times 10^{-8}$ & $9.30 \times 10^{-8}$ & $1.20 \times 10^{-7}$ & $1.57 \times 10^{-7}$ & $2.21 \times 10^{-7}$ & $2.80 \times 10^{-7}$ \\
 $^{53}$Mn & $3.43 \times 10^{-6}$ & $1.61 \times 10^{-4}$ & $3.67 \times 10^{-4}$ & $4.60 \times 10^{-4}$ & $5.51 \times 10^{-4}$ & $5.96 \times 10^{-4}$ & $6.99 \times 10^{-4}$ \\
 $^{60}$Fe & $2.48 \times 10^{-19}$ & $6.76 \times 10^{-16}$ & $6.14 \times 10^{-10}$ & $6.36 \times 10^{-8}$ & $1.52 \times 10^{-7}$ & $1.59 \times 10^{-6}$ & $2.25 \times 10^{-4}$ \\
 $^{56}$Co & $1.96 \times 10^{-5}$ & $7.1 \times 10^{-5}$ & $1.0 \times 10^{-4}$ & $1.9 \times 10^{-4}$ & $1.11 \times 10^{-4}$ & $1.23 \times 10^{-4}$ & $1.30 \times 10^{-4}$ \\
 $^{57}$Co & $2.39 \times 10^{-5}$ & $6.48 \times 10^{-4}$ & $1.26 \times 10^{-3}$ & $1.45 \times 10^{-3}$ & $1.57 \times 10^{-3}$ & $1.71 \times 10^{-3}$ & $1.93 \times 10^{-3}$ \\
 $^{60}$Co & $9.31 \times 10^{-14}$ & $5.13 \times 10^{-11}$ & $1.94 \times 10^{-8}$ & $3.3 \times 10^{-7}$ & $6.8 \times 10^{-7}$ & $1.58 \times 10^{-6}$ & $4.3 \times 10^{-6}$ \\
 $^{56}$Ni & $2.32 \times 10^{-1}$ & $3.9 \times 10^{-1}$ & $3.21 \times 10^{-1}$ & $3.8 \times 10^{-1}$ & $3.1 \times 10^{-1}$ & $3.18 \times 10^{-1}$ & $3.39 \times 10^{-1}$ \\
 $^{57}$Ni & $7.39 \times 10^{-3}$ & $1.4 \times 10^{-2}$ & $1.9 \times 10^{-2}$ & $1.6 \times 10^{-2}$ & $1.5 \times 10^{-2}$ & $1.11 \times 10^{-2}$ & $1.21 \times 10^{-2}$ \\
 $^{59}$Ni & $1.64 \times 10^{-5}$ & $2.36 \times 10^{-4}$ & $4.36 \times 10^{-4}$ & $4.89 \times 10^{-4}$ & $5.13 \times 10^{-4}$ & $5.63 \times 10^{-4}$ & $6.31 \times 10^{-4}$ \\
 $^{63}$Ni & $1.96 \times 10^{-15}$ & $3.20 \times 10^{-12}$ & $1.41 \times 10^{-8}$ & $3.73 \times 10^{-7}$ & $8.4 \times 10^{-7}$ & $2.84 \times 10^{-6}$ & $1.70 \times 10^{-5}$ \\ \hline
 
\end{tabular}
\end{center}
\end{table*}

\begin{table*}

\begin{center}
\caption{Mass of major isotopes in the ejecta after all 
short-lived radioactive isotopes have decayed. The 
isotope masses are in units of solar mass.}
\label{table:Isotope2}
\begin{tabular}{|c| c c c c c c c|}
\hline
Isotopes &  100-043-1-c3 & 200-045-1-c3 & 300-047-1-c3 & 500-048-1-c3 & 550-048-1-c3 & 750-049-1-c3 & 900-050-1-c3  \\ \hline
 
$^{12}$C & $1.32 \times 10^{-1}$ & $1.39 \times 10^{-1}$ & $1.34 \times 10^{-1}$ & $1.18 \times 10^{-1}$ & $1.23 \times 10^{-1}$ & $1.4 \times 10^{-1}$ & $9.35 \times 10^{-2}$ \\
 $^{13}$C & $7.21 \times 10^{-12}$ & $1.95 \times 10^{-11}$ & $2.42 \times 10^{-11}$ & $3.0 \times 10^{-11}$ & $2.20 \times 10^{-11}$ & $3.16 \times 10^{-11}$ & $2.63 \times 10^{-11}$ \\
 $^{14}$N & $2.76 \times 10^{-9}$ & $7.6 \times 10^{-9}$ & $8.29 \times 10^{-9}$ & $1.1 \times 10^{-8}$ & $7.29 \times 10^{-9}$ & $1.7 \times 10^{-8}$ & $1.8 \times 10^{-8}$ \\
 $^{15}$N & $3.31 \times 10^{-7}$ & $5.52 \times 10^{-7}$ & $5.86 \times 10^{-7}$ & $6.42 \times 10^{-7}$ & $5.66 \times 10^{-7}$ & $7.33 \times 10^{-7}$ & $8.57 \times 10^{-7}$ \\
 $^{16}$O & $3.63 \times 10^{-1}$ & $3.92 \times 10^{-1}$ & $3.80 \times 10^{-1}$ & $3.41 \times 10^{-1}$ & $3.52 \times 10^{-1}$ & $3.8 \times 10^{-1}$ & $2.85 \times 10^{-1}$ \\
 $^{17}$O & $6.1 \times 10^{-11}$ & $1.67 \times 10^{-10}$ & $2.4 \times 10^{-10}$ & $2.53 \times 10^{-10}$ & $1.77 \times 10^{-10}$ & $2.54 \times 10^{-10}$ & $2.30 \times 10^{-10}$ \\
 $^{18}$O & $2.89 \times 10^{-12}$ & $7.78 \times 10^{-12}$ & $9.52 \times 10^{-12}$ & $1.16 \times 10^{-11}$ & $8.20 \times 10^{-12}$ & $1.17 \times 10^{-11}$ & $1.5 \times 10^{-11}$ \\
 $^{19}$F & $1.8 \times 10^{-10}$ & $2.68 \times 10^{-10}$ & $3.24 \times 10^{-10}$ & $4.1 \times 10^{-10}$ & $2.88 \times 10^{-10}$ & $4.1 \times 10^{-10}$ & $3.74 \times 10^{-10}$ \\
 $^{20}$Ne & $2.16 \times 10^{-1}$ & $2.29 \times 10^{-1}$ & $2.22 \times 10^{-1}$ & $1.97 \times 10^{-1}$ & $2.5 \times 10^{-1}$ & $1.75 \times 10^{-1}$ & $1.57 \times 10^{-1}$ \\
 $^{21}$Ne & $4.79 \times 10^{-9}$ & $1.6 \times 10^{-8}$ & $1.28 \times 10^{-8}$ & $1.57 \times 10^{-8}$ & $1.17 \times 10^{-8}$ & $1.55 \times 10^{-8}$ & $1.64 \times 10^{-8}$ \\
 $^{22}$Ne & $8.38 \times 10^{-9}$ & $1.74 \times 10^{-8}$ & $2.6 \times 10^{-8}$ & $2.52 \times 10^{-8}$ & $1.96 \times 10^{-8}$ & $2.49 \times 10^{-8}$ & $2.79 \times 10^{-8}$ \\
 $^{23}$Na & $2.68 \times 10^{-6}$ & $4.68 \times 10^{-6}$ & $5.61 \times 10^{-6}$ & $6.65 \times 10^{-6}$ & $5.65 \times 10^{-6}$ & $6.72 \times 10^{-6}$ & $7.73 \times 10^{-6}$ \\
 $^{24}$Mg & $3.61 \times 10^{-3}$ & $5.60 \times 10^{-3}$ & $5.90 \times 10^{-3}$ & $6.90 \times 10^{-3}$ & $5.75 \times 10^{-3}$ & $7.22 \times 10^{-3}$ & $7.65 \times 10^{-3}$ \\
 $^{25}$Mg & $8.84 \times 10^{-8}$ & $1.64 \times 10^{-7}$ & $1.92 \times 10^{-7}$ & $2.30 \times 10^{-7}$ & $1.86 \times 10^{-7}$ & $2.28 \times 10^{-7}$ & $2.58 \times 10^{-7}$ \\
 $^{26}$Mg & $7.73 \times 10^{-7}$ & $1.34 \times 10^{-6}$ & $1.61 \times 10^{-6}$ & $1.89 \times 10^{-6}$ & $1.72 \times 10^{-6}$ & $1.93 \times 10^{-6}$ & $2.22 \times 10^{-6}$ \\
 $^{26}$Al & $2.49 \times 10^{-29}$ & $2.88 \times 10^{-29}$ & $2.99 \times 10^{-29}$ & $4.65 \times 10^{-28}$ & $3.27 \times 10^{-29}$ & $4.44 \times 10^{-28}$ & $3.45 \times 10^{-29}$ \\
 $^{27}$Al & $2.34 \times 10^{-5}$ & $3.45 \times 10^{-5}$ & $3.76 \times 10^{-5}$ & $4.74 \times 10^{-5}$ & $3.91 \times 10^{-5}$ & $4.87 \times 10^{-5}$ & $4.57 \times 10^{-5}$ \\
 $^{28}$Si & $2.42 \times 10^{-2}$ & $3.40 \times 10^{-2}$ & $3.61 \times 10^{-2}$ & $3.76 \times 10^{-2}$ & $3.36 \times 10^{-2}$ & $3.73 \times 10^{-2}$ & $3.66 \times 10^{-2}$ \\
 $^{29}$Si & $6.79 \times 10^{-6}$ & $1.11 \times 10^{-5}$ & $1.11 \times 10^{-5}$ & $1.21 \times 10^{-5}$ & $1.6 \times 10^{-5}$ & $1.32 \times 10^{-5}$ & $1.55 \times 10^{-5}$ \\
 $^{30}$Si & $4.38 \times 10^{-6}$ & $6.99 \times 10^{-6}$ & $7.42 \times 10^{-6}$ & $8.52 \times 10^{-6}$ & $7.17 \times 10^{-6}$ & $9.24 \times 10^{-6}$ & $1.3 \times 10^{-5}$ \\
 $^{31}$P & $1.45 \times 10^{-5}$ & $2.23 \times 10^{-5}$ & $2.35 \times 10^{-5}$ & $2.59 \times 10^{-5}$ & $2.22 \times 10^{-5}$ & $2.84 \times 10^{-5}$ & $2.95 \times 10^{-5}$ \\
 $^{32}$S & $1.32 \times 10^{-2}$ & $1.78 \times 10^{-2}$ & $1.89 \times 10^{-2}$ & $2.1 \times 10^{-2}$ & $1.80 \times 10^{-2}$ & $1.95 \times 10^{-2}$ & $1.81 \times 10^{-2}$ \\
 $^{33}$S & $4.24 \times 10^{-6}$ & $7.37 \times 10^{-6}$ & $7.31 \times 10^{-6}$ & $7.78 \times 10^{-6}$ & $6.87 \times 10^{-6}$ & $9.28 \times 10^{-6}$ & $1.19 \times 10^{-5}$ \\
 $^{34}$S & $1.75 \times 10^{-7}$ & $3.7 \times 10^{-7}$ & $3.61 \times 10^{-7}$ & $4.0 \times 10^{-7}$ & $3.34 \times 10^{-7}$ & $5.5 \times 10^{-7}$ & $6.83 \times 10^{-7}$ \\
 $^{36}$S & $2.84 \times 10^{-13}$ & $4.9 \times 10^{-13}$ & $4.78 \times 10^{-13}$ & $3.15 \times 10^{-12}$ & $4.30 \times 10^{-12}$ & $1.32 \times 10^{-10}$ & $3.94 \times 10^{-10}$ \\
 $^{35}$Cl & $7.15 \times 10^{-7}$ & $1.6 \times 10^{-6}$ & $1.16 \times 10^{-6}$ & $1.29 \times 10^{-6}$ & $1.13 \times 10^{-6}$ & $1.50 \times 10^{-6}$ & $1.66 \times 10^{-6}$ \\
 $^{37}$Cl & $2.86 \times 10^{-7}$ & $5.3 \times 10^{-7}$ & $5.38 \times 10^{-7}$ & $5.22 \times 10^{-7}$ & $5.7 \times 10^{-7}$ & $7.12 \times 10^{-7}$ & $8.88 \times 10^{-7}$ \\
 $^{36}$Ar & $2.59 \times 10^{-3}$ & $3.45 \times 10^{-3}$ & $3.63 \times 10^{-3}$ & $3.98 \times 10^{-3}$ & $3.56 \times 10^{-3}$ & $3.90 \times 10^{-3}$ & $3.58 \times 10^{-3}$ \\
 $^{38}$Ar & $2.42 \times 10^{-8}$ & $4.18 \times 10^{-8}$ & $4.88 \times 10^{-8}$ & $6.49 \times 10^{-8}$ & $6.66 \times 10^{-8}$ & $1.13 \times 10^{-7}$ & $1.48 \times 10^{-7}$ \\
 $^{40}$Ar & $7.5 \times 10^{-17}$ & $1.28 \times 10^{-16}$ & $1.49 \times 10^{-16}$ & $4.89 \times 10^{-13}$ & $7.4 \times 10^{-13}$ & $1.45 \times 10^{-11}$ & $3.79 \times 10^{-11}$ \\
 $^{39}$K & $1.29 \times 10^{-7}$ & $1.82 \times 10^{-7}$ & $1.99 \times 10^{-7}$ & $2.43 \times 10^{-7}$ & $2.17 \times 10^{-7}$ & $2.80 \times 10^{-7}$ & $3.28 \times 10^{-7}$ \\
 $^{40}$K & $9.7 \times 10^{-14}$ & $1.67 \times 10^{-13}$ & $1.94 \times 10^{-13}$ & $3.55 \times 10^{-13}$ & $3.70 \times 10^{-13}$ & $6.85 \times 10^{-13}$ & $1.1 \times 10^{-12}$ \\
 $^{41}$K & $4.3 \times 10^{-8}$ & $7.64 \times 10^{-8}$ & $9.20 \times 10^{-8}$ & $7.1 \times 10^{-8}$ & $8.9 \times 10^{-8}$ & $9.62 \times 10^{-8}$ & $1.3 \times 10^{-7}$ \\
 $^{40}$Ca & $2.37 \times 10^{-3}$ & $3.10 \times 10^{-3}$ & $3.23 \times 10^{-3}$ & $3.66 \times 10^{-3}$ & $3.29 \times 10^{-3}$ & $3.62 \times 10^{-3}$ & $3.28 \times 10^{-3}$ \\
 $^{42}$Ca & $1.38 \times 10^{-10}$ & $2.44 \times 10^{-10}$ & $3.59 \times 10^{-10}$ & $3.81 \times 10^{-9}$ & $4.8 \times 10^{-9}$ & $7.81 \times 10^{-9}$ & $1.1 \times 10^{-8}$ \\
 $^{43}$Ca & $2.18 \times 10^{-9}$ & $2.30 \times 10^{-9}$ & $2.52 \times 10^{-9}$ & $3.38 \times 10^{-9}$ & $2.80 \times 10^{-9}$ & $3.35 \times 10^{-9}$ & $7.76 \times 10^{-9}$ \\
 $^{44}$Ca & $1.47 \times 10^{-6}$ & $2.4 \times 10^{-6}$ & $2.20 \times 10^{-6}$ & $2.91 \times 10^{-6}$ & $2.55 \times 10^{-6}$ & $2.99 \times 10^{-6}$ & $2.93 \times 10^{-6}$ \\
 $^{46}$Ca & $1.36 \times 10^{-23}$ & $5.99 \times 10^{-22}$ & $4.97 \times 10^{-22}$ & $3.4 \times 10^{-11}$ & $4.80 \times 10^{-11}$ & $4.23 \times 10^{-9}$ & $2.3 \times 10^{-8}$ \\
 $^{48}$Ca & $1.84 \times 10^{-25}$ & $2.13 \times 10^{-25}$ & $2.21 \times 10^{-25}$ & $1.41 \times 10^{-12}$ & $3.0 \times 10^{-12}$ & $3.45 \times 10^{-9}$ & $6.64 \times 10^{-7}$ \\
 $^{45}$Sc & $3.18 \times 10^{-9}$ & $6.8 \times 10^{-9}$ & $8.62 \times 10^{-9}$ & $1.4 \times 10^{-8}$ & $1.5 \times 10^{-8}$ & $1.25 \times 10^{-8}$ & $1.34 \times 10^{-8}$ \\
 $^{46}$Ti & $7.17 \times 10^{-9}$ & $2.27 \times 10^{-8}$ & $3.85 \times 10^{-8}$ & $1.50 \times 10^{-7}$ & $1.58 \times 10^{-7}$ & $2.52 \times 10^{-7}$ & $3.5 \times 10^{-7}$ \\
 $^{47}$Ti & $6.93 \times 10^{-9}$ & $1.15 \times 10^{-8}$ & $1.53 \times 10^{-8}$ & $2.77 \times 10^{-8}$ & $2.66 \times 10^{-8}$ & $3.74 \times 10^{-8}$ & $5.16 \times 10^{-8}$ \\
 $^{48}$Ti & $3.97 \times 10^{-5}$ & $5.47 \times 10^{-5}$ & $5.89 \times 10^{-5}$ & $7.91 \times 10^{-5}$ & $7.27 \times 10^{-5}$ & $8.43 \times 10^{-5}$ & $8.19 \times 10^{-5}$ \\
 $^{49}$Ti & $9.54 \times 10^{-7}$ & $2.4 \times 10^{-6}$ & $2.85 \times 10^{-6}$ & $4.68 \times 10^{-6}$ & $4.48 \times 10^{-6}$ & $5.67 \times 10^{-6}$ & $8.63 \times 10^{-6}$ \\
 $^{50}$Ti & $1.18 \times 10^{-19}$ & $8.2 \times 10^{-15}$ & $7.8 \times 10^{-15}$ & $6.73 \times 10^{-6}$ & $1.0 \times 10^{-5}$ & $4.65 \times 10^{-4}$ & $1.89 \times 10^{-3}$ \\
 $^{50}$V & $3.45 \times 10^{-13}$ & $2.17 \times 10^{-12}$ & $4.39 \times 10^{-12}$ & $4.64 \times 10^{-9}$ & $7.17 \times 10^{-9}$ & $4.23 \times 10^{-8}$ & $7.73 \times 10^{-8}$ \\
 $^{51}$V & $3.25 \times 10^{-6}$ & $9.61 \times 10^{-6}$ & $1.48 \times 10^{-5}$ & $3.59 \times 10^{-5}$ & $3.82 \times 10^{-5}$ & $1.60 \times 10^{-4}$ & $3.45 \times 10^{-4}$ \\
 
 \end{tabular}
\end{center}
\end{table*}
 
\begin{table*}

\begin{center}
\caption{(cont'd) Mass of major isotopes in the ejecta after all 
short-lived radioactive isotopes have decayed. The 
isotope masses are in units of solar mass.}
\label{table:Isotope2b}
\begin{tabular}{|c| c c c c c c c|}
\hline
Isotopes &  100-043-1-c3 & 200-045-1-c3 & 300-047-1-c3 & 500-048-1-c3 & 550-048-1-c3 & 750-049-1-c3 & 900-050-1-c3  \\ \hline

$^{50}$Cr & $8.78 \times 10^{-6}$ & $3.0 \times 10^{-5}$ & $5.36 \times 10^{-5}$ & $1.58 \times 10^{-4}$ & $1.65 \times 10^{-4}$ & $2.24 \times 10^{-4}$ & $2.51 \times 10^{-4}$ \\
 $^{52}$Cr & $8.67 \times 10^{-4}$ & $1.47 \times 10^{-3}$ & $1.80 \times 10^{-3}$ & $3.69 \times 10^{-3}$ & $4.2 \times 10^{-3}$ & $9.85 \times 10^{-3}$ & $1.48 \times 10^{-2}$ \\
 $^{53}$Cr & $6.70 \times 10^{-5}$ & $1.64 \times 10^{-4}$ & $2.34 \times 10^{-4}$ & $5.77 \times 10^{-4}$ & $6.3 \times 10^{-4}$ & $1.13 \times 10^{-3}$ & $1.66 \times 10^{-3}$ \\
 $^{54}$Cr & $4.87 \times 10^{-11}$ & $2.61 \times 10^{-9}$ & $5.32 \times 10^{-9}$ & $9.13 \times 10^{-5}$ & $1.32 \times 10^{-4}$ & $3.53 \times 10^{-3}$ & $1.13 \times 10^{-2}$ \\
 $^{55}$Mn & $1.5 \times 10^{-3}$ & $2.52 \times 10^{-3}$ & $3.64 \times 10^{-3}$ & $7.85 \times 10^{-3}$ & $7.93 \times 10^{-3}$ & $1.13 \times 10^{-2}$ & $1.37 \times 10^{-2}$ \\
 $^{54}$Fe & $7.18 \times 10^{-3}$ & $2.6 \times 10^{-2}$ & $3.50 \times 10^{-2}$ & $8.58 \times 10^{-2}$ & $8.94 \times 10^{-2}$ & $1.11 \times 10^{-1}$ & $1.18 \times 10^{-1}$ \\
 $^{56}$Fe & $1.82 \times 10^{-1}$ & $2.42 \times 10^{-1}$ & $2.78 \times 10^{-1}$ & $3.78 \times 10^{-1}$ & $3.55 \times 10^{-1}$ & $4.13 \times 10^{-1}$ & $4.56 \times 10^{-1}$ \\
 $^{57}$Fe & $3.40 \times 10^{-3}$ & $4.99 \times 10^{-3}$ & $6.8 \times 10^{-3}$ & $9.29 \times 10^{-3}$ & $8.86 \times 10^{-3}$ & $1.5 \times 10^{-2}$ & $1.18 \times 10^{-2}$ \\
 $^{58}$Fe & $1.19 \times 10^{-10}$ & $8.82 \times 10^{-9}$ & $2.6 \times 10^{-8}$ & $3.73 \times 10^{-4}$ & $5.34 \times 10^{-4}$ & $9.16 \times 10^{-3}$ & $2.45 \times 10^{-2}$ \\
 $^{60}$Fe & $2.71 \times 10^{-26}$ & $9.96 \times 10^{-22}$ & $6.64 \times 10^{-22}$ & $5.21 \times 10^{-10}$ & $7.15 \times 10^{-10}$ & $9.73 \times 10^{-8}$ & $6.26 \times 10^{-6}$ \\
 $^{59}$Co & $2.43 \times 10^{-5}$ & $3.38 \times 10^{-5}$ & $4.88 \times 10^{-5}$ & $3.59 \times 10^{-4}$ & $3.94 \times 10^{-4}$ & $7.49 \times 10^{-4}$ & $9.97 \times 10^{-4}$ \\
 $^{58}$Ni & $5.35 \times 10^{-3}$ & $1.10 \times 10^{-2}$ & $1.74 \times 10^{-2}$ & $3.85 \times 10^{-2}$ & $4.0 \times 10^{-2}$ & $4.74 \times 10^{-2}$ & $4.96 \times 10^{-2}$ \\
 $^{60}$Ni & $8.3 \times 10^{-4}$ & $8.46 \times 10^{-4}$ & $9.47 \times 10^{-4}$ & $3.88 \times 10^{-3}$ & $4.17 \times 10^{-3}$ & $7.81 \times 10^{-3}$ & $1.0 \times 10^{-2}$ \\
 $^{61}$Ni & $1.50 \times 10^{-5}$ & $1.54 \times 10^{-5}$ & $1.69 \times 10^{-5}$ & $2.42 \times 10^{-5}$ & $2.36 \times 10^{-5}$ & $5.79 \times 10^{-5}$ & $1.9 \times 10^{-4}$ \\
 $^{62}$Ni & $2.67 \times 10^{-5}$ & $2.24 \times 10^{-5}$ & $2.30 \times 10^{-5}$ & $1.74 \times 10^{-4}$ & $2.40 \times 10^{-4}$ & $2.56 \times 10^{-3}$ & $5.86 \times 10^{-3}$ \\
 $^{64}$Ni & $7.9 \times 10^{-18}$ & $2.33 \times 10^{-15}$ & $3.55 \times 10^{-15}$ & $3.9 \times 10^{-7}$ & $4.61 \times 10^{-7}$ & $3.44 \times 10^{-5}$ & $5.98 \times 10^{-4}$ \\
 $^{63}$Cu & $6.64 \times 10^{-8}$ & $6.7 \times 10^{-8}$ & $7.31 \times 10^{-8}$ & $1.99 \times 10^{-7}$ & $2.37 \times 10^{-7}$ & $2.43 \times 10^{-6}$ & $1.1 \times 10^{-5}$ \\
 $^{65}$Cu & $3.87 \times 10^{-8}$ & $4.7 \times 10^{-8}$ & $4.43 \times 10^{-8}$ & $6.7 \times 10^{-8}$ & $5.74 \times 10^{-8}$ & $3.17 \times 10^{-7}$ & $1.49 \times 10^{-6}$ \\
 $^{64}$Zn & $4.53 \times 10^{-6}$ & $5.42 \times 10^{-6}$ & $6.10 \times 10^{-6}$ & $7.93 \times 10^{-6}$ & $6.72 \times 10^{-6}$ & $7.68 \times 10^{-6}$ & $8.76 \times 10^{-6}$ \\
 $^{66}$Zn & $1.34 \times 10^{-7}$ & $1.18 \times 10^{-7}$ & $1.28 \times 10^{-7}$ & $1.44 \times 10^{-7}$ & $1.27 \times 10^{-7}$ & $2.12 \times 10^{-7}$ & $8.84 \times 10^{-6}$ \\
 $^{67}$Zn & $2.17 \times 10^{-10}$ & $2.81 \times 10^{-10}$ & $3.13 \times 10^{-10}$ & $4.43 \times 10^{-10}$ & $3.83 \times 10^{-10}$ & $2.51 \times 10^{-9}$ & $4.56 \times 10^{-7}$ \\
 $^{68}$Zn & $7.61 \times 10^{-9}$ & $1.19 \times 10^{-8}$ & $1.26 \times 10^{-8}$ & $1.77 \times 10^{-8}$ & $1.56 \times 10^{-8}$ & $3.61 \times 10^{-8}$ & $2.6 \times 10^{-7}$ \\
 $^{70}$Zn & $1.94 \times 10^{-27}$ & $7.30 \times 10^{-27}$ & $7.19 \times 10^{-27}$ & $2.29 \times 10^{-14}$ & $3.85 \times 10^{-14}$ & $2.38 \times 10^{-11}$ & $2.88 \times 10^{-8}$ \\ \hline

\end{tabular}
\end{center}
\end{table*}

\begin{table*}

\begin{center}
\caption{Masses of the radioactive isotopes in the ejecta after
the explosion. The isotope masses are in units of solar mass.}
\label{table:Decay2}
\begin{tabular}{|c| c c c c c c c|}
\hline
Isotopes &  100-043-1-c3 & 200-045-1-c3 & 300-047-1-c3 & 500-048-1-c3 & 550-048-1-c3 & 750-049-1-c3 & 900-050-1-c3  \\ \hline
 
$^{22}$Na & $7.3 \times 10^{-9}$ & $1.43 \times 10^{-8}$ & $1.69 \times 10^{-8}$ & $2.5 \times 10^{-8}$ & $1.59 \times 10^{-8}$ & $2.2 \times 10^{-8}$ & $2.27 \times 10^{-8}$ \\
 $^{26}$Al & $6.45 \times 10^{-7}$ & $1.10 \times 10^{-6}$ & $1.32 \times 10^{-6}$ & $1.55 \times 10^{-6}$ & $1.35 \times 10^{-6}$ & $1.58 \times 10^{-6}$ & $1.84 \times 10^{-6}$ \\
 $^{39}$Ar & $7.45 \times 10^{-16}$ & $1.26 \times 10^{-15}$ & $1.47 \times 10^{-15}$ & $4.96 \times 10^{-14}$ & $7.28 \times 10^{-14}$ & $5.30 \times 10^{-13}$ & $1.4 \times 10^{-12}$ \\
 $^{40}$K & $9.12 \times 10^{-14}$ & $1.68 \times 10^{-13}$ & $1.95 \times 10^{-13}$ & $3.57 \times 10^{-13}$ & $3.72 \times 10^{-13}$ & $6.89 \times 10^{-13}$ & $1.1 \times 10^{-12}$ \\
 $^{41}$Ca & $3.69 \times 10^{-8}$ & $6.54 \times 10^{-8}$ & $7.86 \times 10^{-8}$ & $7.11 \times 10^{-8}$ & $7.61 \times 10^{-8}$ & $9.73 \times 10^{-8}$ & $1.4 \times 10^{-7}$ \\
 $^{44}$Ti & $1.48 \times 10^{-6}$ & $2.1 \times 10^{-6}$ & $2.17 \times 10^{-6}$ & $2.71 \times 10^{-6}$ & $2.44 \times 10^{-6}$ & $2.80 \times 10^{-6}$ & $2.73 \times 10^{-6}$ \\
 $^{48}$V & $5.9 \times 10^{-10}$ & $9.69 \times 10^{-10}$ & $1.32 \times 10^{-9}$ & $2.54 \times 10^{-9}$ & $2.54 \times 10^{-9}$ & $3.17 \times 10^{-9}$ & $3.51 \times 10^{-9}$ \\
 $^{49}$V & $6.40 \times 10^{-11}$ & $2.53 \times 10^{-10}$ & $4.74 \times 10^{-10}$ & $4.2 \times 10^{-8}$ & $5.36 \times 10^{-8}$ & $1.35 \times 10^{-7}$ & $1.99 \times 10^{-7}$ \\
 $^{53}$Mn & $9.19 \times 10^{-8}$ & $1.16 \times 10^{-6}$ & $3.15 \times 10^{-6}$ & $2.23 \times 10^{-4}$ & $2.58 \times 10^{-4}$ & $5.35 \times 10^{-4}$ & $7.8 \times 10^{-4}$ \\
 $^{60}$Fe & $1.32 \times 10^{-25}$ & $1.47 \times 10^{-20}$ & $9.87 \times 10^{-21}$ & $7.47 \times 10^{-9}$ & $1.15 \times 10^{-8}$ & $1.39 \times 10^{-6}$ & $9.9 \times 10^{-5}$ \\
 $^{56}$Co & $6.46 \times 10^{-6}$ & $1.81 \times 10^{-5}$ & $3.28 \times 10^{-5}$ & $8.57 \times 10^{-5}$ & $9.8 \times 10^{-5}$ & $1.11 \times 10^{-4}$ & $1.19 \times 10^{-4}$ \\
 $^{57}$Co & $1.18 \times 10^{-6}$ & $1.6 \times 10^{-5}$ & $3.15 \times 10^{-5}$ & $8.72 \times 10^{-4}$ & $9.10 \times 10^{-4}$ & $1.59 \times 10^{-3}$ & $1.88 \times 10^{-3}$ \\
 $^{60}$Co & $3.96 \times 10^{-17}$ & $2.68 \times 10^{-14}$ & $3.67 \times 10^{-14}$ & $4.90 \times 10^{-8}$ & $7.2 \times 10^{-8}$ & $9.86 \times 10^{-7}$ & $2.60 \times 10^{-6}$ \\
 $^{56}$Ni & $1.82 \times 10^{-1}$ & $2.42 \times 10^{-1}$ & $2.78 \times 10^{-1}$ & $3.58 \times 10^{-1}$ & $3.30 \times 10^{-1}$ & $3.44 \times 10^{-1}$ & $3.56 \times 10^{-1}$ \\
 $^{57}$Ni & $3.40 \times 10^{-3}$ & $4.98 \times 10^{-3}$ & $6.5 \times 10^{-3}$ & $8.37 \times 10^{-3}$ & $7.87 \times 10^{-3}$ & $8.38 \times 10^{-3}$ & $8.77 \times 10^{-3}$ \\
 $^{59}$Ni & $2.8 \times 10^{-6}$ & $9.1 \times 10^{-6}$ & $2.21 \times 10^{-5}$ & $3.8 \times 10^{-4}$ & $3.18 \times 10^{-4}$ & $5.18 \times 10^{-4}$ & $5.91 \times 10^{-4}$ \\
 $^{63}$Ni & $1.33 \times 10^{-20}$ & $6.32 \times 10^{-16}$ & $6.26 \times 10^{-16}$ & $5.77 \times 10^{-8}$ & $8.25 \times 10^{-8}$ & $1.94 \times 10^{-6}$ & $9.8 \times 10^{-6}$ \\ \hline

\end{tabular}
\end{center}
\end{table*}

\begin{table}

\begin{center}
\caption{Mass of major isotopes in the ejecta after all 
short-lived radioactive isotopes have decayed. The 
isotope masses are in units of solar mass.}
\label{table:Isotope3}
\begin{tabular}{|c| c c c|}
\hline
Isotopes &  800-000-1-c3-1 & 800-000-1-b1-1 & 900-000-1-c3-1 \\ \hline

$^{12}$C & $1.53 \times 10^{-7}$ & $1.29 \times 10^{-7}$ & $1.92 \times 10^{-7}$ \\
 $^{13}$C & $9.46 \times 10^{-14}$ & $7.52 \times 10^{-14}$ & $9.42 \times 10^{-14}$ \\
 $^{14}$N & $1.77 \times 10^{-9}$ & $8.72 \times 10^{-10}$ & $1.98 \times 10^{-9}$ \\
 $^{15}$N & $3.6 \times 10^{-7}$ & $1.40 \times 10^{-7}$ & $3.27 \times 10^{-7}$ \\
 $^{16}$O & $2.56 \times 10^{-1}$ & $1.37 \times 10^{-1}$ & $1.96 \times 10^{-1}$ \\
 $^{17}$O & $4.86 \times 10^{-12}$ & $1.82 \times 10^{-12}$ & $1.20 \times 10^{-12}$ \\
 $^{18}$O & $3.62 \times 10^{-14}$ & $1.78 \times 10^{-14}$ & $2.80 \times 10^{-14}$ \\
 $^{19}$F & $2.18 \times 10^{-12}$ & $1.12 \times 10^{-12}$ & $1.5 \times 10^{-12}$ \\
 $^{20}$Ne & $1.93 \times 10^{-1}$ & $1.5 \times 10^{-1}$ & $1.46 \times 10^{-1}$ \\
 $^{21}$Ne & $1.72 \times 10^{-8}$ & $7.44 \times 10^{-9}$ & $7.44 \times 10^{-9}$ \\
 $^{22}$Ne & $4.72 \times 10^{-8}$ & $2.83 \times 10^{-8}$ & $2.65 \times 10^{-8}$ \\
 $^{23}$Na & $3.53 \times 10^{-8}$ & $1.56 \times 10^{-8}$ & $1.68 \times 10^{-8}$ \\
 $^{24}$Mg & $2.8 \times 10^{-3}$ & $9.22 \times 10^{-4}$ & $1.32 \times 10^{-3}$ \\
 $^{25}$Mg & $4.53 \times 10^{-9}$ & $1.99 \times 10^{-9}$ & $2.3 \times 10^{-9}$ \\
 $^{26}$Mg & $8.90 \times 10^{-9}$ & $6.26 \times 10^{-9}$ & $6.19 \times 10^{-9}$ \\
 $^{26}$Al & $3.38 \times 10^{-29}$ & $5.29 \times 10^{-13}$ & $5.21 \times 10^{-13}$ \\
 $^{27}$Al & $1.90 \times 10^{-6}$ & $7.24 \times 10^{-7}$ & $1.26 \times 10^{-6}$ \\
 $^{28}$Si & $2.58 \times 10^{-2}$ & $2.7 \times 10^{-2}$ & $2.43 \times 10^{-2}$ \\
 $^{29}$Si & $6.14 \times 10^{-6}$ & $3.46 \times 10^{-6}$ & $6.32 \times 10^{-6}$ \\
 $^{30}$Si & $2.66 \times 10^{-6}$ & $1.74 \times 10^{-6}$ & $3.15 \times 10^{-6}$ \\
 $^{31}$P & $1.1 \times 10^{-5}$ & $6.8 \times 10^{-6}$ & $9.98 \times 10^{-6}$ \\
 $^{32}$S & $1.61 \times 10^{-2}$ & $1.36 \times 10^{-2}$ & $1.52 \times 10^{-2}$ \\
 $^{33}$S & $4.8 \times 10^{-6}$ & $2.35 \times 10^{-6}$ & $4.21 \times 10^{-6}$ \\
 $^{34}$S & $9.9 \times 10^{-7}$ & $9.7 \times 10^{-7}$ & $1.19 \times 10^{-6}$ \\
 $^{36}$S & $2.30 \times 10^{-10}$ & $1.11 \times 10^{-9}$ & $3.48 \times 10^{-10}$ \\
 $^{35}$Cl & $9.93 \times 10^{-7}$ & $7.87 \times 10^{-7}$ & $9.92 \times 10^{-7}$ \\
 $^{37}$Cl & $7.12 \times 10^{-7}$ & $4.97 \times 10^{-7}$ & $8.9 \times 10^{-7}$ \\
 $^{36}$Ar & $3.57 \times 10^{-3}$ & $3.4 \times 10^{-3}$ & $3.38 \times 10^{-3}$ \\
 $^{38}$Ar & $4.60 \times 10^{-7}$ & $3.56 \times 10^{-7}$ & $4.17 \times 10^{-7}$ \\
 $^{40}$Ar & $2.45 \times 10^{-11}$ & $1.1 \times 10^{-10}$ & $3.61 \times 10^{-11}$ \\
 $^{39}$K & $4.86 \times 10^{-7}$ & $3.54 \times 10^{-7}$ & $4.84 \times 10^{-7}$ \\
 $^{40}$K & $9.60 \times 10^{-13}$ & $1.42 \times 10^{-12}$ & $1.34 \times 10^{-12}$ \\
 $^{41}$K & $1.60 \times 10^{-7}$ & $1.20 \times 10^{-7}$ & $1.75 \times 10^{-7}$ \\
 $^{40}$Ca & $3.61 \times 10^{-3}$ & $3.4 \times 10^{-3}$ & $3.41 \times 10^{-3}$ \\
 $^{42}$Ca & $4.74 \times 10^{-8}$ & $5.78 \times 10^{-7}$ & $1.82 \times 10^{-7}$ \\
 $^{43}$Ca & $3.68 \times 10^{-9}$ & $2.10 \times 10^{-8}$ & $5.77 \times 10^{-9}$ \\
 $^{44}$Ca & $3.47 \times 10^{-6}$ & $3.0 \times 10^{-6}$ & $3.41 \times 10^{-6}$ \\
 $^{46}$Ca & $1.2 \times 10^{-8}$ & $7.39 \times 10^{-8}$ & $1.72 \times 10^{-8}$ \\
 $^{48}$Ca & $2.26 \times 10^{-7}$ & $1.13 \times 10^{-5}$ & $6.94 \times 10^{-7}$ \\
 $^{45}$Sc & $3.39 \times 10^{-8}$ & $1.5 \times 10^{-7}$ & $7.95 \times 10^{-8}$ \\
 $^{46}$Ti & $4.18 \times 10^{-7}$ & $5.7 \times 10^{-7}$ & $4.83 \times 10^{-7}$ \\
 $^{47}$Ti & $5.76 \times 10^{-8}$ & $8.28 \times 10^{-8}$ & $6.37 \times 10^{-8}$ \\
 $^{48}$Ti & $1.6 \times 10^{-4}$ & $9.43 \times 10^{-5}$ & $1.3 \times 10^{-4}$ \\
 $^{49}$Ti & $9.34 \times 10^{-6}$ & $2.47 \times 10^{-5}$ & $1.9 \times 10^{-5}$ \\
 $^{50}$Ti & $9.98 \times 10^{-4}$ & $6.66 \times 10^{-3}$ & $1.62 \times 10^{-3}$ \\
 $^{50}$V & $7.33 \times 10^{-8}$ & $1.73 \times 10^{-7}$ & $1.1 \times 10^{-7}$ \\
 $^{51}$V & $2.63 \times 10^{-4}$ & $8.73 \times 10^{-4}$ & $3.59 \times 10^{-4}$ \\
 
\end{tabular}
\end{center}
\end{table} 
 
\begin{table}

\begin{center}
\caption{(cont'd) Mass of major isotopes in the ejecta after all 
short-lived radioactive isotopes have decayed. The 
isotope masses are in units of solar mass.}
\label{table:Isotope3b}
\begin{tabular}{|c| c c c|}
\hline
Isotopes &  800-000-1-c3-1 & 800-000-1-b1-1 & 900-000-1-c3-1 \\ \hline

$^{50}$Cr & $3.33 \times 10^{-4}$ & $3.69 \times 10^{-4}$ & $3.50 \times 10^{-4}$ \\
 $^{52}$Cr & $1.57 \times 10^{-2}$ & $2.82 \times 10^{-2}$ & $1.94 \times 10^{-2}$ \\
 $^{53}$Cr & $1.76 \times 10^{-3}$ & $3.11 \times 10^{-3}$ & $2.6 \times 10^{-3}$ \\
 $^{54}$Cr & $6.65 \times 10^{-3}$ & $3.59 \times 10^{-2}$ & $1.2 \times 10^{-2}$ \\
 $^{55}$Mn & $1.60 \times 10^{-2}$ & $2.14 \times 10^{-2}$ & $1.74 \times 10^{-2}$ \\
 $^{54}$Fe & $1.50 \times 10^{-1}$ & $1.60 \times 10^{-1}$ & $1.55 \times 10^{-1}$ \\
 $^{56}$Fe & $4.97 \times 10^{-1}$ & $5.0 \times 10^{-1}$ & $5.9 \times 10^{-1}$ \\
 $^{57}$Fe & $1.28 \times 10^{-2}$ & $1.45 \times 10^{-2}$ & $1.34 \times 10^{-2}$ \\
 $^{58}$Fe & $1.61 \times 10^{-2}$ & $7.12 \times 10^{-2}$ & $2.40 \times 10^{-2}$ \\
 $^{60}$Fe & $2.38 \times 10^{-6}$ & $4.87 \times 10^{-5}$ & $6.4 \times 10^{-6}$ \\
 $^{59}$Co & $1.11 \times 10^{-3}$ & $1.89 \times 10^{-3}$ & $1.30 \times 10^{-3}$ \\
 $^{58}$Ni & $6.6 \times 10^{-2}$ & $6.49 \times 10^{-2}$ & $6.29 \times 10^{-2}$ \\
 $^{60}$Ni & $1.15 \times 10^{-2}$ & $1.55 \times 10^{-2}$ & $1.30 \times 10^{-2}$ \\
 $^{61}$Ni & $9.27 \times 10^{-5}$ & $2.89 \times 10^{-4}$ & $1.25 \times 10^{-4}$ \\
 $^{62}$Ni & $4.41 \times 10^{-3}$ & $1.57 \times 10^{-2}$ & $6.41 \times 10^{-3}$ \\
 $^{64}$Ni & $2.52 \times 10^{-4}$ & $3.32 \times 10^{-3}$ & $4.77 \times 10^{-4}$ \\
 $^{63}$Cu & $6.2 \times 10^{-6}$ & $4.41 \times 10^{-5}$ & $9.63 \times 10^{-6}$ \\
 $^{65}$Cu & $7.99 \times 10^{-7}$ & $7.29 \times 10^{-6}$ & $1.37 \times 10^{-6}$ \\
 $^{64}$Zn & $7.72 \times 10^{-6}$ & $5.86 \times 10^{-6}$ & $7.32 \times 10^{-6}$ \\
 $^{66}$Zn & $3.24 \times 10^{-6}$ & $1.37 \times 10^{-4}$ & $1.0 \times 10^{-5}$ \\
 $^{67}$Zn & $1.75 \times 10^{-7}$ & $3.68 \times 10^{-6}$ & $4.27 \times 10^{-7}$ \\
 $^{68}$Zn & $1.4 \times 10^{-7}$ & $1.2 \times 10^{-6}$ & $1.68 \times 10^{-7}$ \\
 $^{70}$Zn & $1.5 \times 10^{-8}$ & $3.42 \times 10^{-7}$ & $3.15 \times 10^{-8}$ \\ \hline
 
 \end{tabular}
\end{center}
\end{table}
 
 \begin{table}

\begin{center}
\caption{Masses of the radioactive isotopes in the ejecta after
the explosion. The isotope masses are in units of solar mass.}
\label{table:Decay3}
\begin{tabular}{|c| c c c|}
\hline
Isotopes &  800-000-1-c3-1 & 800-000-1-b1-1 & 900-000-1-c3-1 \\ \hline

 $^{22}$Na & $1.71 \times 10^{-10}$ & $1.27 \times 10^{-10}$ & $1.20 \times 10^{-10}$ \\
 $^{26}$Al & $2.76 \times 10^{-9}$ & $1.46 \times 10^{-9}$ & $2.58 \times 10^{-9}$ \\
 $^{39}$Ar & $8.96 \times 10^{-13}$ & $2.43 \times 10^{-12}$ & $1.25 \times 10^{-12}$ \\
 $^{40}$K & $9.65 \times 10^{-13}$ & $1.43 \times 10^{-12}$ & $1.35 \times 10^{-12}$ \\
 $^{41}$Ca & $1.54 \times 10^{-7}$ & $1.19 \times 10^{-7}$ & $1.74 \times 10^{-7}$ \\
 $^{44}$Ti & $3.27 \times 10^{-6}$ & $2.82 \times 10^{-6}$ & $3.22 \times 10^{-6}$ \\
 $^{48}$V & $5.33 \times 10^{-9}$ & $5.25 \times 10^{-9}$ & $5.40 \times 10^{-9}$ \\
 $^{49}$V & $2.36 \times 10^{-7}$ & $3.37 \times 10^{-7}$ & $2.82 \times 10^{-7}$ \\
 $^{53}$Mn & $8.60 \times 10^{-4}$ & $1.12 \times 10^{-3}$ & $9.63 \times 10^{-4}$ \\
 $^{60}$Fe & $3.40 \times 10^{-5}$ & $6.84 \times 10^{-4}$ & $8.50 \times 10^{-5}$ \\
 $^{56}$Co & $1.54 \times 10^{-4}$ & $1.72 \times 10^{-4}$ & $1.63 \times 10^{-4}$ \\
 $^{57}$Co & $2.32 \times 10^{-3}$ & $2.89 \times 10^{-3}$ & $2.50 \times 10^{-3}$ \\
 $^{60}$Co & $1.82 \times 10^{-6}$ & $7.75 \times 10^{-6}$ & $2.73 \times 10^{-6}$ \\
 $^{56}$Ni & $3.83 \times 10^{-1}$ & $3.22 \times 10^{-1}$ & $3.70 \times 10^{-1}$ \\
 $^{57}$Ni & $9.47 \times 10^{-3}$ & $8.72 \times 10^{-3}$ & $9.43 \times 10^{-3}$ \\
 $^{59}$Ni & $7.38 \times 10^{-4}$ & $9.3 \times 10^{-4}$ & $7.91 \times 10^{-4}$ \\
 $^{63}$Ni & $4.95 \times 10^{-6}$ & $3.65 \times 10^{-5}$ & $7.92 \times 10^{-6}$ \\ \hline

\end{tabular}
\end{center}
\end{table}

\end{document}